\documentclass[aps,prl,showpacs, floatfix]{revtex4}
\usepackage{amsmath}
\usepackage{graphicx}

\begin{document}

\title{The rheology of dense, polydisperse granular fluids under shear.}
\author{James F. \surname{Lutsko}}
\affiliation{Center for Nonlinear Phenomena and Complex Systems\\
Universit\'{e} Libre de Bruxelles\\
Campus Plaine, CP 231, 1050 Bruxelles, Belgium}
\email{jlutsko@ulb.ac.be}
\date{\today }

\begin{abstract}
The solution of the Enskog equation for the one-body velocity distribution
of a moderately dense, arbitrary mixture of inelastic hard spheres
undergoing planar shear flow is described. A generalization of the Grad
moment method , implemented by means of a novel generating function
technique, is used so as to avoid any assumptions concerning the size of the
shear rate. The result is illustrated by using it to calculate the pressure,
normal stresses and shear viscosity of a model polydisperse granular fluid
in which grain size, mass and coefficient of restitution varies amoungst the
grains. The results are compared to a numerical solution of the Enskog
equation as well as molecular dynamics simulations. Most bulk properties are
well described by the Enskog theory and it is shown that the generalized
moment method is more accurate than the simple (Grad) moment method.
However, the description of the distribution of temperatures in the mixture
predicted by Enskog theory does not compare well to simulation, even at
relatively modest densities.
\end{abstract}

\maketitle

\section*{Introduction}

Granular systems under rapid flow can be modelled as a fluid of inelastic
hard spheres for which a variety of theoretical and simulation methods may
be use to explore and understand the rich phenomenology that they exhibit%
\cite{GranPhysToday},\cite{GranRMP},\cite{CampbellReview}. Of particular
interest are sheared granular flows, in which the velocity in the direction
of flow varies with position in an orthogonal direction, due to their
practical relevance, accessibility to experiment and theoretical elegance.
For a steady rate of shearing, such systems typically reach a steady state
in which viscous heating, due to the shear, balances collisional cooling,
due to the inelastic collisions, thus giving an example of a steady state
nonequilibrium system. A number of papers have discussed the rheology of
single component sheared granular fluids\cite{JenkinsRichman},\cite%
{SelaGoldhirsch},\cite{ChouRichman1},\cite{ChouRichman2} in which all
particles are mechanically identical. However, all real fluids can be
expected to contain a distribution of particle sizes and degrees of
inelasticity. The purpose of this paper is to extend these studies to dense
fluids composed of an arbitrary mixture of particle sizes and inelasticities.

The usual model for granular fluids consists of hard spheres which lose
energy when they collide. Different models for the energy loss are possible
and here, attention will be restricted to the simplest case in which the
energy loss is proportional to the contribution to the kinetic energy of the
normal velocities in the rest frame. This model is amenable to the same
theoretical tools used to study elastic hard-sphere systems. In particular,
it is possible to construct the exact Liouville equation describing the time
evolution of the N-body distribution function\cite{HCSLiouville},\cite%
{ErnstHCS},\cite{LutskoJCP} from which the Enskog and Boltzmann approximate
kinetic equations follow. The latter are closed equations for the one-body
distribution: the Enskog equation involves only the assumption of
''molecular chaos''\cite{Lutsko96},\cite{LutskoHCS}, while the Boltzmann
equation is its low-density limit. One of the attractions of the hard-sphere
models is the existence of the Enskog equation which allows for the
description of finite density fluids outside the domain of the validity of
the Boltzmann equation. Since one of the purposes of the present work is to
provide a foundation for the study of transport properties in realistic
systems, the Enskog equation is used as a starting point. The price paid for
this is that the results obtained must be evaluated numerically, but as
discussed below, it is always possible, and quite trivial, to take the
Boltzmann limit of the final expressions and to thereby proceed analytically
in this special case.

The specific state studied here is that of uniform shear flow in which the
flow is described by a velocity field $\overrightarrow{v}\left( 
\overrightarrow{r}\right) =\overleftrightarrow{a}\cdot \overleftrightarrow{r}%
=ay\widehat{x}$ where $a$ is the shear rate. As discussed below, this flow
admits of a uniform state with spatially constant density and temperature.
The shear rate, temperature and degree of inelasticity are related in the
steady state by the requirement that the viscous heating and collisional
cooling balance. Although the Enskog equation could be solved perturbatively
in the shear rate, it is difficult to carry out the expansion to
sufficiently high order to describe physically interesting effects such as
normal stresses and shear thinning (although see ref.\cite{SelaGoldhirsch}
where this is done for the Boltzmann equation). It is instead simpler to use
a moment method without making any assumptions about the smallness of the
shear rate as has been done for elastic hard spheres\cite{Lutsko_EnskogPRL},%
\cite{LutskoEnskog} and for the Boltzmann equation for sheared binary fluids%
\cite{GarzoDiff}. A similarly method that has been used is the so-called
''Generalized Maxwellian'' of Chou and Richman\cite{ChouRichman1},\cite%
{ChouRichman2}. In fact, as shown below, these two methods can actually be
viewed as special cases of a generalized moment expansion about an arbitrary
Gaussian state.

An objection to using the moment method with the Enskog equation is that the
calculations are technically difficult. In particular, the collision
integrals which occur in the study of sheared fluids can be challenging to
evaluate even in the case of the simpler Boltzmann theory (see, e.g. ref.%
\cite{GarzoDiff}). One contribution of this work is to present a generating
function technique which greatly simplifies the calculations. It is shown
that all integrals of interest can be obtained by differentiating and taking
appropriate limits of a single generating function which itself simply
involves the evaluation of a few Gaussian integrals. With this technique, it
is straightforward to evaluate all results for the most general case of
differing particle sizes and coefficients of restitution in $D$-dimensions.
The final results in fact turn out to be as simple in structure as the
equivalent model of sheared single-component elastic hard spheres\cite%
{Lutsko_EnskogPRL},\cite{LutskoEnskog}.

A question which has aroused considerable interest is the degree to which
mean-field theories, like the Enskog-Boltzmann kinetic theory, are
applicable to granular systems\cite{TanGold}. This provides another reason
for studying the particular case of USF as it is possible to simulate USF
with the use of modified periodic boundaries by means of the Lees-Edwards
simulation technique\cite{LeesEdwards} so as to compare the approximate
kinetic theory to numerical experiments. Furthermore, the Enskog equation
may be solved numerically by means of the Direct Simulation Monte Carlo
(DSMC) method\cite{DSMC}. Both methods are used here in order to elucidate
the accuracy of the analytic calculations as a method of solving the Enskog
equation and of the Enskog approximation itself compared to simulation.

The organization of this paper is as follows. In Section II, the Enskog
theory is reviewed and the moment method for solving it is presented. It is
shown how the moment method can be extended to allow for an arbitrary
Gaussian reference state and the generating function formalism is
introduced. The lowest non-trivial moment solutions of the Enskog equation
are then described and used to calculate the pressure tensor for USF which
thus describes the pressure, normal stresses and shear viscosity of the
fluid in the steady state. In Section III, the solution of the moment
equations is compared to DSMC and MD results for a model polydisperse fluid
for a range of applied shear rates. The generalized moment method is shown
to be superior to the simple (Grad) moment method and the Enskog theory is
shown to give a good description of many rheological properties over a wide
range of densities. However, the description of the distribution of
temperatures as a function of grain size is shown to poor raising questions
as to the accuracy of Enskog theory. The paper concludes with a discussion
of the use and applicability of the results.

\section{Theory}

\subsection{Enskog approximation}

Consider a system of $N$ grains which are modeled as hard spheres. Each
sphere is described by a position, $\overrightarrow{q}$, velocity, $%
\overrightarrow{v}$, and a species label $r$. A grain of species $r$ has
mass $m_{r}$ while two grains of species $r$ and $s$ collide when they are
separated by a distance $\sigma _{rs}$. This array of hard-sphere diameters
may be specified arbitrarily but an important special case is that of
additive hard sphere diameters wherein each species has a fixed diameter $%
\sigma _{r}$ and $\sigma _{rs}=\frac{1}{2}\left( \sigma _{r}+\sigma
_{s}\right) $. When grains $i$ and $j$ collide, their relative velocity
after collision, $\overrightarrow{v}_{ij}^{\prime }=\overrightarrow{v}%
_{i}^{\prime }-\overrightarrow{v}_{j}^{\prime }$ is given by%
\begin{equation}
\overrightarrow{v}_{ij}^{\prime }=\overrightarrow{v}_{ij}-\left( 1+\alpha
_{r_{i}r_{j}}\right) \widehat{q}_{ij}\left( \widehat{q}_{ij}\cdot 
\overrightarrow{v}_{ij}\right)
\end{equation}%
where $\alpha _{rs}$ is the coefficient of restitution for collisions
between grains of species $r$ and $s$. The collisions are elastic if $\alpha
_{rs}=1$ while $\alpha _{rs}<1$ leads to an irreversible loss of energy in
each collision. Between collisions, the grains stream freely so that their
velocities are constant. This model is a particular case of endothermic
hard-sphere collisions in which the energy loss is proportional to the
kinetic energy along the line of collision in the rest frame $E_{ij}^{\prime
}-E_{ij}=-\left( 1-\alpha _{r_{i}r_{j}}^{2}\right) \frac{1}{2}\mu
_{r_{i}r_{j}}\left( \widehat{q}_{ij}\cdot \overrightarrow{v}_{ij}\right)
^{2} $ where the reduced mass is $\mu _{rs}=m_{r}m_{s}/\left(
m_{r}+m_{s}\right) $. Finally, it is useful to define the momentum exchange
operator for any function of the relative velocities $g\left( 
\overrightarrow{v}_{ij}\right) $ as%
\begin{equation}
\widehat{b}_{ij}g\left( \overrightarrow{v}_{ij}\right) =g\left( 
\overrightarrow{v}_{ij}^{\prime }\right) =g\left( \overrightarrow{v}%
_{ij}-\left( 1+\alpha _{r_{i}r_{j}}\right) \widehat{q}_{ij}\left( \widehat{q}%
_{ij}\cdot \overrightarrow{v}_{ij}\right) \right)  \label{2}
\end{equation}%
and all other velocities are left unchanged. Its inverse is 
\begin{equation}
\widehat{b}_{ij}^{-1}g\left( \overrightarrow{v}_{ij}\right) =g\left( 
\overrightarrow{v}_{ij}-\frac{1+\alpha _{r_{i}r_{j}}}{\alpha _{r_{i}r_{j}}}%
\widehat{q}_{ij}\left( \widehat{q}_{ij}\cdot \overrightarrow{v}_{ij}\right)
\right) .  \label{3}
\end{equation}

The statistical properties of the system are determined by one and two body
distribution functions, $f_{r}\left( \overrightarrow{q},\overrightarrow{v}%
;t\right) $ and $f_{r_{1}r_{2}}\left( \overrightarrow{q}_{1},\overrightarrow{%
v}_{1};\overrightarrow{q}_{2},\overrightarrow{v}_{2};t\right) $
respectively. The former gives the probability density to of finding a grain
of species $r$ with the given position and velocity at time $t$ and the
latter gives the joint probability for two grains. The one-body distribution
satisfies an exact equation (the first of the
Born-Bogoliubov-Green-Kirkwood-Yvon (BBGKY) hierarchy)%
\begin{equation}
\left( \frac{\partial }{\partial t}+\overrightarrow{v}_{1}\cdot \frac{%
\partial }{\partial \overrightarrow{q}_{1}}\right) f_{r_{1}}\left( 
\overrightarrow{q}_{1},\overrightarrow{v}_{1};t\right) =-\sum_{r_{1}}\int d%
\overrightarrow{q}_{2}d\overrightarrow{v}_{2}\overline{T}_{-}\left(
12\right) f_{r_{1}r_{2}}\left( \overrightarrow{q}_{1},\overrightarrow{v}_{1};%
\overrightarrow{q}_{2},\overrightarrow{v}_{2};t\right)  \label{4}
\end{equation}%
where the binary collision operator is%
\begin{equation}
\overline{T}_{-}\left( ij\right) =-\delta \left( q_{ij}-\sigma
_{r_{i}r_{j}}\right) \left[ \frac{1}{\alpha _{r_{i}r_{j}}}\widehat{b}%
_{ij}^{-1}-1\right] \Theta \left( -\overrightarrow{v}_{ij}\cdot \widehat{q}%
_{ij}\right) \overrightarrow{v}_{ij}\cdot \widehat{q}_{ij}  \label{5}
\end{equation}%
where $\Theta \left( x\right) =1$ if $x>0$ and $0$ otherwise. A similar
equation relates the two-body distribution to the three body distribution,
and so on. The Enskog approximation results from noting that the combination 
$\delta \left( q_{12}-\sigma _{r_{1}r_{2}}\right) \Theta \left( -%
\overrightarrow{v}_{12}\cdot \widehat{q}_{12}\right) f_{r_{1}r_{2}}\left( 
\overrightarrow{q}_{1},\overrightarrow{v}_{1};\overrightarrow{q}_{2},%
\overrightarrow{v}_{2};t\right) $ picks out the pre-collisional part of the
distribution and assuming that, \emph{prior to collision and at the moment
of contact}, the grains are uncorrelated. The specific assumption is that%
\begin{eqnarray}
&&\delta \left( q_{12}-\sigma _{r_{1}r_{2}}\right) \Theta \left( -%
\overrightarrow{v}_{12}\cdot \widehat{q}_{12}\right) f_{r_{1}r_{2}}\left( 
\overrightarrow{q}_{1},\overrightarrow{v}_{1};\overrightarrow{q}_{2},%
\overrightarrow{v}_{2};t\right)  \label{6} \\
&\simeq &\delta \left( q_{12}-\sigma _{r_{1}r_{2}}\right) \Theta \left( -%
\overrightarrow{v}_{12}\cdot \widehat{q}_{12}\right) f_{r_{1}}\left( 
\overrightarrow{q}_{1},\overrightarrow{v}_{1};t\right) f_{rr_{2}}\left( 
\overrightarrow{q}_{2},\overrightarrow{v}_{2};t\right) \chi
_{r_{1}r_{2}}\left( \overrightarrow{q}_{1},\overrightarrow{q}_{2};t\right) 
\notag
\end{eqnarray}%
where the term $\chi _{r_{1}r_{2}}\left( \overrightarrow{q}_{1},%
\overrightarrow{q}_{2};t\right) ,$the spatial pair distribution function
(pdf), accounts for spatial correlations as exist even in equilibrium. If it
is taken to be the local-equilibrium functional of the nonequilibrium
densities then the approximation is completely specified and is known as the
Generalized Enskog Approximation\cite{RET}.

\subsection{Hydrodynamic fields}

The local hydrodynamic fields of partial number densities $n_{r}\left( 
\overrightarrow{q},t\right) $, local velocity $\overrightarrow{u}\left( 
\overrightarrow{q},t\right) $ , and temperature $T\left( \overrightarrow{q}%
,t\right) $ are defined as%
\begin{eqnarray}
n_{r}\left( \overrightarrow{q},t\right) &=&\int d\overrightarrow{v}%
\;f_{r}\left( \overrightarrow{q},\overrightarrow{v};t\right)  \label{7} \\
\rho \left( \overrightarrow{q},t\right) \overrightarrow{u}\left( 
\overrightarrow{q},t\right) &=&\sum_{r}\int d\overrightarrow{v}\;m_{r}%
\overrightarrow{v}f_{r}\left( \overrightarrow{q},\overrightarrow{v};t\right)
\notag \\
\frac{D}{2}k_{B}T\left( \overrightarrow{q},t\right) &=&\sum_{r}\int d%
\overrightarrow{v}\;\frac{1}{2}m_{r}v^{2}f_{r}\left( \overrightarrow{q},%
\overrightarrow{v};t\right)  \notag \\
&&\;  \notag
\end{eqnarray}%
where $\rho \left( \overrightarrow{q},t\right) =\sum_{r}m_{r}n_{r}\left( 
\overrightarrow{q},t\right) $ is the total mass density, $D$ is the
dimensionality of the system and $k_{B}$ is Boltzmann's constant. The exact
time evolution of these fields follows from Eq.(\ref{4}) which gives\cite%
{LutskoJCP}%
\begin{eqnarray}
\frac{\partial }{\partial t}n_{r}+\overrightarrow{\nabla }\cdot \left( 
\overrightarrow{u}n_{r}\right) +\overrightarrow{\nabla }\cdot 
\overrightarrow{j}_{r}^{K} &=&0  \label{8} \\
\frac{\partial }{\partial t}\overrightarrow{u}+\overrightarrow{u}\cdot 
\overrightarrow{\nabla }\overrightarrow{u}+\rho ^{-1}\overrightarrow{\nabla }%
\cdot \overleftrightarrow{P} &=&0  \notag \\
\left( \frac{\partial }{\partial t}+\overrightarrow{u}\cdot \overrightarrow{%
\nabla }\right) T-\frac{T}{n}\overrightarrow{\nabla }\cdot \sum_{l}%
\overrightarrow{j}_{l}^{K}+\frac{2}{Dnk_{B}}\left[ \overleftrightarrow{P}:%
\overrightarrow{\nabla }\overrightarrow{u}+\overrightarrow{\nabla }\cdot 
\overrightarrow{q}\right] &=&\frac{2}{Dnk_{B}}\xi  \notag
\end{eqnarray}%
with the number current%
\begin{equation}
\overrightarrow{j}_{r}^{K}=\int d\overrightarrow{v}_{1}\;f_{r}\left( 
\overrightarrow{r},\overrightarrow{v}_{1},t\right) \overrightarrow{V}_{1}
\label{9}
\end{equation}%
where $\overrightarrow{V}_{1}=\overrightarrow{v}_{1}-\overrightarrow{u}%
\left( \overrightarrow{q}_{1},t\right) $, the pressure tensor $%
\overleftrightarrow{P}=\overleftrightarrow{P}^{K}+\overleftrightarrow{P}^{V}$
with kinetic and collisional contributions%
\begin{eqnarray}
\overleftrightarrow{P}^{K}\left( \overrightarrow{r},t\right)
&=&\sum_{r}m_{r}\int d\overrightarrow{v}_{1}\;f_{r}\left( \overrightarrow{r},%
\overrightarrow{v}_{1},t\right) \overrightarrow{V}_{1}\overrightarrow{V}_{1}
\label{10} \\
\overleftrightarrow{P}^{V}\left( \overrightarrow{r},t\right) &=&\frac{1}{4}%
\sum_{r_{1}r_{2}}\mu _{r_{1i}r_{2}}\left( 1+\alpha _{r_{1}r_{2}}\right) \int
dx_{1}dx_{2}\;\widehat{q}_{12}\overrightarrow{q}_{12}\left( \widehat{q}%
_{12}\cdot \overrightarrow{v}_{12}\right) ^{2}\delta \left( q_{12}-\sigma
_{l_{1}l_{2}}\right) \Theta \left( -\widehat{q}_{12}\cdot \overrightarrow{v}%
_{12}\right)  \notag \\
&&\times f_{r_{1}r_{2}}\left( x_{1},x_{2};t\right) \int_{0}^{1}dx\;\delta
\left( \overrightarrow{r}-x\overrightarrow{q}_{1}-\left( 1-x\right) 
\overrightarrow{q}_{2}\right) ,  \notag
\end{eqnarray}%
the heat flux vector $\overrightarrow{q}=\overrightarrow{q}^{K}+%
\overrightarrow{q}^{V}+\overrightarrow{q}^{\delta E}$ with%
\begin{eqnarray}
\overrightarrow{q}^{K}\left( \overrightarrow{r},t\right) &=&\sum_{r}\frac{1}{%
2}m_{r}\int d\overrightarrow{v}_{1}\;f_{r}\left( \overrightarrow{r},%
\overrightarrow{v}_{1},t\right) \overrightarrow{V}_{1}V_{1}^{2}  \label{11}
\\
\overrightarrow{q}^{V}\left( \overrightarrow{r},t\right) &=&\frac{1}{2}%
\sum_{r_{1}r_{2}}\mu _{r_{1i}r_{2}}\left( 1+\alpha _{r_{1}r_{2}}\right) \int
dx_{1}dx_{2}\;\overrightarrow{q}_{12}\left( \widehat{q}_{12}\cdot 
\overrightarrow{v}_{12}\right) ^{2}\delta \left( q_{12}-\sigma
_{l_{1}l_{2}}\right) \Theta \left( -\widehat{q}_{12}\cdot \overrightarrow{v}%
_{12}\right)  \notag \\
&&\times f_{r_{1}r_{2}}\left( x_{1},x_{2};t\right) \left( \overrightarrow{V}-%
\overrightarrow{u}\left( \overrightarrow{q}_{1},t\right) \right) \cdot 
\widehat{q}_{12}\int_{0}^{1}dx\;\delta \left( \overrightarrow{r}-x%
\overrightarrow{q}_{1}-\left( 1-x\right) \overrightarrow{q}_{2}\right) 
\notag \\
\overrightarrow{q}^{\delta E}\left( \overrightarrow{r},t\right) &=&-\frac{1}{%
4}\sum_{r_{1}r_{2}}\left( 1-\alpha _{r_{i}r_{j}}^{2}\right) \frac{1}{2}\mu
_{r_{1i}r_{2}}\frac{m_{r_{1}}-m_{r_{2}}}{m_{r_{1}}+m_{r_{2}}}\int
dx_{1}dx_{2}\;\overrightarrow{q}_{12}\left( \widehat{q}_{12}\cdot 
\overrightarrow{v}_{12}\right) ^{3}\delta \left( q_{12}-\sigma
_{l_{1}l_{2}}\right) \Theta \left( -\widehat{q}_{12}\cdot \overrightarrow{v}%
_{12}\right)  \notag \\
&&\times f_{r_{1}r_{2}}\left( x_{1},x_{2};t\right) \int_{0}^{1}dx\;\delta
\left( \overrightarrow{r}-x\overrightarrow{q}_{1}-\left( 1-x\right) 
\overrightarrow{q}_{2}\right)  \notag
\end{eqnarray}%
where the center of mass velocity $\overrightarrow{V}=\frac{m_{r_{1}}%
\overrightarrow{v}_{1}+m_{r_{2}}\overrightarrow{v}_{2}}{m_{r_{1}}+m_{r_{2}}}$%
, and the energy sink term 
\begin{eqnarray}
\xi \left( \overrightarrow{r},t\right) &=&\frac{1}{4}\sum_{r_{1}r_{2}}\left(
1-\alpha _{r_{i}r_{j}}^{2}\right) \mu _{r_{1i}r_{2}}\int
dx_{1}dx_{2}\;\left( \overrightarrow{q}_{12}\cdot \overrightarrow{v}%
_{12}\right) ^{3}\delta \left( q_{12}-\sigma _{l_{1}l_{2}}\right) \Theta
\left( -\widehat{q}_{12}\cdot \overrightarrow{v}_{12}\right)  \label{12} \\
&&\times f_{r_{1}r_{2}}\left( x_{1},x_{2};t\right) \delta \left( 
\overrightarrow{r}-\overrightarrow{q}_{1}\right) ,  \notag
\end{eqnarray}%
Using the approximation given in Eq.(\ref{6}) gives expressions for the
fluxes and the heat sink which only require knowledge of the one-body
distribution function.

\subsection{Uniform Shear Flow}

The Enskog equation is indeterminate until some boundary condition is
specified. If periodic boundary conditions are imposed, then it is easy to
see that the Enskog equation admits of a spatially homogeneous solution
which is, however, time-dependent due to the cooling resulting from the
dissipative collisions. This is the well known Homogeneous Cooling State
(HCS). Uniform Shear Flow (USF) is another simple nonequilibrium state
supported by this system. In USF, the density and temperature are spatially
homogeneous while the velocity field varies linearly with position, viz $%
\overrightarrow{u}\left( \overrightarrow{r}\right) =\overleftrightarrow{a}%
\cdot \overrightarrow{r}$ where the shear tensor, $\overleftrightarrow{a}$,
will be taken to be $a\widehat{x}\widehat{y}$ in a Cartesian coordinate
system. There is therefore a flow in the x-direction which varies linearly
in the y-direction. We hypothesize that in this case the distribution will
only depend on the peculiar velocity $f_{r}\left( \overrightarrow{q},%
\overrightarrow{v};t\right) =f_{r}\left( \overrightarrow{V}=\overrightarrow{v%
}-\overleftrightarrow{a}\cdot \overrightarrow{q};t\right) $. The Enskog
equation then becomes%
\begin{equation}
\left( \frac{\partial }{\partial t}-\overrightarrow{V}_{1}\cdot 
\overleftrightarrow{a}^{T}\cdot \frac{\partial }{\partial \overrightarrow{V}%
_{1}}\right) f_{r_{1}}\left( \overrightarrow{V}_{1};t\right)
=-\sum_{r_{2}}\int d\overrightarrow{q}_{2}d\overrightarrow{v}_{2}\;\overline{%
T}_{-}\left( 12\right) f_{r_{1}}\left( \overrightarrow{V}_{1};t\right)
f_{r_{2}}\left( \overrightarrow{V}_{2};t\right) \chi _{r_{1}r_{2}}\left( 
\overrightarrow{q}_{1},\overrightarrow{q}_{2};t\right) .  \label{13}
\end{equation}%
In fact, the linear flow field is the only one that makes the collisional
term on the right independent of position as follows from the observation
that it will be independent of position only if $\overrightarrow{u}\left( 
\overrightarrow{q}_{1}\right) -\overrightarrow{u}\left( \overrightarrow{q}%
_{2}\right) $ is a function of $\overrightarrow{q}_{12}$. (The function $%
\chi _{r_{1}r_{2}}\left( \overrightarrow{q}_{1},\overrightarrow{q}%
_{2};t\right) $, evaluated in the local equilibrium approximation, will only
depend on $\overrightarrow{q}_{12}$ if the density is uniform.) In fact, the
requirement that $\overrightarrow{u}\left( \overrightarrow{q}_{1}\right) -%
\overrightarrow{u}\left( \overrightarrow{q}_{2}\right) =\overrightarrow{U}%
\left( \overrightarrow{q}_{12}\right) $ for some field $\overrightarrow{U}%
\left( \overrightarrow{q}_{12}\right) $ is only satisfied for $%
\overrightarrow{U}\left( \overrightarrow{q}_{12}\right) =\overrightarrow{u}%
\left( \overrightarrow{q}_{12}\right) $ (demonstrated by take $%
\overrightarrow{q}_{2}=\overrightarrow{0}$). One must also have that $%
\overrightarrow{u}\left( \overrightarrow{q}_{1}\right) $ is odd, as shown by
reversing the arguments, and that $\overrightarrow{u}\left( \frac{p}{q}%
\overrightarrow{q}_{1}\right) =\frac{p}{q}\overrightarrow{u}\left( 
\overrightarrow{q}_{1}\right) ,$for arbitrary integers $p$ and $q$, as
follows by iterating with $\overrightarrow{q}_{2}=-\overrightarrow{q}_{1}$.
The only continuous function satisfying these constraints is one linear in $%
\overrightarrow{q}_{1}$, which is to say USF.

Another consequence of the assumption of a uniform state is that the density
and temperature are spatially uniform and it may be verified that the
pressure tensor, heat-flux vector, number flux and heating rate are all
spatially uniform as well. The equations for the hydrodynamic fields then
become%
\begin{eqnarray}
\frac{\partial }{\partial t}n_{r} &=&0  \label{14} \\
\frac{\partial }{\partial t}\overrightarrow{u} &=&0  \notag \\
\frac{\partial }{\partial t}T+\frac{2}{Dnk_{B}}\overleftrightarrow{P} &:&%
\overleftrightarrow{a}=\frac{2}{Dnk_{B}}\xi  \notag
\end{eqnarray}%
which shows that the hydrodynamic fields will also be independent of time if
the viscous heating, characterized by the second term on the left in the
temperature equation, balances the cooling described by the source term on
the right. It is therefore consistent to hypothesize not only a spatially
uniform solution to the Enskog equation, but that a time-independent
solution exists. In this steady state, the only two quantities having the
units of time are the temperature and the shear rate. It will therefore be
the case that the relevant dimensionless control parameter is $a^{\ast }=a%
\sqrt{\frac{\left\langle m\right\rangle \left\langle \sigma \right\rangle
^{2}}{k_{B}T}}$, where the average diameter is $\left\langle \sigma
\right\rangle =\sum_{rs}x_{r}x_{s}\sigma _{rs}$ and the average mass is $%
\left\langle m\right\rangle =\sum_{r}x_{r}m_{r}$. For a given set of
material parameters, the steady state will be unique in that the value of $%
a^{\ast }$ as determined from eq.(\ref{14}) will be independent of the shear
rate $a$ applied through the Lees-Edwards boundary conditions:\ in other
words, for a given value of $a$ the temperature will relax to a value such
that the same value of $a^{\ast }$ is always achieved.

Does this mean that once the fluid is moving according to the linear
velocity profile, it will continue to do so forever? The answer is that it
depends on the boundary conditions which have so far not been specified. The
steady-state distribution hypothesized, and the sustained USF it implies, is
only possible if the distribution function is compatible with some set of
boundary conditions. For example, if the system is bounded with rigid moving
walls, then the final distribution will depend on the detailed dynamics of
collisions of grains with the wall. Here, however, we will assume the
imposition of Lees-Edwards boundary conditions which are periodic boundaries
in the co-moving frame. A time-independent, spatially homogeneous
distribution function is in fact compatible with these boundary conditions
and they are also amenable to application in molecular dynamics computer
simulations, as discussed in more detail below.

Before turning to the construction of a spatially uniform, time-independent
solution of the Enskog equation, some comments can be made about the
generality of these results. First, the only properties of the flow state
used so far are that the flow field is a linear function of the coordinate
and that the shear tensor satisfies $\overleftrightarrow{a}\cdot 
\overleftrightarrow{a}=0$ (needed to convert the spatial derivative on the
left in Eq.(\ref{4}) into a velocity derivative in Eq.(\ref{13})). Second,
the Boltzmann equation results from taking the lowest order term in an
expansion of the integral in Eq.(\ref{13}) in terms of the hard-sphere
diameter. This results in the replacement of $\overrightarrow{v}%
_{12}\rightarrow \overrightarrow{V}_{12}$ so that, in this approximation,
the collision integral is independent of position for any choice of the flow
field.

\section{Moment Solutions to the Enskog equation}

\subsection{Moment Expansion}

In order to develop an approximate solution of the Enskog equation without
making any restrictive assumptions about the size of either the shear rate
or the degree of inelasticity, the distribution function is expanded in a
complete set of polynomials about some suitable reference state\cite%
{ChapmanCowling},\cite{Truesdell}. Here, more generally than is usually
done, the reference state will be taken to be an arbitrary Gaussian so that
the expansion takes the form%
\begin{equation}
f_{r}\left( \overrightarrow{V};\left\{ n_{s}\right\} ,T\right) =n_{r}\det
\left( \overleftrightarrow{\Gamma }\right) ^{1/2}\pi ^{-D/2}\exp \left( -%
\overrightarrow{V}\cdot \overleftrightarrow{\Gamma }^{r}\cdot 
\overrightarrow{V}\right) \left( \sum_{n=0}\frac{1}{n!}%
\sum_{I_{n}}A_{I_{n}}^{r}H_{I_{n}}\left( \sqrt{2}\overleftrightarrow{\Gamma }%
^{r1/2}\cdot \overrightarrow{V}\right) \right)  \label{M0a}
\end{equation}%
where $\overleftrightarrow{\Gamma }^{r}$ is a positive-definite, real,
symmetric matrix and an abbreviated notation is used whereby $I_{n}\equiv
i_{1}...i_{n}$. As such, it can be written, via Cholesky decomposition, as $%
\overleftrightarrow{\Gamma }^{r}=\overleftrightarrow{\Gamma }^{r1/2}\cdot
\left( \overleftrightarrow{\Gamma }^{r1/2}\right) ^{T}$ for some matrix $%
\overleftrightarrow{\Gamma }^{r1/2}$. The functions used in the expansion
are the Hermite polynomials\cite{Truesdell} given by%
\begin{equation}
H_{I_{n}}\left( \overrightarrow{c}\right) =\left( -1\right) ^{n}e^{c^{2}/2}%
\frac{\partial }{\partial c_{i_{1}}}...\frac{\partial }{\partial c_{i_{n}}}%
e^{-c^{2}/2}
\end{equation}%
so that, e.g., $H_{ij}\left( \overrightarrow{c}\right) =c_{i}c_{j}-\delta
_{ij}$. They are orthonormal in $D$-dimensions because%
\begin{equation}
\left( \frac{1}{2\pi }\right) ^{D/2}\int d\overrightarrow{c}%
\;e^{-c^{2}/2}H_{I_{n}}\left( \overrightarrow{c}\right) H_{J_{m}}\left( 
\overrightarrow{c}\right) =\delta _{mn}\sum_{P\left(
j_{1}j_{2}...j_{m}\right) }\delta _{i_{1}j_{1}}...\delta _{i_{n}j_{n}}.
\end{equation}%
so that the coefficients of the expansion are related to the velocity
moments via%
\begin{equation}
\int d\overrightarrow{V}\;f_{r}\left( \overrightarrow{V};\left\{
n_{s}\right\} ,T\right) H_{I_{n}}^{r}\left( \overrightarrow{C}_{r}\right)
=n_{r}A_{I_{n}}^{r}
\end{equation}%
where $\overrightarrow{C}_{r}=\sqrt{2}\overleftrightarrow{\Gamma }%
^{r1/2}\cdot \overrightarrow{V}.$ Evaluating equation (\ref{7}) relates the
lower order coefficients to the hydrodynamic fields as%
\begin{eqnarray}
n_{r}\left( \overrightarrow{q},t\right) &=&n_{r}A^{r}  \label{temp} \\
\rho \left( \overrightarrow{q},t\right) u_{i}\left( \overrightarrow{q}%
,t\right) &=&\sum_{r}m_{r}n_{r}A^{r}u_{i}+\sum_{r}\sum_{j}m_{r}n_{r}\left( 
\sqrt{2}\overleftrightarrow{\Gamma }^{r1/2}\right) _{ij}^{-1}A_{j}^{r} 
\notag \\
Dnk_{B}T &=&\frac{1}{2}\sum_{r}\left( m_{r}n_{r}Tr\left( \left( 
\overleftrightarrow{\Gamma }^{r}\right) ^{-1}\right) +m_{r}n_{r}Tr\left(
\left( \overleftrightarrow{\Gamma }^{r}\right) ^{-1/2}\cdot 
\overleftrightarrow{A}^{r}\cdot \left( \overleftrightarrow{\Gamma }%
^{rT}\right) ^{-1/2}\right) \right)  \notag
\end{eqnarray}%
implying $A^{r}=1$ and $\sum_{r}\sum_{j}m_{r}n_{r}\left( \sqrt{2}%
\overleftrightarrow{\Gamma }^{r1/2}\right) _{ij}^{-1}A_{j}^{r}=0$. The
kinetic contribution to the stress tensor is%
\begin{equation}
P_{ij}^{K}=\sum_{r}m_{r}\int d\overrightarrow{V}\;f_{r}\left( 
\overrightarrow{V};\left\{ n_{s}\right\} ,T\right) V_{i}V_{j}=\frac{1}{2}%
\sum_{r}m_{r}n_{r}\left( \left( \overleftrightarrow{\Gamma }^{r}\right)
^{-1/2}\cdot \left( \overrightarrow{1}+\overleftrightarrow{A}^{r}\right)
\cdot \left( \overleftrightarrow{\Gamma }^{rT}\right) ^{-1/2}\right) _{ij}.
\end{equation}%
Each species has a temperature given by%
\begin{equation}
Dn_{r}k_{B}T_{r}=m_{r}\int d\overrightarrow{V}\;f_{r}\left( \overrightarrow{V%
};\left\{ n_{s}\right\} ,T\right) V^{2}=\frac{1}{2}\left( m_{r}n_{r}Tr\left(
\left( \overleftrightarrow{\Gamma }^{r}\right) ^{-1}\right)
+m_{r}n_{r}Tr\left( \left( \overleftrightarrow{\Gamma }^{r}\right)
^{-1/2}\cdot \overleftrightarrow{A}^{r}\cdot \left( \overleftrightarrow{%
\Gamma }^{rT}\right) ^{-1/2}\right) \right) .  \label{partial_temps}
\end{equation}

Calculation of the velocity moments of Eq.(\ref{M0a}) shows that there are
actually redundant degrees of freedom in the sense that any change in $%
\overleftrightarrow{\Gamma }^{r}$ can always be compensated by changes in
the coefficients of the expansion. These redundant degrees of freedom can
most conveniently be eliminated by restricting the form of either $%
\overleftrightarrow{\Gamma }^{r}$ or $A_{I_{2}}^{r}$ although one could
imagine applying the restrictions to higher moments. There are two cases of
particular interest. In the first, $\overleftrightarrow{\Gamma }$ is left
unspecified and we set $A_{i_{1}i_{2}}^{r}=0$ so that all information about
the second moments comes from the Gaussian. This will be referred to below
as the Generalized Moment Expansion or GME. In the second case, $%
\overleftrightarrow{\Gamma }$ is specialized to a diagonal matrix by setting 
\begin{equation}
\overleftrightarrow{\Gamma }^{r}=\frac{m_{r}}{2k_{B}T_{r}}%
\overleftrightarrow{1}
\end{equation}%
where the parameters $T_{r}$ are to be determined. In this case, only the
trace of the second order coefficients is set to zero so that 
\begin{equation}
\sum_{i}A_{ii}^{r}=0.
\end{equation}%
This represents an expansion about a local equilibrium state in which the
temperatures of the different species are allowed to vary and will be
referred to below as the Simple Moment Expansion or SME. Further trade-offs
between the degrees of freedom in the reference state and those in the
moment expansion are possible, but do not appear to offer any qualitative
advantages. For example, one could use the restrictions%
\begin{eqnarray}
\overleftrightarrow{\Gamma }^{r} &=&\frac{m_{r}}{2k_{B}T}\overleftrightarrow{%
1} \\
\sum_{r}\sum_{i}A_{ii}^{r} &=&0  \notag
\end{eqnarray}%
so that the reference state is simple equilibrium. This possibility will not
be considered here, although there is nothing to rule it out in principle.
Note that in no case can we force the temperatures of the subspecies to be
equal since that would eliminate certain degrees of freedom altogether and
this would lead to inconsistencies when we use the expansion to solve the
Enskog equation.

Substituting the general expansion given in Eq.(\ref{M0a}) into the Enskog
equation, multiplying by $H_{I_{n}}^{r_{1}}\left( \sqrt{2}%
\overleftrightarrow{\Gamma }^{r_{1}1/2}\cdot \overrightarrow{V}_{1}\right) $
and integrating over velocities yields an infinite hierarchy of coupled
equations for the moments which are given explicitly in the appendix \ref%
{MomentEquations}. The $n$-th order approximation is usually taken to
consist of truncating the expansion in Eq.(\ref{M0a}) to the first $n$ terms
and using the first $n$ equations of this hierarchy to determine the
moments. The remainder of this paper will focus on the simplest nontrivial
approximations, which are in both cases the second order approximation. For
the GME, some simplification allows the moment approximation to be written
as 
\begin{equation}
\frac{\partial \Gamma _{i_{1}i_{2}}^{r_{1}-1}}{\partial t}+a\left( \Gamma
_{yi_{2}}^{r_{1}-1}\delta _{i_{1}x}+\Gamma _{yi_{1}}^{r_{1}-1}\delta
_{i_{2}x}\right) =\sum_{r_{2}}n_{r_{2}}\chi
_{r_{1}r_{2}}E_{i_{1}i_{2}}^{r_{1}r_{2}}
\label{Generalized Moment Equations}
\end{equation}%
with 
\begin{equation}
E_{i_{1}i_{2}}^{r_{1}r_{2}}=-2\pi ^{-D}\det \left( \overleftrightarrow{%
\Gamma }^{r_{1}}\overleftrightarrow{\Gamma }^{r_{2}}\right) ^{1/2}\int d%
\overrightarrow{V}_{1}d\overrightarrow{V}_{2}d\overrightarrow{q}%
_{2}\;V_{1i_{1}}V_{1i_{2}}\overline{T}_{-}(12)\exp \left( -\overrightarrow{V}%
_{1}\cdot \overleftrightarrow{\Gamma }^{r_{1}}\cdot \overrightarrow{V}_{1}-%
\overrightarrow{V}_{2}\cdot \overleftrightarrow{\Gamma }^{r_{1}}\cdot 
\overrightarrow{V}_{2}\right)  \label{E-def}
\end{equation}%
and where $\chi _{r_{1}r_{2}}\equiv \chi _{r_{1}r_{2}}\left( \overrightarrow{%
q}_{1},\overrightarrow{q}_{1}+\sigma _{rs}\widehat{q}_{12};t\right) $ is
independent of $\overrightarrow{q}_{1}$ and $\widehat{q}_{12}$ in a
homogeneous system. This system of equations suffices to determine the
matrices $\overleftrightarrow{\Gamma }^{r_{1}}$. For a given applied shear
rate $a$, the steady-state temperature, and hence the dimensionless shear
rate $a^{\ast }$, is then determined from the second moments from Eq.(\ref%
{temp}). Explicitly, the temperature is $T=\sum_{r}x_{r}T_{r}$ where the
partial temperatures are given by%
\begin{equation}
Dk_{B}T_{r}=\frac{m_{r}}{2}Tr\left( \left( \overleftrightarrow{\Gamma }%
^{r_{1}}\right) ^{-1}\right) .  \label{GME2}
\end{equation}%
This lowest order GME corresponds to the ''Generalized Maxwellian''\
approximation of Chou et al\cite{ChouRichman1},\cite{ChouRichman2}. For the
SME, the moment equations are%
\begin{eqnarray}
&&\frac{\partial }{\partial t}A_{i_{1}i_{2}}^{r_{1}}+\frac{\partial \ln
T_{r_{1}}}{\partial t}\left( A_{i_{1}i_{2}}^{r_{1}}+\delta
_{i_{1}i_{2}}\right) +a\left( \delta _{xi_{1}}A_{i_{2}y}^{r_{1}}+\delta
_{xi_{2}}A_{i_{1}y}^{r_{1}}+\delta _{xi_{1}}\delta _{i_{2}y}+\delta
_{xi_{2}}\delta _{i_{1}y}\right)  \label{SME} \\
&=&\sum_{r_{2}}n_{r_{2}}\chi _{r_{1}r_{2}}\left[ B_{i_{1}i_{2}}^{r_{1}r_{2}}+%
\frac{1}{2}\sum_{j_{1}j_{2}}\left(
C_{i_{1}i_{2},j_{1}j_{2}}^{r_{1}r_{2}}A_{j_{1}j_{2}}^{r_{1}}+D_{i_{1}i_{2},j_{1}j_{2}}^{r_{1}r_{2}}A_{j_{1}j_{2}}^{r_{2}}\right) %
\right]  \notag \\
Tr\left( \overleftrightarrow{A}^{r_{1}}\right) &=&0  \notag
\end{eqnarray}%
with%
\begin{eqnarray}
B_{i_{1}i_{2}}^{r_{1}r_{2}} &=&-\int d\overrightarrow{V}_{1}d\overrightarrow{%
V}_{2}d\overrightarrow{q}_{2}\;H_{i_{1}i_{2}}\left( \overrightarrow{C}%
_{1}\right) \overline{T}_{-}(12)\Phi _{r_{1}}\left( \overrightarrow{V}%
_{1};T_{r_{1}}\right) \Phi _{r_{2}}\left( \overrightarrow{V}%
_{2};T_{r_{2}}\right)  \label{BCD-def} \\
C_{i_{1}i_{2},j_{1}j_{2}}^{r_{1}r_{2}} &=&-\int d\overrightarrow{V}_{1}d%
\overrightarrow{V}_{2}d\overrightarrow{q}_{2}\;H_{i_{1}i_{2}}\left( 
\overrightarrow{C}_{1}\right) \overline{T}_{-}(12)\Phi _{r_{1}}\left( 
\overrightarrow{V}_{1};T_{r_{1}}\right) H_{j_{1}j_{2}}\left( \overrightarrow{%
C}_{1}\right) \Phi _{r_{2}}\left( \overrightarrow{V}_{2};T_{r_{2}}\right) 
\notag \\
D_{i_{1}i_{2},j_{1}j_{2}}^{r_{1}r_{2}} &=&-\int d\overrightarrow{V}_{1}d%
\overrightarrow{V}_{2}d\overrightarrow{q}_{2}\;H_{i_{1}i_{2}}\left( 
\overrightarrow{C}_{1}\right) \overline{T}_{-}(12)\Phi _{r_{1}}\left( 
\overrightarrow{V}_{1};T_{r_{1}}\right) \Phi _{r_{2}}\left( \overrightarrow{V%
}_{2};T_{r_{2}}\right) H_{j_{1}j_{2}}\left( \overrightarrow{C}_{2}\right) 
\notag \\
\overrightarrow{C}_{j} &=&\left( \frac{m_{r_{j}}}{k_{B}T_{r_{j}}}\right)
^{1/2}\overrightarrow{V}_{j}.  \notag \\
\Phi _{r}\left( \overrightarrow{V}_{1};T_{r}\right) &=&\left( \frac{m_{r}}{%
2k_{B}T_{r}\pi }\right) ^{D/2}\exp \left( -\frac{m_{r}}{2k_{B}T_{r}}%
V_{1}^{2}\right)  \notag
\end{eqnarray}%
The first of Eqs.(\ref{SME}) are a set of linear equations for the
coefficients $A_{i_{1}i_{2}}^{r_{1}}$ whereas the second can be thought of
as a set of constraints that serve to determine the partial temperatures.

Finally, one problem with the moment expansion in general is that the
truncated distributions are not necessarily positive definite. An ad hoc
procedure to rectify this problem is to resum the truncated moment expansion
so that one writes, in the general case, 
\begin{eqnarray}
f_{r}\left( \overrightarrow{V};\left\{ n_{s}\right\} ,T\right) &=&n_{r}\det
\left( \overleftrightarrow{\Gamma }\right) ^{1/2}\pi ^{-D/2}\exp \left( -%
\overrightarrow{V}\cdot \overleftrightarrow{\Gamma }^{r}\cdot 
\overrightarrow{V}\right) \left( 1+\sum_{n}\frac{1}{n!}%
\sum_{I_{n}}A_{I_{n}}^{r}H_{I_{n}}\left( \sqrt{2}\overleftrightarrow{\Gamma }%
^{r1/2}\cdot \overrightarrow{V}\right) \right) \\
&=&n_{r}Z^{-1}\pi ^{-D/2}\exp \left( -\overrightarrow{V}\cdot 
\overleftrightarrow{\Gamma }^{r}\cdot \overrightarrow{V}+\sum_{n}\frac{1}{n!}%
\sum_{I_{n}}\overline{A}_{I_{n}}^{r}H_{I_{n}}\left( \sqrt{2}%
\overleftrightarrow{\Gamma }^{r1/2}\cdot \overrightarrow{V}\right) \right) 
\notag
\end{eqnarray}%
where the new coefficients $\overline{A}_{I_{n}}^{r}$ are chosen so that the
two series agree, term by term, up to the desired order. In general, the
normalization constant $Z$ must also be determined. For the second order
GME, this is not an issue since the approximate distribution is Gaussian.
For the SME, one has that%
\begin{eqnarray}
f_{r}\left( \overrightarrow{V};\left\{ n_{s}\right\} ,T\right) &\simeq
&n_{r}\left( \frac{m_{r}}{2k_{B}T_{r}}\right) ^{D/2}\pi ^{-D/2}\exp \left( -%
\frac{m_{r}}{2k_{B}T_{r}}V^{2}\right) \left( 1+\frac{1}{2!}%
\sum_{i_{1}i_{2}}A_{i_{1}i_{2}}^{r}H_{i_{1}i_{2}}\left( \sqrt{2}%
\overleftrightarrow{\Gamma }^{r1/2}\cdot \overrightarrow{V}\right) \right) \\
&\simeq &n_{r}\det \left( \overleftrightarrow{1}+\overleftrightarrow{A}%
^{r}\right) ^{-1/2}\pi ^{-D/2}\exp \left( -\frac{m_{r}}{2k_{B}T_{r}}%
\sum_{i_{1}i_{2}}\overrightarrow{V}\cdot \left( \overleftrightarrow{1}+%
\overleftrightarrow{A}^{r}\right) ^{-1}\cdot \overrightarrow{V}\right) 
\notag
\end{eqnarray}%
which is structurally the same as the GME\ except that the matrix of second
moments is determined through the linearized equations (\ref{SME})-(\ref%
{BCD-def}). Thus, in this formulation, the second order GME and SME\ are
virtually identical except for the approximations used to determine the
second moments.

\subsection{Generating Function for Collision Integrals}

All of the collision integrals that will be needed can be obtained from the
generating function 
\begin{equation}
Z_{I_{n}}^{r_{1}r_{2}}=-\pi ^{-D}\int d\overrightarrow{V}_{1}d%
\overrightarrow{V}_{2}d\overrightarrow{q}_{2}\;\left( \prod_{j=1}^{n}\left( 
\overrightarrow{q}_{12}\right) _{i_{j}}\right) \exp \left( \overrightarrow{%
\Lambda }\cdot \overrightarrow{V}_{1}\right) \overline{T}_{-}(12)\exp \left(
-\overrightarrow{V}_{1}\cdot \overleftrightarrow{\Gamma }^{r_{1}}\cdot 
\overrightarrow{V}_{1}-\overrightarrow{V}_{2}\cdot \overleftrightarrow{%
\Gamma }^{r_{2}}\cdot \overrightarrow{V}_{2}\right)  \label{M12}
\end{equation}%
by differentiating with respect to the matrices $\Gamma _{ij}^{r_{1}}$ and $%
\Gamma _{ij}^{r2}$ and the vector $\Lambda _{i}$ and taking appropriate
limits (such as $\overrightarrow{\Lambda }\rightarrow 0$). For example, by
inspection, one has that%
\begin{eqnarray}
E_{i_{1}i_{2}}^{r_{1}r_{2}} &=&2\det \left( \overleftrightarrow{\Gamma }%
^{r_{1}}\overleftrightarrow{\Gamma }^{r_{2}}\right) ^{1/2}\lim_{%
\overleftrightarrow{\Lambda }\rightarrow \overleftrightarrow{0}}\frac{%
\partial ^{2}}{\partial \Lambda _{i_{1}}\partial \Lambda _{i_{2}}}%
Z^{r_{1}r_{2}} \\
B_{i_{1}i_{2}}^{r_{1}r_{2}} &=&\frac{m_{r_{1}}}{2k_{B}T_{r_{1}}}\lim_{%
\overleftrightarrow{\Gamma }^{x}\rightarrow \frac{m_{x}}{2k_{B}T_{x}}%
\overleftrightarrow{1}}E_{i_{1}i_{2}}^{r_{1}r_{2}}  \notag \\
C_{i_{1}i_{2},j_{1}j_{2}}^{r_{1}r_{2}} &=&-\frac{m_{r_{1}}}{2k_{B}T_{r_{1}}}%
\lim_{\overleftrightarrow{\Gamma }^{x}\rightarrow \frac{m_{x}}{2k_{B}T_{x}}%
\overleftrightarrow{1}}\det \left( \overleftrightarrow{\Gamma }^{r_{1}}%
\overleftrightarrow{\Gamma }^{r_{2}}\right) ^{1/2}\left( \frac{m_{r_{1}}}{%
k_{B}T_{r_{1}}}\frac{\partial }{\partial \Gamma _{i_{j1}j_{2}}^{r_{1}}}%
+\delta _{j_{1}j_{2}}\right) \lim_{\overleftrightarrow{\Lambda }\rightarrow 
\overleftrightarrow{0}}\frac{\partial ^{2}}{\partial \Lambda
_{i_{1}}\partial \Lambda _{i_{2}}}Z^{r_{1}r_{2}}  \notag \\
&=&-\frac{1}{2}\left( \frac{m_{r_{1}}}{k_{B}T_{r_{1}}}\right) ^{2}\lim_{%
\overleftrightarrow{\Gamma }^{x}\rightarrow \frac{m_{x}}{2k_{B}T_{x}}%
\overleftrightarrow{1}}\frac{\partial }{\partial \Gamma _{j_{1}j_{2}}^{r_{1}}%
}E_{i_{1}i_{2}}^{r_{1}r_{2}}  \notag \\
D_{i_{1}i_{2},j_{1}j_{2}}^{r_{1}r_{2}} &=&-\frac{m_{r_{1}}}{2k_{B}T_{r_{1}}}%
\frac{m_{r_{2}}}{k_{B}T_{r_{2}}}\lim_{\overleftrightarrow{\Gamma }%
^{x}\rightarrow \frac{m_{x}}{2k_{B}T_{x}}\overleftrightarrow{1}}\frac{%
\partial }{\partial \Gamma _{j_{1}j_{2}}^{r_{2}}}E_{i_{1}i_{2}}^{r_{1}r_{2}}
\notag
\end{eqnarray}%
while comparison of Eqs.(\ref{10}), evaluated for a uniform system, and (\ref%
{A4}) the collisional contribution to the pressure is found to be 
\begin{equation}
P_{i_{1}i_{2}}^{V}=\frac{1}{4}\sum_{r_{1}r_{2}}n_{r_{1}}n_{r_{2}}\chi
_{r_{1}r_{2}}\det \left( \overleftrightarrow{\Gamma }^{r_{1}}%
\overleftrightarrow{\Gamma }^{r_{2}}\right) ^{1/2}m_{r_{1}}\lim_{%
\overleftrightarrow{\Lambda }\rightarrow \overleftrightarrow{0}}\frac{%
\partial }{\partial \Lambda _{i_{2}}}Z_{i_{1}}^{r_{1}r_{2}}.
\end{equation}%
The generating function is shown in appendix \ref{GeneratingFunction} to be%
\begin{equation}
Z_{I_{n}}^{r_{1}r_{2}}=\frac{1}{2}\det \left( \overleftrightarrow{\Gamma }%
^{r_{1}}\overleftrightarrow{\Gamma }^{r_{2}}\right) ^{-1/2}\sigma
_{r_{1}r_{2}}^{D-1}\int d\widehat{q}\;\left( \prod_{j=1}^{n}\sigma
_{r_{1}r_{2}}\widehat{q}_{i_{j}}\right) \left[ \widetilde{Z}%
^{r_{1}r_{2}}\left( \left( 1+\alpha _{r_{1}r_{2}}\right) \frac{\mu
_{r_{1}r_{2}}}{m_{r_{1}}}\right) -\widetilde{Z}^{r_{1}r_{2}}\left( 0\right) %
\right]
\end{equation}%
with%
\begin{eqnarray}
\widetilde{Z}^{r_{1}r_{2}}\left( x\right) &=&X_{r_{1}r_{2}}F_{1}\left( \frac{%
2w_{r_{1}r_{2}}+\overrightarrow{\Lambda }\cdot \overleftrightarrow{\Gamma }%
^{r_{1}-1}\cdot \widehat{q}_{12}}{2X_{r_{1}r_{2}}}-\frac{1}{2}x\left( 
\overrightarrow{\Lambda }\cdot \widehat{q}\right) X_{r_{1}r_{2}}\right) \\
&&\times \exp \left( \frac{1}{4}\overrightarrow{\Lambda }\cdot 
\overleftrightarrow{G}^{r_{1}r_{2}}\left( x\right) \cdot \overrightarrow{%
\Lambda }-xw_{r_{1}r_{2}}\overrightarrow{\Lambda }\cdot \widehat{q}\right) 
\notag \\
F_{n}(x) &=&-\frac{2}{\sqrt{\pi }}\int_{-\infty }^{-x}\left( u+x\right)
^{n}\exp \left( -u^{2}\right) du  \notag \\
X_{r_{1}r_{2}} &=&\sqrt{\widehat{q}\cdot \left( \overleftrightarrow{\Gamma }%
^{r_{1-1}}+\overleftrightarrow{\Gamma }^{r_{2-1}}\right) \cdot \widehat{q}} 
\notag \\
\overleftrightarrow{G}^{r_{1}r_{2}}\left( x\right) &=&\overleftrightarrow{%
\Gamma }^{r_{1}-1}+2\left( \overleftrightarrow{\Gamma }^{r_{1}}+%
\overleftrightarrow{\Gamma }^{r_{2}}\right) ^{-1}-2x\overleftrightarrow{%
\Gamma }^{r_{1}-1}\cdot \widehat{q}\widehat{q}+x^{2}X_{r_{1}r_{2}}^{2}%
\widehat{q}\widehat{q}  \notag \\
w_{r_{1}r_{2}} &=&\sigma _{r_{1}r_{2}}\widehat{q}\cdot \overleftrightarrow{a}%
\cdot \widehat{q}  \notag
\end{eqnarray}%
and, in particular%
\begin{eqnarray}
F_{0}(x) &=&\text{erf}\left( x\right) -1 \\
F_{1}(x) &=&\frac{1}{\sqrt{\pi }}e^{-x^{2}}+x\left( \text{erf}\left(
x\right) -1\right) \allowbreak \allowbreak .  \notag
\end{eqnarray}%
(The Appendix also discusses the general case of an arbitrary flow state.)
The elements needed for the second order moment equations are worked out in
Appendix \ref{Evaluations} where it is shown that%
\begin{eqnarray}
E_{i_{1}i_{2}}^{r_{1}r_{2}} &=&-\sigma _{r_{1}r_{2}}^{D-1}\left( 1+\alpha
_{r_{1}r_{2}}\right) \frac{\mu _{r_{1}r_{2}}}{m_{r_{1}}}\int d\widehat{q}%
\;X_{r_{1}r_{2}}F_{1}\left( \frac{w_{r_{1}r_{2}}}{X_{r_{1}r_{2}}}\right) %
\left[ \left( \overleftrightarrow{\Gamma }^{r_{1}-1}\cdot \widehat{q}\right)
_{i_{1}}\widehat{q}_{i_{2}}+\left( \overleftrightarrow{\Gamma }%
^{r_{1}-1}\cdot \widehat{q}\right) _{i_{2}}\widehat{q}_{i_{1}}\right]
\label{E-eval} \\
&&+\frac{1}{2}\sigma _{r_{1}r_{2}}^{D-1}\left( 1+\alpha _{r_{1}r_{2}}\right)
^{2}\left( \frac{\mu _{r_{1}r_{2}}}{m_{r_{1}}}\right) ^{2}\int d\widehat{q}\;%
\widehat{q}_{i_{1}}\widehat{q}_{i_{2}}X_{r_{1}r_{2}}^{3}\left[ \left( \frac{%
w_{r_{1}r_{2}}}{X_{r_{1}r_{2}}}\right) F_{0}\left( \frac{w_{r_{1}r_{2}}}{%
X_{r_{1}r_{2}}}\right) +F_{1}\left( \frac{w_{r_{1}r_{2}}}{X_{r_{1}r_{2}}}%
\right) \left( 2\left( \frac{w_{r_{1}r_{2}}}{X_{r_{1}r_{2}}}\right)
^{2}+2\right) \right]  \notag
\end{eqnarray}%
from which immediately follows the coefficients for the simple moment
approximation%
\begin{eqnarray}
B_{i_{1}i_{2}}^{r_{1}r_{2}} &=&-2\sigma _{r_{1}r_{2}}^{D-1}\left( 1+\alpha
_{r_{1}r_{2}}\right) \frac{\mu _{r_{1}r_{2}}}{m_{r_{1}}}Y_{r_{1}r_{2}}\int d%
\widehat{q}\;\widehat{q}_{i_{1}}\widehat{q}_{i_{2}}F_{1}\left( \frac{%
w_{r_{1}r_{2}}}{Y_{r_{1}r2}}\right)  \label{BCD-eval1} \\
&&+\frac{1}{4}\sigma _{r_{1}r_{2}}^{D-1}\left( 1+\alpha _{r_{1}r_{2}}\right)
^{2}\left( \frac{\mu _{r_{1}r_{2}}}{m_{r_{1}}}\right) ^{2}Y_{r_{1}r_{2}}^{3}%
\frac{m_{r_{1}}}{k_{B}T_{r_{1}}}  \notag \\
&&\times \int d\widehat{q}\;\widehat{q}_{i_{1}}\widehat{q}_{i_{2}}\left[
\left( \frac{w_{r_{1}r_{2}}}{Y_{r_{1}r_{2}}}\right) F_{0}\left( \frac{%
w_{r_{1}r_{2}}}{Y_{r_{1}r_{2}}}\right) +F_{1}\left( \frac{w_{r_{1}r_{2}}}{%
Y_{r_{1}r_{2}}}\right) \left( 2\left( \frac{w_{r_{1}r_{2}}}{Y_{r_{1}r_{2}}}%
\right) ^{2}+2\right) \right]  \notag \\
C_{i_{1}i_{2},j_{1}j_{2}}^{r_{1}r_{2}} &=&\frac{T_{r_{1}}}{T_{r_{2}}}\frac{%
m_{r_{2}}}{m_{r_{1}}}D_{i_{1}i_{2},j_{1}j_{2}}^{r_{1}r_{2}}-2\sigma
_{r_{1}r_{2}}^{D-1}\left( 1+\alpha _{r_{1}r_{2}}\right) \frac{\mu
_{r_{1}r_{2}}}{m_{r_{1}}}Y_{r_{1}r_{2}}\int d\widehat{q}\;\left( \delta
_{i_{1}j_{1}}\widehat{q}_{i_{2}}\widehat{q}_{j_{2}}+\delta _{i_{2}j_{1}}%
\widehat{q}_{i_{1}}\widehat{q}_{j_{2}}\right) F_{1}\left( \frac{%
w_{r_{1}r_{2}}}{Y_{r_{1}r_{2}}}\right)  \notag \\
D_{i_{1}i_{2},j_{1}j_{2}}^{r_{1}r_{2}} &=&-4\sigma _{r_{1}r_{2}}^{D-1}\left(
1+\alpha _{r_{1}r_{2}}\right) \frac{\mu _{r_{1}r_{2}}}{m_{r_{1}}}%
Y_{r_{1}r_{2}}^{-1}\left( \frac{k_{B}T_{r_{2}}}{m_{r_{2}}}\right) \int d%
\widehat{q}\;\widehat{q}_{i_{1}}\widehat{q}_{i_{2}}\widehat{q}_{j_{1}}%
\widehat{q}_{j_{2}}\left( F_{1}\left( \frac{w_{r_{1}r_{2}}}{Y_{r_{1}r_{2}}}%
\right) -\left( \frac{w_{r_{1}r_{2}}}{Y_{r_{1}r_{2}}}\right) F_{0}\left( 
\frac{w_{r_{1}r_{2}}}{Y_{r_{1}r_{2}}}\right) \right)  \notag \\
&&+3\sigma _{r_{1}r_{2}}^{D-1}\left( 1+\alpha _{r_{1}r_{2}}\right)
^{2}\left( \frac{\mu _{r_{1}r_{2}}}{m_{r_{1}}}\right) ^{2}\frac{m_{r_{1}}}{%
k_{B}T_{r_{1}}}\left( \frac{k_{B}T_{r_{2}}}{m_{r_{2}}}\right)
Y_{r_{1}r_{2}}\int d\widehat{q}\;\widehat{q}_{i_{1}}\widehat{q}_{i_{2}}%
\widehat{q}_{j_{1}}\widehat{q}_{j_{2}}F_{1}\left( \frac{w_{r_{1}r_{2}}}{%
Y_{r_{1}r_{2}}}\right)  \notag
\end{eqnarray}%
where%
\begin{equation}
Y_{r_{1}r_{2}}=\sqrt{2\frac{k_{B}T_{r_{1}}}{m_{r_{1}}}+2\frac{k_{B}T_{r_{2}}%
}{m_{r_{2}}}}.
\end{equation}%
which, together with Eqs.(\ref{SME}), Eq.(\ref{Generalized Moment Equations}%
), Eq.(\ref{partial_temps}) and the requirement that $\sum_{r}n_{r}T_{r}=nT$
complete the specification of the second-order moment approximations. The
collisional contribution to the pressure is%
\begin{equation}
P_{i_{1}i_{2}}^{V}=-\frac{1}{8}\sum_{r_{1}r_{2}}n_{r_{1}}n_{r_{2}}\sigma
_{r_{1}r_{2}}^{D}\chi _{r_{1}r_{2}}\left( 1+\alpha _{r_{1}r_{2}}\right) \mu
_{r_{1}r_{2}}\int d\widehat{q}\;\widehat{q}_{i_{1}}\widehat{q}%
_{i_{2}}X_{r_{1}r_{2}}^{2}\left( F_{0}\left( \frac{w_{r_{1}r_{2}}}{%
X_{r_{1}r_{2}}}\right) +2\frac{w_{r_{1}r_{2}}}{X_{r_{1}r_{2}}}F_{1}\left( 
\frac{w_{r_{1}r_{2}}}{X_{r_{1}r_{2}}}\right) \right)
\end{equation}%
The evaluation of the SME model in the Boltzmann limit, obtained by
expanding in the hard-sphere diameters and keeping only the leading order
(which here amounts to taking the limit $w_{r_{1}r_{2}}\rightarrow 0$ and
using $F_{0}\left( 0\right) =-1$ and $F_{1}(0)=\frac{1}{\sqrt{\pi }}$) is
performed in Appendix \ref{BoltzmannLimit} . For a one component system, the
GME results are in agreement with Chou and Richman\cite{ChouRichman1},\cite%
{ChouRichman2} while the Boltzmann limit of the SME is in agreement, for a
one-component system, with the expressions given by Garzo\cite{GarzoDiff}.
In this simple case, the angular integrals can be performed analytically
(see appendix \ref{BoltzmannLimit}). It is remarkable that the structure of
the these terms is virtually identical to the equivalent quantities which
occur in the elementary case of the moment solution of the Enskog equation
for USF of elastic hard spheres\cite{Lutsko_EnskogPRL},\cite{LutskoEnskog}.
The practical result is that it is technically no more difficult to work
with an arbitrary mixture than with a single species.

\subsection{Polydisperse Granular Fluids}

As an extreme example, these results can be generalized to describe a
polydisperse granular fluid in which there is a continuous distribution of
grain sizes, masses and coefficients of restitution. This is equivalent to
having an infinite number of species and in general one must supply the
distribution of grains amongst the species, i.e. $x_{r}$ as well as the
hard-sphere diameters, $\sigma _{r_{i}r_{j}}$ and coefficients of
restitution $\alpha _{r_{i}r_{j}}$. In fact, formally, the species label can
be replaced by a continuous index over some interval, say $[0,1]$, and sums
over species replaced by integrals over this index. In the event that each
species has a unique hard sphere diameter $\sigma _{r}$, and the hard-sphere
diameters are additive, i.e. 
\begin{equation}
\sigma _{rr^{\prime }}=\frac{1}{2}\left( \sigma _{r}+\sigma _{r^{\prime
}}\right)
\end{equation}%
it makes sense to replace the integrals over species labels by integrals
over the distribution of hard sphere diameters. Specifically, the measures $%
x_{r}dr$ become $x_{r}\frac{dr}{d\sigma }d\sigma \equiv x\left( \sigma
\right) d\sigma $ where $x\left( \sigma \right) $ is the fraction of grains
having diameter $\sigma $. The moment equations then become%
\begin{eqnarray}
&&n^{-1}\left( \sigma _{1}\right) \frac{\partial }{\partial t}n\left( \sigma
_{1}\right) \left( \overleftrightarrow{\Gamma }\left( \sigma _{1}\right)
\right) _{i_{1}i_{2}}^{-1}-a\left( \delta _{i_{1}x}\left( 
\overleftrightarrow{\Gamma }\left( \sigma _{1}\right) \right)
_{yi_{2}}^{-1}+\delta _{i_{2}x}\left( \overleftrightarrow{\Gamma }\left(
\sigma _{1}\right) \right) _{yi_{1}}^{-1}\right) \\
&=&-n\int d\sigma _{2}\;x\left( \sigma _{2}\right) \chi \left( \frac{\sigma
_{1}+\sigma _{2}}{2}\right) \left( \frac{\sigma _{1}+\sigma _{2}}{2}\right)
^{D-1}\left( 1+\alpha \left( \sigma _{1},\sigma _{2}\right) \right) \frac{%
\mu \left( \sigma _{1},\sigma _{2}\right) }{m\left( \sigma _{1}\right) }%
\widetilde{E}_{i_{1}i_{2}}\left( \sigma _{1},\sigma _{2}\right)  \notag
\end{eqnarray}%
with%
\begin{eqnarray}
\widetilde{E}_{i_{1}i_{2}}\left( \sigma _{1},\sigma _{2}\right) &=&-\int d%
\widehat{q}\;X\left( \sigma _{1},\sigma _{2}\right) F_{1}\left( \frac{%
w\left( \sigma _{1},\sigma _{2}\right) }{X\left( \sigma _{1},\sigma
_{2}\right) }\right) \left[ \left( \overleftrightarrow{\Gamma }^{-1}\left(
\sigma _{1}\right) \cdot \widehat{q}\right) _{i_{1}}\widehat{q}%
_{i_{2}}+\left( \overleftrightarrow{\Gamma }^{-1}\left( \sigma _{1}\right)
\cdot \widehat{q}\right) _{i_{2}}\widehat{q}_{i_{1}}\right] \\
&&+\frac{1}{2}\left( 1+\alpha \left( \sigma _{1},\sigma _{2}\right) \right) 
\frac{\mu \left( \sigma _{1},\sigma _{2}\right) }{m\left( \sigma _{1}\right) 
}  \notag \\
&&\times \int d\widehat{q}\;\widehat{q}_{i_{1}}\widehat{q}%
_{i_{2}}X^{3}\left( \sigma _{1},\sigma _{2}\right) \left[ \left( \frac{%
w\left( \sigma _{1},\sigma _{2}\right) }{X\left( \sigma _{1},\sigma
_{2}\right) }\right) F_{0}\left( \frac{w\left( \sigma _{1},\sigma
_{2}\right) }{X\left( \sigma _{1},\sigma _{2}\right) }\right) +F_{1}\left( 
\frac{w\left( \sigma _{1},\sigma _{2}\right) }{X\left( \sigma _{1},\sigma
_{2}\right) }\right) \left( 2\left( \frac{w\left( \sigma _{1},\sigma
_{2}\right) }{X\left( \sigma _{1},\sigma _{2}\right) }\right) ^{2}+2\right) %
\right]  \notag \\
X\left( \sigma _{1},\sigma _{2}\right) &=&\sqrt{\widehat{q}\cdot \left( 
\overleftrightarrow{\Gamma }^{_{-1}}\left( \sigma _{1}\right) +%
\overleftrightarrow{\Gamma }^{_{-1}}\left( \sigma _{2}\right) \right) \cdot 
\widehat{q}}  \notag \\
w\left( \sigma _{1},\sigma _{2}\right) &=&\left( \frac{\sigma _{1}+\sigma
_{2}}{2}\right) \widehat{q}\cdot \overleftrightarrow{a}\cdot \widehat{q} 
\notag
\end{eqnarray}%
and the contributions to the pressure becomes%
\begin{eqnarray}
P_{i_{1}i_{2}}^{K} &=&n\int d\sigma _{1}\;x\left( \sigma _{1}\right) \left( 
\overleftrightarrow{\Gamma }\left( \sigma _{1}\right) \right)
_{i_{1}i_{2}}^{-1} \\
P_{i_{1}i_{2}}^{V} &=&-\frac{1}{4}n^{2}\int d\sigma _{1}d\sigma
_{2}\;x\left( \sigma _{1}\right) x\left( \sigma _{2}\right) \left( \frac{%
\sigma _{1}+\sigma _{2}}{2}\right) ^{D}\chi \left( \frac{\sigma _{1}+\sigma
_{2}}{2}\right) \left( 1+\alpha \left( \sigma _{1},\sigma _{2}\right)
\right) \mu \left( \sigma _{1},\sigma _{2}\right)  \notag \\
&&\times \int d\widehat{q}\;\widehat{q}_{i_{1}}\widehat{q}%
_{i_{2}}X^{2}\left( \sigma _{1},\sigma _{2}\right) \left( F_{0}\left( \frac{%
w\left( \sigma _{1},\sigma _{2}\right) }{X\left( \sigma _{1},\sigma
_{2}\right) }\right) +2\left( \frac{w\left( \sigma _{1},\sigma _{2}\right) }{%
X\left( \sigma _{1},\sigma _{2}\right) }\right) F_{1}\left( \frac{w\left(
\sigma _{1},\sigma _{2}\right) }{X\left( \sigma _{1},\sigma _{2}\right) }%
\right) \right)  \notag
\end{eqnarray}%
where some model for $\alpha \left( \sigma _{1},\sigma _{2}\right) $ and the
masses $m\left( \sigma _{1}\right) $ must be supplied. The generalization of
the SME expressions is similarly straightforward. These expressions appear
formidable to implement, but if the $\sigma -$integrals are performed using
n-point Gaussian quadratures, then $\int d\sigma x\left( \sigma \right) 
\mathcal{F}\left( \sigma \right) \rightarrow \sum_{i=1}^{n}w_{i}x\left(
\sigma _{i}\right) \mathcal{F}\left( \sigma _{i}\right) $, where $w_{i}$ are
the Gaussian weights and the $\sigma _{i}$ are determined by the Gaussian
abscissas. In this form, the calculation is identical to that for $n$%
-species with $x_{r_{i}}=w_{i}x\left( \sigma _{i}\right) $. Thus,
numerically, there is no practical difference between the polydisperse fluid
and a mixture with $n$-species.

\section{Comparison to simulation}

\subsection{Simulation and numerical methods}

In order to evaluate the models presented here, three types of calculations
were performed for three dimensional systems: numerical solution of the
second order moment expansions, numerical solution of the Enskog equation by
means of direct simulation monte carlo (DSMC) and molecular dynamics (MD)\
simulations. Comparison of the first two elucidates the accuracy of the
second order moment approximations while comparison of both to the MD
indicates the accuracy of the underlying assumptions:\ that the state
obtained is indeed USF and the assumption of molecular chaos.

The focus here will be on the steady-state properties of the systems, so
that when evaluating the SME and GME models, the time-derivatives are set
identically to zero. Implementation of the (static) SME requires the
numerical evaluation of the coefficients given in Eq.(\ref{BCD-eval1}) and
the solution of equations (\ref{SME}) for the partial temperatures and the
shear rate as a function of the global temperature and coefficient of
restitution (the second order moments are determined from Eq.(\ref{SME})
which are linear so that moments may be taken as given functions of the
other parameters). The GME\ requires a similar numerical evaluation of the
function $E_{ij}^{r_{1}r_{2}}$and solution of the nonlinear moment
equations, Eqs.(\ref{Generalized Moment Equations}) and (\ref{GME2}). All
numerical calculations were performed using the Gnu Scientific Library\cite%
{GSL}. In all cases, the (two dimensional) numerical integrals were
calculated using the GSL routine ''qags'' (Gauss-Kronrod 21-point
integration rule applied adaptively until the desired accuracy is achieved)
with a specification of relative accuracy of $10^{-4}$and absolute accuracy
of $10^{-6}$for the inner integral and $10^{-3}$and $10^{-6}$ for the outer
integrals. The linear equations for the SME moments were solved by LU
decomposition and the nonlinear equations for the partial temperatures and
shear rate were solved with the GSL\ ''hybrids'' algorithm (Powell's Hybrid
method with numerical evaluation of the Jacobian). Convergence was
considered to be achieved when the sum of the absolute value of the
residuals was less than $10^{-7}$. The same methods and tolerances were used
to solve the nonlinear moment equations for the GME.

The second set of calculations performed consisted of the numerical solution
of the Enskog equation by means of the Direct Simulation Monte Carlo (DSMC)
method\cite{DSMC}. These calculations were performed using a cubic cell with
sides equal to the maximum of the hard sphere diameters, with $10^{5}$
points and a time step of $\Delta t=0.0117\tau _{mft}$ where $\tau _{mft}$
is the mean free time. All calculations began from an initial configuration
corresponding to the equilibrium hard sphere fluid. Shear flow was imposed
by means of Lees-Edwards boundary conditions which are periodic boundaries
in the Lagrangian frame\cite{LeesEdwards}. For each combination of
temperature, shear rates and coefficients of restitution, the initial
configuration was relaxed over a period of $100\tau _{mft}$ and steady-state
statistics, reported below, were then obtained by averaging over another $%
100\tau _{mft}$.

Finally, these calculations are compared below to molecular dynamics
simulations. In all cases, the systems consisted of $500$ grains and the
starting configuration was the equilibrium fluid. Shear flow was again
imposed by means of Lees-Edwards boundary conditions. After turning on the
shear flow and collisional dissipation, the systems were allowed to relax
for a period of $5\times 10^{7}$ collisions after which statistics were
gathered for another $5\times 10^{7}$ collisions. Errors were computed by
estimating the desired statistics using data from each period of $10^{5}$
collisions and calculating the standard error between the estimates (the
same method was used in the DSMC calculation). In the figures shown below,
error bars are in general not given because in most cases, the estimated
errors are comparable to or smaller than the size of the symbols. Exceptions
to this (in the case of the temperature distributions) are explicitly
commented upon in the text. Larger systems were not simulated as they are
subject to various hydrodynamic instabilities which violate the assumption
that the state is USF\cite{GoddardAlam},\cite{LutskoShearInstab}.

\subsection{Binary mixtures}

One check on the expressions given here is to compare to the results of
Montanero and Garzo who have evaluated the SME in the Boltzmann limit for a
binary mixture and compared to DSMC simulations for a variety of
combinations of mass, diameter and density ratios. The expressions for the
SME when evaluated for $n\left\langle \sigma \right\rangle ^{3}=0.001$, so
as to achieve the Boltzmann limit, do indeed agree well with the data given
in ref.\cite{GarzoShearMixture}. A particular case is for $\sigma
_{11}=\sigma _{12}=\sigma _{22}=1$, $m_{1}=10m_{2}$, $\alpha _{11}=\alpha
_{12}=\alpha _{22}=0.75$ and $x_{1}=x_{2}=0.5$ for which Montanero and Garzo
report $P_{xy}^{K}=-0.498$ and $P_{yy}^{K}=0.723$ from DSMC simulations and $%
P_{xy}^{K}=-0.498$ and $P_{yy}^{K}=0.743$ from their evaluation of the SME
in the Boltzmann limit. I find that the (very low density) SME gives $%
P_{xy}^{K}=-0.4981$ and $P_{yy}^{K}=0.7435$ in excellent agreement. By
comparison, the GME gives $P_{xy}^{K}=-0.496$ and $P_{yy}^{K}=0.726$ and so
is in even better agreement with the DSMC\ numerical solution of the
Boltzmann equation. In addition, in the same limit, the GME is able to
account for at least some of the normal stress differences, $%
P_{yy}^{K}-P_{zz}^{K}$, that the SME\ misses (in the SME in the Boltzmann
limit, $P_{yy}^{K}=P_{zz}^{K}$) but which are clearly nonzero in the DSMC
calculations.

\subsection{Polydisperse model}

A simple model was used in which the diameters are additive, the masses
scale with the diameters in the usual way%
\begin{equation}
m\left( \sigma \right) =\frac{4\pi }{3}\rho _{0}\sigma ^{3}
\end{equation}%
and the coefficients of restitution are also additive%
\begin{equation}
\alpha \left( \sigma _{1},\sigma _{2}\right) =\frac{1}{2}\left( \alpha
\left( \sigma _{1}\right) +\alpha \left( \sigma _{2}\right) \right) .
\end{equation}%
The distribution of diameters was taken to be a simple triangular
distribution%
\begin{equation}
x\left( \sigma \right) =\left\{ 
\begin{array}{c}
u^{-2}\left( \sigma -1+u\right) ,\;1-u<\sigma <1 \\ 
u^{-2}\left( 1+u-\sigma \right) ,\;1<\sigma <1+u%
\end{array}%
\right.
\end{equation}%
so that the average diameter is $1$ and the polydispersity, defined as the
variance divided by the average of the sizes, is $\delta =\frac{1}{\sqrt{6}}%
u $. The coefficients of restitution were assumed to scale linearly with the
diameter with the smallest grains being hard, $\alpha \left( \sigma
=1-u\right) =1$, and the largest being soft, $\alpha \left( \sigma
=1+u\right) =\alpha _{0}<1$, where $\alpha _{0}$ is a free parameter, so
that the average value is $\left\langle \alpha \right\rangle =\frac{1+\alpha
_{0}}{2}$. The equilibrium pair structure function $\chi \left( \sigma
_{1},\sigma _{2}\right) $ was evaluated using the approximation of ref (\cite%
{SantosChi}) and the accuracy of this approximation was verified in the
equilibrium ($\alpha _{0}=1$) simulations. In all of the calculations
reported here, the integrals over the distribution of hard-sphere diameters
were performed using a Gauss-Legendre integration scheme with 10 points.
Using 5 points, the results differed by about 10\%. When evaluating the
equations for an equilibrium ($\alpha _{0}=1$) mixture, the difference
between the 5 and 10 point schemes was also about 10\% and the absolute
accuracy of the 10 point scheme compared to the known exact results was 1\%.
The MD and DSMC simulations were performed with a system obtained by a
random sampling over the distribution of hard-sphere diameters. A variety of
simulations were also performed with other samplings and it was confirmed
that the results reported below do not vary significantly from sample to
sample.

In the following, comparisons will be made for systems at three densities:\
a low density fluid, $n^{\ast }\equiv n\left\langle \sigma ^{3}\right\rangle
=0.1$, a moderately dense fluid $n^{\ast }=0.25$ and a dense fluid $n^{\ast
}=0.5$. In all cases, a value of $u=0.5$ or polydispersity of $\delta
=\allowbreak 20.\,\allowbreak 4\%$ was used. All results are from a single
random sampling of this distribution. For the DSMC calculations, the large
number of points used means that the distribution is well sampled. For the
MD, however, the relatively small number of atoms might mean that the
results reported below are influenced by the particular realization used. To
control against this, I have checked a number of data points using multiple
indpendent samplings from the distribution and confirmed that the variation
induced by different samplings is negligable, at least for the properties
discussed below.

\subsection{Accuracy of the second moments}

\begin{figure*}[tbp]
\includegraphics[angle=0,scale=0.4]{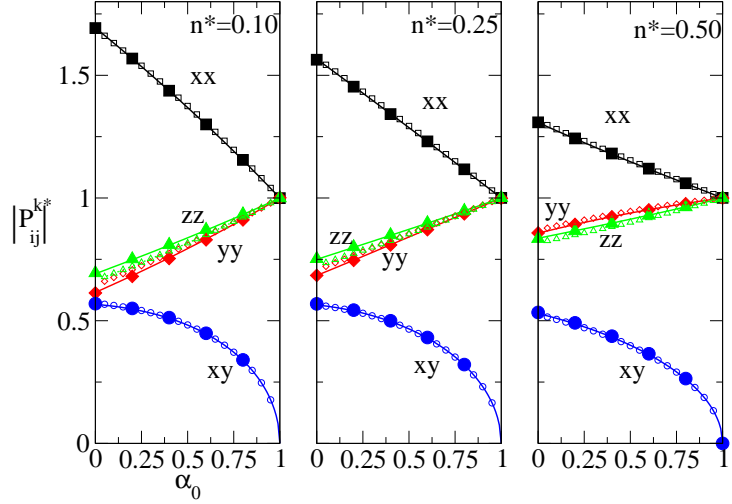}
\caption{The absolute values of the kinetic part of the stress tensor (i.e.,
the second moments) normalized to $nk_{B}T$ for three densities as
determined by the GME (solid lines), SME (open symbols) and DSMC (filled symbols).
Note that the xy moments are actually
negative.}
\label{fig1}
\end{figure*}

\begin{figure*}[tbp]
\includegraphics[angle=0,scale=0.4]{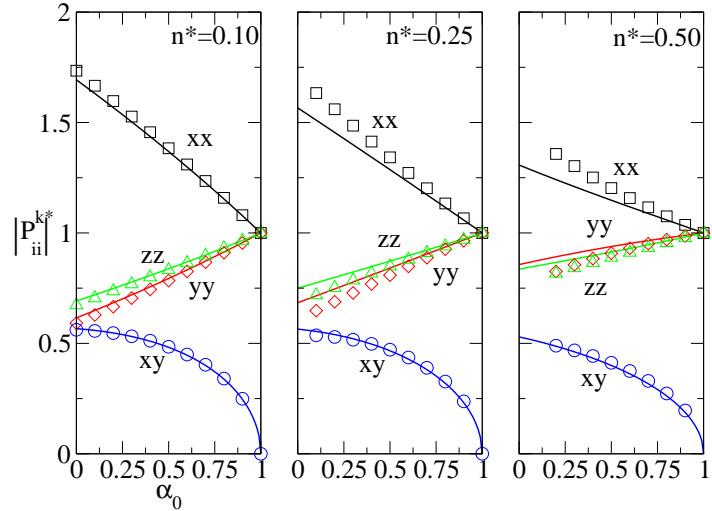}
\caption{Same as \ref{fig1}, but showing results from the GME (solid lines)
and MD simulations (symbols).}
\label{fig2}
\end{figure*}

Figure \ref{fig1} shows the kinetic part of the stress tensor, or
equivalently the second moments, as obtained from the SME, the GME and the
DSMC. Comparison with the numerical solution of the Enskog equation, i.e.
the DSMC results, shows that the GME gives a virtually exact estimate of the
second moments at all densities and degrees of inelasticity. It is
interesting to note that the difference between the $yy$ and $zz$ moments,
which is zero in the Boltzmann limit (see Appendix \ref{BoltzmannLimit}), is
never very great and actually changes sign at high density. The SME is in
close agreement with the GME. The only significant difference is in the $yy$
and $zz$ moments where the SME tends to underestimate the difference between
them. Figure \ref{fig2} compares the GME\ calculation to the MD\ results for
the same systems. The calculations are in excellent agreement with the
simulations at low density and remain reasonable even at the highest
density. In particular, the $xy$ moments are in good agreement at all
densities. These results show that the GME gives an accurate estimate of the
second velocity moments as determined by the Enskog equation and that the
Enskog equation gives a reasonable approximation to the second moments at
all densities investigated.

\subsection{Accuracy of the second moment approximation}

\begin{figure*}[tbp]
\includegraphics[angle=0,scale=0.4]{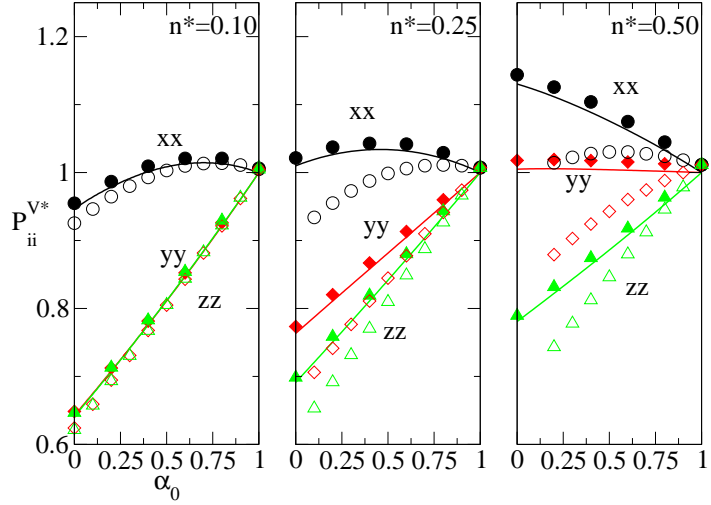}
\caption{The diagonal components of the collisional contribution to the
stress tensor as a function of $\protect\alpha _{0}$ normalized to their
equilibrium ($\protect\alpha _{0}=1$) values. The lines are GME, the filled
symbols DSMC and the open symbols MD.}
\label{fig3}
\end{figure*}

The next question is whether stopping at second moments is sufficient to
accurately approximate the full solution of the Enskog equation. One measure
of this is the calculation of the collisional contribution to the stress
tensor. Figure \ref{fig3} shows the diagonal components of this quantity as
calculated from the GME and DSMC and measured in the MD simulations. At low
density, the agreement between the GME and DSMC is good, although not quite
as good as for the moments themselves. This shows that although higher order
moments will give some small contribution, the GME appears, in this case at
least, to be a good approximation to the solution of the Enskog equation.
However, comparison to the MD shows the shortcomings of the Enskog equation
itself. At low density, agreement is good but even at moderate density,
considerable differences between MD and the Enskog approximation are
apparent although the latter remains a reasonable semi-quantitative
approximation. At the highest density, the differences become qualitative in
nature. In the MD, the $xx$ component changes non-monotonically with $\alpha
_{0}$ whereas the Enskog theory predicts a monotonic increase with
increasing inelasticity. Enskog predicts little change in the $yy$ component
whereas in fact it drops rapidly. Only the $zz$ component is represented at
all reasonably.

\subsection{Viscoelastic properties}

\begin{figure*}[tbp]
\includegraphics[angle=0,scale=0.4]{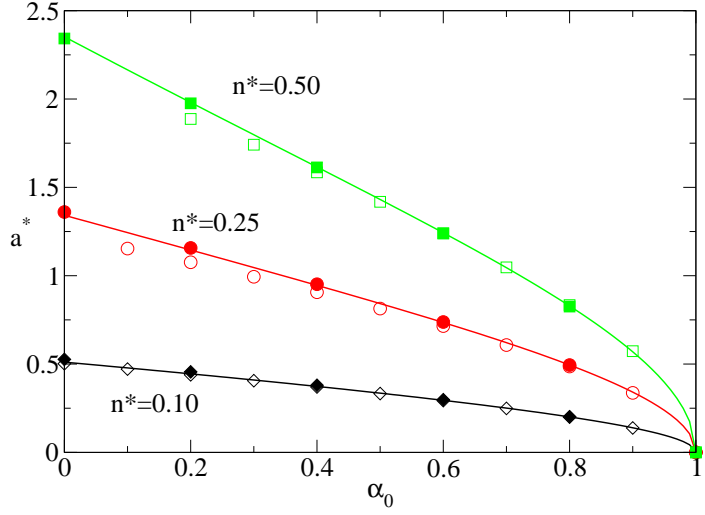}
\caption{The reduced shear rate $a^{*}$ as a function of $\protect\alpha_{0}$
as determined from the GME (lines) DSMC (filled symbols) and MD (open
symbols).}
\label{fig4}
\end{figure*}

Figure \ref{fig4} shows the dimensionless shear rate $a^{\ast }$ as a
function of $\alpha _{0}$ according to the DSMC, GME and MD. All of these
are in good agreement at all densities and values of inelasticity. This
agreement is also fortunate since it means that any differences between
Enskog and MD are not attributable to a misestimated shear rate.

\begin{figure*}[tbp]
\includegraphics[angle=0,scale=0.4]{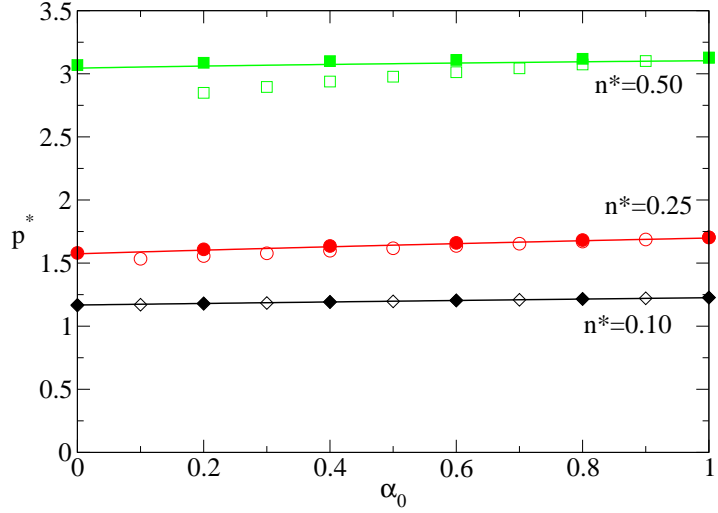}
\caption{Same as \ref{fig5} but showing the reduced pressure $%
p^{*}=p/nk_{B}T $.}
\label{fig5}
\end{figure*}

\begin{figure*}[tbp]
\includegraphics[angle=0,scale=0.4]{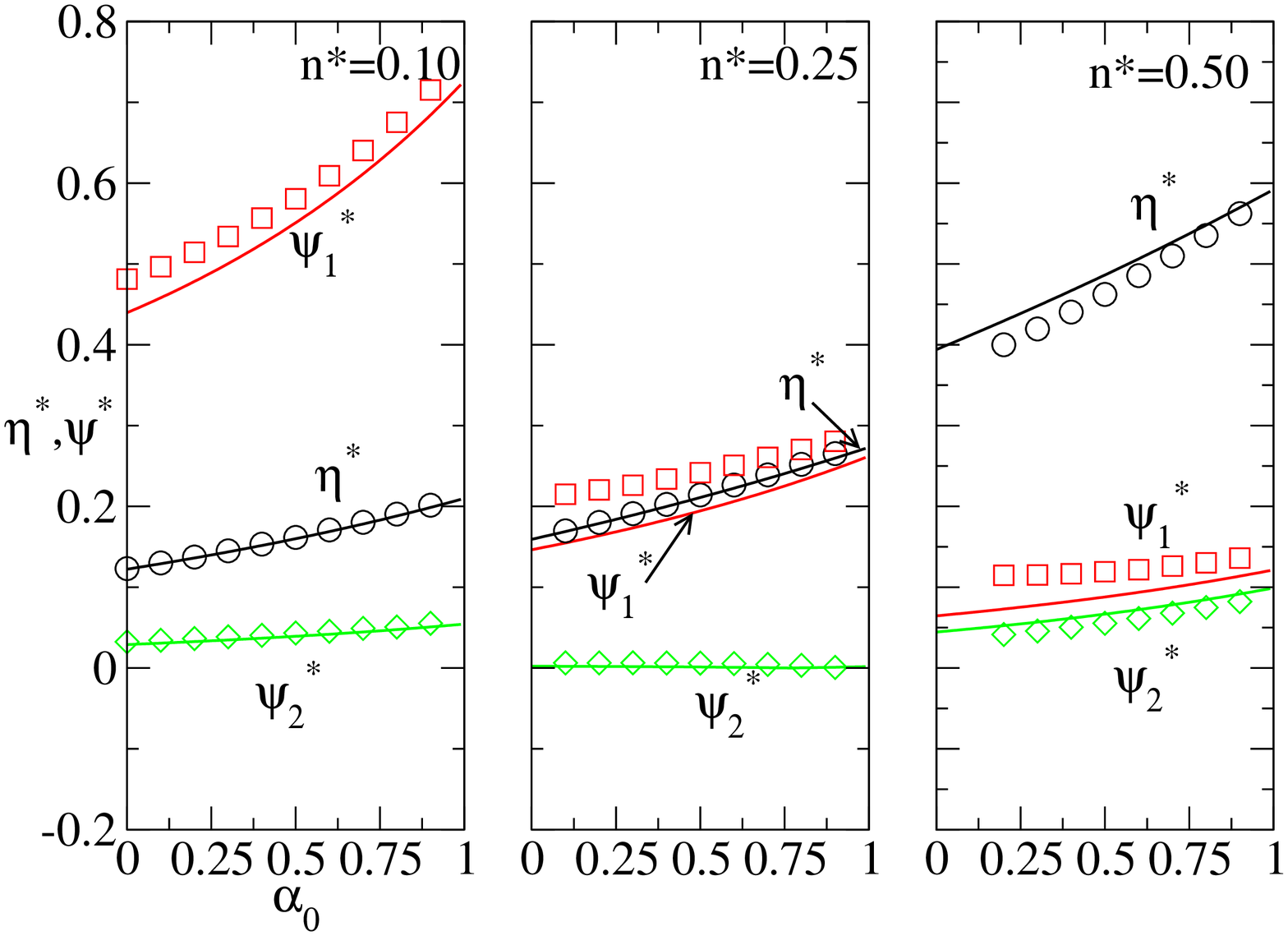}
\caption{The reduced shear viscosity and viscometric functions, as defined
in Eqs.(\ref{shearvisc})-(\ref{visco}) as functions of $\protect\alpha_{0}$.
The lines are from the GME and the symbols from MD. }
\label{fig6}
\end{figure*}

Figure \ref{fig5} shows the pressure (trace of the stress tensor). In
contrast to elastic hard spheres, for which the pressure increases with
increasing shear rate\cite{LutskoEnskog}, the pressure is nearly constant.
The calculations are again all in reasonably good agreement with the MD.
Figure \ref{fig6} shows the dimensionless shear viscosity%
\begin{equation}
\eta ^{\ast }=\frac{P_{xy}}{a}\sqrt{\frac{\left\langle \sigma \right\rangle
^{4}}{k_{B}T\left\langle m\right\rangle }}  \label{shearvisc}
\end{equation}%
and the viscometric functions 
\begin{eqnarray}
\psi _{1}^{\ast } &=&\frac{P_{xx}-P_{yy}}{a^{2}}\frac{\left\langle \sigma
\right\rangle }{\left\langle m\right\rangle }  \label{visco} \\
\psi _{2}^{\ast } &=&\frac{P_{yy}-P_{zz}}{a^{2}}\frac{\left\langle \sigma
\right\rangle }{\left\langle m\right\rangle }  \notag
\end{eqnarray}%
which measure the normal stresses. The Enskog theory gives a very reasonable
estimate for all of the viscoelastic properties. Although $\psi _{1}^{\ast }$
is systematically underestimated, $\psi _{2}^{\ast }$ and the shear
viscosity are well approximated at all densities. In all cases, the errors
grow with density and decreasing $\alpha _{0}$.

\subsection{Temperature distribution}

\begin{figure*}[tbp]
\includegraphics[angle=0,scale=0.4]{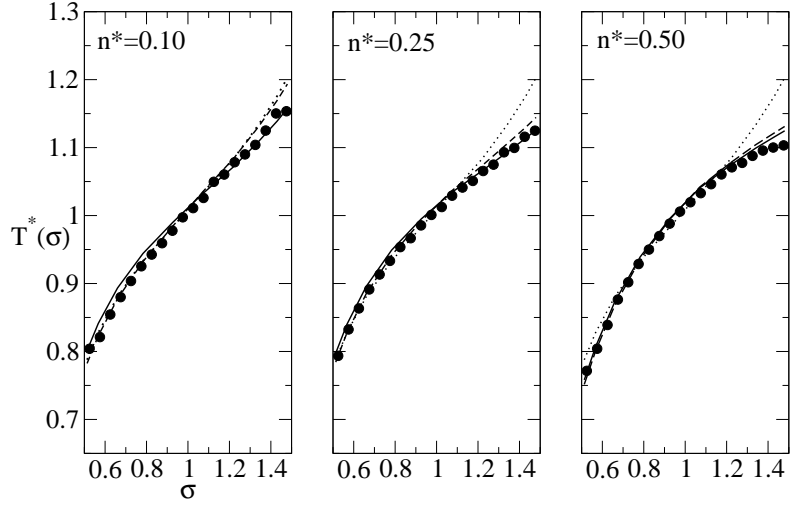}
\caption{The reduced temperature distribution $T^{\ast }(\protect\sigma )=T(%
\protect\sigma )/T$ as a function of grain size, $\protect\sigma $. The line
is from the GME, the filled symbols from DSMC, the dashed line is the SME
and the dotted line is from the Boltzmann equation (also in the GME
approximation).}
\label{fig7}
\end{figure*}

\begin{figure*}[tbp]
\includegraphics[angle=0,scale=0.4]{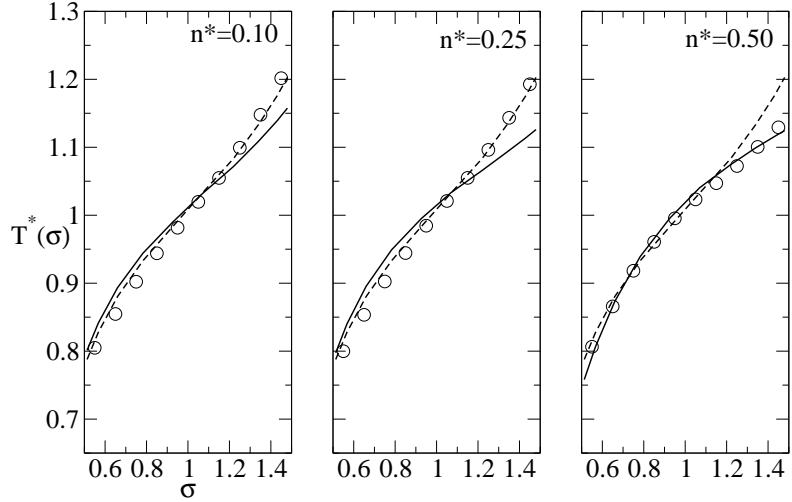}
\caption{The reduced temperature distribution $T^{\ast }(\protect\sigma )=T(%
\protect\sigma )/T$ as a function of grain size, $\protect\sigma $. The full
line is from the GME, the open symbols are from MD, and the dashed line is
from the Boltzmann equation (also in the GME approximation).}
\label{fig8}
\end{figure*}

So far, the comparisons have shown the Enskog theory and the MD to be in
good agreement for bulk properties up to $n^{\ast }\leq 0.25$. Even above
this density, the physically interesting quantities - the pressure, shear
viscosity and viscometric functions, are well approximated. This picture
changes when attention focuses on variations of properties with grain
species. Figure \ref{fig7} shows a comparison of the predicted temperature
distribution as a function of grain size according to the SME, GME and DSMC
for the particular value $\alpha =0.4$ as well as the zero density,
Boltzmann limit , prediction. The SME and GME\ are again very good
approximations to the numerical results with the former being slightly more
accurate for the smaller grains and the latter more accurate for the larger
grains for which the SME\ deviates from the Boltzmann result too slowly. The
surprising result is shown in Fig. \ref{fig8} which compares the
distributions obtained from the GME and MD simulations. Although reasonable,
the Enskog results are in poor agreement with the MD for the largest grains,
especially at the \emph{lower} densities and most especially for $n^{\ast
}=0.25$. Even more surprisingly, the MD\ results at lower densities are in
good agreement with the GME approximation to the \emph{Boltzmann} equation.
The two differences between the Boltzmann and Enskog theories are that (a)
the Enskog theory has a higher collision frequency due to the prefactor of
the pair distribution function which occurs in the collision term and (b)
the Enskog theory accounts for the non-locality of the interactions of the
grains in the collision term. It is hard to imagine that the second point is
in error, so it seems most likely that the Enskog theory is overestimating
the collision rate for large grains. Some support for this hypothesis comes
from the fact that setting the pdf to its Boltzmann limit (ie. unity)
increases the temperature of the largest grains by about a third of the
difference between the Boltzmann and Enskog results for $n^{\ast }=0.25$.
This suggests that even at low density, the Enskog theory is based on a poor
estimate of the collision rates and so that the assumption of molecular
chaos, Eq.(\ref{6}), is in error. This error is not apparent when
considering the bulk properties because the distribution of grain sizes is
such that the largest grains make a relatively small contribution to most
properties: the largest contributions come from grains near the middle of
the distribution where the Enskog theory is relatively accurate.

\section{Conclusions}

In this paper, the moment approximation to the solution of the
Boltzmann-Enskog kinetic theory can be generalized so as to represent an
expansion about an arbitrary Gaussian state. This framework encompasses both
the Generalized Maxwellian approximation as well as the simple moment
expansion about local equilibrium as special cases. It shows in particular
how corrections to the Generalized Maxwellian approximation might be
calculated.

A generating function technique was also presented as a simplified means of
calculating collision integrals for the particular case of uniform shear
flow. Although the present calculation were only performed to second order,
the generating function technique would make higher order calculations much
more feasible than more straightforward methods. The technique is based on
the observation that the post-collisional velocities of hard spheres are
linear functions of the pre-collisional velocities so that pre-collisional
Gaussians remain Gaussian and integrals over such functions are relatively
straightforward to perform. This technique is particularly valuable in
anisotropic states, such as USF, where the usual approach to evaluating
collision integrals becomes very messy. The method should be applicable to
many other types of kinetic theory calculations.

These general methods were applied to the particular case of arbitrary
mixtures of granular fluids. It was shown, by comparison to DSMC
simulations, that both the SME and the GME are very good approximations to
the exact solutions to the Enskog equation for a model polydisperse granular
fluid. The GME tends to be slightly more accurate than the SME and has the
additional advantage that the approximate distribution is positive definite. 

Comparison to MD simulations showed that the Enskog equation gives a good
estimate of bulk properties such as the temperature, pressure, shear
viscosity and viscometric functions (i.e., normal stresses) over a wide
range of coefficients of restitution and densities. Shear thinning is
particularly well predicted. However, a more detailed examination shows that
part of this agreement (particularly in the case of the viscometric
functions) is due to a cancellation of errors while the description of the
variation of temperature with grain size is in fact rather poor. The fact
that this agreement is so poor even at relatively low densities raises the
question of whether the approximate kinetic theory is fundamentally lacking
in some way. Possible explanations of the errors are that the local
equilibrium pair distribution function is simply inaccurate, that the
assumption of molecular chaos is violated or that the systems are not
actually in a state of USF due, e.g., to some sort of segregation process.
The exploration of these possibilities will be the subject of a later work.

\section{Acknowledgments}

This work is supported, in part, by the European Space Agency under contract
number C90105. The author is grateful for useful comments from Vicente Garzo
concerning an early version of this work.

\bigskip

\bigskip

\appendix

\section{Moment Equations}

\label{MomentEquations}

In this appendix, the left hand side of the moment equations is developed,
first for a general Gaussian state and then specialized to uniform shear
flow. The kinetic equations take the form%
\begin{equation}
\left( \frac{\partial }{\partial t}+\overrightarrow{v}\cdot \frac{\partial }{%
\partial \overrightarrow{q}}\right) f_{r}\left( \overrightarrow{q}_{1},%
\overrightarrow{v}_{1};t\right) =\sum_{s}J\left[ f_{r},f_{s}\right]
\end{equation}%
and the distribution is expanded as%
\begin{equation}
f_{r}\left( \overrightarrow{q},\overrightarrow{V};\left\{ n_{s}\right\}
,T\right) =n_{r}\det \left( \overleftrightarrow{\Gamma }^{r}\right)
^{1/2}\pi ^{-D/2}\exp \left( -\overrightarrow{V}^{r}\cdot 
\overleftrightarrow{\Gamma }^{r}\cdot \overrightarrow{V}^{r}\right) \left(
\sum_{n=0}\frac{1}{n!}\sum_{I_{n}}A_{I_{n}}^{r}H_{I_{n}}\left( \sqrt{2}%
\overleftrightarrow{\Gamma }^{r1/2}\cdot \overrightarrow{V}\right) \right)
\end{equation}%
where $\overrightarrow{V}^{r}=\overrightarrow{v}-\overrightarrow{u}%
^{r}\left( q\right) $. This is slightly more general than the form given in
the text as we allow here for an arbitrary, species-dependent, linear
contribution to the Gaussian. In the following, all dependence on space and
time will not be indicated explicitly, although all quantities do in fact
have such dependence. Furthermore, since we are only interested in the left
hand side of the equation, which only involves a single species, the species
label will also be suppressed until the end of the calculation.

The first step is to switch variables from $\left\{ \overrightarrow{q},%
\overrightarrow{v},t\right\} $ to $\left\{ \overrightarrow{q}^{\prime }=%
\overrightarrow{q},C_{i}=\sqrt{2}\Gamma _{ij}^{1/2}\left( v_{j}-u_{j}\right)
,t^{\prime }=t\right\} $using%
\begin{eqnarray}
\frac{\partial }{\partial t} &=&\frac{\partial }{\partial t^{\prime }}+\frac{%
\partial C_{i}}{\partial t}\frac{\partial }{\partial C_{i}}=\frac{\partial }{%
\partial t^{\prime }}+\left( \frac{\partial \Gamma _{ij}^{1/2}}{\partial
t^{\prime }}\Gamma _{jl}^{-1/2}C_{l}-\sqrt{2}\Gamma _{ij}^{1/2}\frac{%
\partial u_{j}}{\partial t^{\prime }}\right) \frac{\partial }{\partial C_{i}}
\\
\frac{\partial }{\partial q_{l}} &=&\frac{\partial }{\partial q_{l}^{\prime }%
}+\frac{\partial C_{i}}{\partial q_{l}^{\prime }}\frac{\partial }{\partial
C_{i}}=\frac{\partial }{\partial q_{l}^{\prime }}+\left( \frac{\partial
\Gamma _{ij}^{1/2}}{\partial q_{l}^{\prime }}\Gamma _{jk}^{-1/2}C_{k}-\sqrt{2%
}\Gamma _{ij}^{1/2}\frac{\partial u_{j}}{\partial q_{l}^{\prime }}\right) 
\frac{\partial }{\partial C_{i}}  \notag
\end{eqnarray}%
so that 
\begin{eqnarray}
\frac{\partial }{\partial t}+\overrightarrow{v}\cdot \frac{\partial }{%
\partial \overrightarrow{q}} &=&\frac{\partial }{\partial t^{\prime }}%
+\left( \frac{\partial \Gamma _{ij}^{1/2}}{\partial t}\Gamma
_{jk}^{-1/2}C_{k}-\sqrt{2}\Gamma _{ij}^{1/2}\frac{\partial u_{j}}{\partial t}%
\right) \frac{\partial }{\partial C_{i}} \\
&&+\left( \frac{1}{\sqrt{2}}\Gamma _{lm}^{-1/2}C_{m}+u_{l}\right) \left( 
\frac{\partial }{\partial q_{l}^{\prime }}+\left( \frac{\partial \Gamma
_{ij}^{1/2}}{\partial q_{l}^{\prime }}\Gamma _{jk}^{-1/2}C_{k}-\sqrt{2}%
\Gamma _{ij}^{1/2}\frac{\partial u_{j}}{\partial q_{l}^{\prime }}\right) 
\frac{\partial }{\partial C_{i}}\right) .  \notag
\end{eqnarray}%
Introducing $\widetilde{f}=\det \left( 2\overleftrightarrow{\Gamma }\right)
^{-1/2}f$, the kinetic equation becomes%
\begin{eqnarray}
&&\frac{\partial }{\partial t^{\prime }}\widetilde{f}+\left( \frac{\partial
\Gamma _{ij}^{1/2}}{\partial t}\Gamma _{jk}^{-1/2}C_{k}-\sqrt{2}\Gamma
_{ij}^{1/2}\frac{\partial u_{j}}{\partial t}\right) \frac{\partial }{%
\partial C_{i}}\widetilde{f} \\
&&+\left( \frac{1}{\sqrt{2}}\Gamma _{lm}^{-1/2}C_{m}+u_{l}\right) \left( 
\frac{\partial }{\partial q_{l}^{\prime }}+\left( \frac{\partial \Gamma
_{ij}^{1/2}}{\partial q_{l}^{\prime }}\Gamma _{jk}^{-1/2}C_{k}-\sqrt{2}%
\Gamma _{ij}^{1/2}\frac{\partial u_{j}}{\partial q_{l}^{\prime }}\right) 
\frac{\partial }{\partial C_{i}}\right) \widetilde{f}  \notag \\
&&+\frac{\partial }{\partial t^{\prime }}\ln \det \left( \overleftrightarrow{%
\Gamma }\right) ^{1/2}+\left( \frac{1}{\sqrt{2}}\Gamma
_{lm}^{-1/2}C_{m}+u_{l}\right) \frac{\partial }{\partial q_{l}^{\prime }}\ln
\det \left( \overleftrightarrow{\Gamma }\right) ^{1/2}  \notag \\
&=&\det \left( 2\overleftrightarrow{\Gamma }^{r}\right) ^{-1/2}\sum_{s}J%
\left[ f_{r},f_{s}\right] .  \notag
\end{eqnarray}

The next step is to multiply through by $H_{I_{n}}\left( \overrightarrow{C}%
\right) $ and to integrate over $\overrightarrow{C}$. These evaluations are
performed using the basic identities, which follow directly from the
definition of the Hermite polynomials,%
\begin{eqnarray}
C_{x}H_{I_{n}}\left( \overrightarrow{C}\right) &=&H_{xI_{n}}\left( 
\overrightarrow{C}\right) +\mathcal{S}_{I_{n}}\delta
_{i_{n}x}H_{I_{n-1}}\left( \overrightarrow{C}\right) \\
\frac{\partial }{\partial C_{x}}H_{I_{n}}\left( \overrightarrow{C}\right) &=&%
\mathcal{S}_{I_{n}}\delta _{i_{n}x}H_{I_{n-1}}\left( \overrightarrow{C}%
\right)  \notag
\end{eqnarray}%
where the operator $\mathcal{S}_{I_{n}}$ indicates a sum over all
inequivalent permutations of the indicated set of indices. Repeated
application of these gives%
\begin{eqnarray}
C_{x}C_{y}H_{I_{n}}\left( \overrightarrow{C}\right) &=&H_{I_{n}xy}\left( 
\overrightarrow{C}\right) +\delta _{xy}H_{I_{n}}\left( \overrightarrow{C}%
\right) \\
&&+\mathcal{S}_{I_{n}}\left( \delta _{xi_{n}}\delta
_{yi_{n-1}}H_{I_{n-2}}\left( \overrightarrow{C}\right) +\delta
_{xi_{n}}H_{I_{n-1}y}\left( \overrightarrow{C}\right) +\delta
_{yi_{n}}H_{I_{n-1}x}\left( \overrightarrow{C}\right) \right)  \notag \\
C_{y}\frac{\partial }{\partial C_{x}}H_{I_{n}}\left( \overrightarrow{C}%
\right) &=&\mathcal{S}_{I_{n}}\delta _{i_{n}x}H_{yI_{n-1}}\left( 
\overrightarrow{C}\right) +\mathcal{S}_{I_{n}}\delta _{i_{n}x}\delta
_{i_{n-1}y}H_{I_{n-2}}\left( \overrightarrow{C}\right)  \notag \\
C_{z}C_{y}\frac{\partial }{\partial C_{x}}H_{I_{n}}\left( \overrightarrow{C}%
\right) &=&\mathcal{S}_{I_{n}}\delta _{i_{n}x}\left( 
\begin{array}{c}
H_{I_{n-1}zy}\left( \overrightarrow{C}\right) +\delta _{zy}H_{I_{n-1}}\left( 
\overrightarrow{C}\right) +\delta _{zi_{n-1}}\delta
_{yi_{n-2}}H_{I_{n-3}}\left( \overrightarrow{C}\right) \\ 
+\delta _{zi_{n-1}}H_{I_{n-2}y}\left( \overrightarrow{C}\right) +\delta
_{yi_{n-1}}H_{I_{n-2}z}\left( \overrightarrow{C}\right)%
\end{array}%
\right) .  \notag
\end{eqnarray}%
Combined with the orthonormality of the Hermite polynomials, and integrating
by parts where needed, one then has that%
\begin{eqnarray}
n^{-1}\int d\overrightarrow{C}\;H_{I_{n}}\left( \overrightarrow{C}\right) 
\widetilde{f}\left( \overrightarrow{C}\right) &=&A_{I_{n}} \\
n^{-1}\int d\overrightarrow{C}\;C_{x}H_{I_{n}}\left( \overrightarrow{C}%
\right) \widetilde{f}\left( \overrightarrow{C}\right) &=&A_{xI_{n}}+\mathcal{%
S}_{I_{n}}\delta _{i_{n}x}A_{I_{n-1}}  \notag \\
n^{-1}\int d\overrightarrow{C}\;C_{y}H_{I_{n}}\left( \overrightarrow{C}%
\right) \frac{\partial }{\partial C_{x}}\widetilde{f}\left( \overrightarrow{C%
}\right) &=&-\delta _{xy}A_{I_{n}}-\mathcal{S}_{I_{n}}\delta _{i_{n}x}\left(
A_{yI_{n-1}}+\delta _{i_{n-1}y}A_{I_{n-2}}\right)  \notag \\
n^{-1}\int d\overrightarrow{C}\;C_{z}C_{y}H_{I_{n}}\left( \overrightarrow{C}%
\right) \frac{\partial }{\partial C_{x}}\widetilde{f}\left( \overrightarrow{C%
}\right) &=&-\mathcal{S}_{yz}\delta _{xy}\left( A_{zI_{n}}+\mathcal{S}%
_{I_{n}}\delta _{i_{n}z}A_{I_{n-1}}\right)  \notag \\
&&-\mathcal{S}_{I_{n}}\delta _{i_{n}x}\left( H_{I_{n-1}zy}\left( 
\overrightarrow{C}\right) +\delta _{zy}H_{I_{n-1}}\left( \overrightarrow{C}%
\right) \right)  \notag \\
&&-\mathcal{S}_{I_{n}}\delta _{i_{n}x}\left( H\delta _{zi_{n-1}}\delta
_{yi_{n-2}}H_{I_{n-3}}\left( \overrightarrow{C}\right) +\mathcal{S}%
_{yz}\delta _{zi_{n-1}}H_{I_{n-2}y}\left( \overrightarrow{C}\right) \right) .
\notag
\end{eqnarray}%
Using these, the kinetic equation becomes%
\begin{eqnarray}
&&n_{r}^{-1}\frac{\partial }{\partial t^{\prime }}%
n_{r}A_{I_{n}}^{r}+n_{r}^{-1}\frac{\partial }{\partial q_{l}^{\prime }}%
n_{r}u_{l}^{r}A_{I_{n}}^{r}-\mathcal{S}_{I_{n}}\left( \frac{\partial \Gamma
_{i_{n}j}^{r1/2}}{\partial t}\Gamma _{jk}^{r-1/2}+u_{l}^{r}\frac{\partial
\Gamma _{i_{n}j}^{r1/2}}{\partial q_{l}^{\prime }}\Gamma
_{jk}^{r-1/2}-\Gamma _{i_{n}j}^{r1/2}\frac{\partial u_{j}^{r}}{\partial
q_{l}^{\prime }}\Gamma _{lk}^{r-1/2}\right) \left( A_{kI_{n-1}}^{r}+\delta
_{ki_{n-1}}A_{I_{n-2}}^{r}\right) \\
&&+\sqrt{2}\mathcal{S}_{I_{n}}\Gamma _{i_{n}j}^{r1/2}\left( \left( \frac{%
\partial u_{j}^{r}}{\partial t}+u_{l}^{r}\frac{\partial u_{j}^{r}}{\partial
q_{l}^{\prime }}\right) A_{I_{n-1}}^{r}+\frac{1}{2}\frac{\partial }{\partial
q_{l}^{\prime }}\left( \Gamma _{jl}^{r-1}A_{I_{n-1}}^{r}\right) \right) 
\notag \\
&&-\frac{1}{\sqrt{2}}\mathcal{S}_{I_{n}}\left( \frac{\partial \Gamma
_{i_{n}j}^{r1/2}}{\partial q_{l}^{\prime }}\Gamma _{ji_{n-2}}^{r-1/2}\Gamma
_{li_{n-1}}^{r-1/2}\right) A_{I_{n-3}}^{r}  \notag \\
&&+n_{r}^{-1}\frac{1}{\sqrt{2}}\frac{\partial }{\partial q_{l}^{\prime }}%
\left( \Gamma _{lm}^{r-1/2}n_{r}A_{mI_{n}}^{r}\right) -\frac{1}{\sqrt{2}}%
\mathcal{S}_{I_{n}}\left( \left( \frac{\partial \Gamma _{i_{n}j}^{r1/2}}{%
\partial q_{l}^{\prime }}\Gamma _{jk}^{r-1/2}\right) \Gamma
_{lm}^{r-1/2}A_{I_{n-1}mk}^{r}\right)  \notag \\
&&-\frac{1}{\sqrt{2}}\mathcal{S}_{I_{n}}\left( \left( \frac{\partial \Gamma
_{i_{n}j}^{r1/2}}{\partial q_{l}^{\prime }}\Gamma _{jk}^{r-1/2}\right)
\Gamma _{li_{n-1}}^{r-1/2}+\left( \frac{\partial \Gamma _{i_{n}j}^{r1/2}}{%
\partial q_{l}^{\prime }}\Gamma _{ji_{n-1}}^{r-1/2}\right) \Gamma
_{lk}^{r-1/2}\right) A_{I_{n-2}k}^{r}  \notag \\
&=&n_{r}^{-1}\det \left( \overleftrightarrow{\Gamma }^{r}\right) ^{-1/2}\int
d\overrightarrow{C}^{r}H_{I_{n}}\left( \overrightarrow{C}^{r}\right)
\sum_{s}J\left[ f_{r},f_{s}\right] .  \notag
\end{eqnarray}%
The zeroth order equation gives%
\begin{equation*}
n_{r}^{-1}\frac{\partial }{\partial t^{\prime }}n_{r}+n_{r}^{-1}\frac{%
\partial }{\partial q_{l}^{\prime }}n_{r}u_{l}^{r}=0
\end{equation*}%
so that the general equation becomes%
\begin{eqnarray}
&&\frac{\partial }{\partial t^{\prime }}A_{I_{n}}^{r}+u_{l}^{r}\frac{%
\partial }{\partial q_{l}^{\prime }}A_{I_{n}}^{r}-\mathcal{S}_{I_{n}}\left( 
\frac{\partial \Gamma _{i_{n}j}^{r1/2}}{\partial t}\Gamma
_{jk}^{r-1/2}+u_{l}^{r}\frac{\partial \Gamma _{i_{n}j}^{r1/2}}{\partial
q_{l}^{\prime }}\Gamma _{jk}^{r-1/2}-\Gamma _{i_{n}j}^{r1/2}\frac{\partial
u_{j}^{r}}{\partial q_{l}^{\prime }}\Gamma _{lk}^{r-1/2}\right) \left(
A_{kI_{n-1}}^{r}+\delta _{ki_{n-1}}A_{I_{n-2}}^{r}\right) \\
&&+\sqrt{2}\mathcal{S}_{I_{n}}\Gamma _{i_{n}j}^{r1/2}\left( \left( \frac{%
\partial u_{j}^{r}}{\partial t}+u_{l}^{r}\frac{\partial u_{j}^{r}}{\partial
q_{l}^{\prime }}\right) A_{I_{n-1}}^{r}+\frac{1}{2}n_{r}^{-1}\frac{\partial 
}{\partial q_{l}^{\prime }}\left( n_{r}\Gamma
_{jl}^{r-1}A_{I_{n-1}}^{r}\right) \right)  \notag \\
&&-\frac{1}{\sqrt{2}}\mathcal{S}_{I_{n}}\left( \frac{\partial \Gamma
_{i_{n}j}^{r1/2}}{\partial q_{l}^{\prime }}\Gamma _{ji_{n-2}}^{r-1/2}\Gamma
_{li_{n-1}}^{r-1/2}\right) A_{I_{n-3}}^{r}  \notag \\
&&+n_{r}^{-1}\frac{1}{\sqrt{2}}\frac{\partial }{\partial q_{l}^{\prime }}%
\left( \Gamma _{lm}^{r-1/2}n_{r}A_{mI_{n}}^{r}\right) -\frac{1}{\sqrt{2}}%
\mathcal{S}_{I_{n}}\left( \left( \frac{\partial \Gamma _{i_{n}j}^{r1/2}}{%
\partial q_{l}^{\prime }}\Gamma _{jk}^{r-1/2}\right) \Gamma
_{lm}^{r-1/2}A_{I_{n-1}mk}^{r}\right)  \notag \\
&&-\frac{1}{\sqrt{2}}\mathcal{S}_{I_{n}}\left( \left( \frac{\partial \Gamma
_{i_{n}j}^{r1/2}}{\partial q_{l}^{\prime }}\Gamma _{jk}^{r-1/2}\right)
\Gamma _{li_{n-1}}^{r-1/2}+\left( \frac{\partial \Gamma _{i_{n}j}^{r1/2}}{%
\partial q_{l}^{\prime }}\Gamma _{ji_{n-1}}^{r-1/2}\right) \Gamma
_{lk}^{r-1/2}\right) A_{I_{n-2}k}^{r}  \notag \\
&=&n_{r}^{-1}\det \left( \overleftrightarrow{\Gamma }^{r}\right) ^{-1/2}\int
d\overrightarrow{C}^{r}H_{I_{n}}\left( \overrightarrow{C}^{r}\right)
\sum_{s}J\left[ f_{r},f_{s}\right] .  \notag
\end{eqnarray}%
Specializing to USF gives%
\begin{eqnarray}
&&\frac{\partial }{\partial t^{\prime }}A_{I_{n}}^{r}-\mathcal{S}%
_{I_{n}}\left( \frac{\partial \Gamma _{i_{n}j}^{r1/2}}{\partial t}\Gamma
_{jk}^{r-1/2}-a_{jl}\Gamma _{i_{n}j}^{r1/2}\Gamma _{lk}^{r-1/2}\right)
\left( A_{kI_{n-1}}^{r}+\delta _{ki_{n-1}}A_{I_{n-2}}^{r}\right) \\
&=&n_{r}^{-1}\det \left( 2\overleftrightarrow{\Gamma }^{r}\right)
^{-1/2}\int d\overrightarrow{C}^{r}H_{I_{n}}\left( \overrightarrow{C}%
^{r}\right) \sum_{s}J\left[ f_{r},f_{s}\right]  \notag \\
&=&.n_{r}^{-1}\int d\overrightarrow{V}^{r}H_{I_{n}}\left( \overrightarrow{C}%
^{r}\right) \sum_{s}J\left[ f_{r},f_{s}\right]  \notag
\end{eqnarray}%
For the second order GME, this gives%
\begin{equation}
-\frac{\partial \Gamma _{i_{2}j}^{r1/2}}{\partial t}\Gamma _{ji_{1}}^{r-1/2}-%
\frac{\partial \Gamma _{i_{1}j}^{r1/2}}{\partial t}\Gamma
_{ji_{2}}^{r-1/2}+a_{jl}\Gamma _{i_{2}j}^{r1/2}\Gamma
_{li_{1}}^{r-1/2}+a_{jl}\Gamma _{i_{1}j}^{r1/2}\Gamma
_{li_{2}}^{r-1/2}=n_{r}^{-1}.\int d\overrightarrow{V}^{r}H_{I_{n}}\left( 
\overrightarrow{C}^{r}\right) \sum_{s}J\left[ f_{r},f_{s}\right]
\end{equation}%
Multiplying through by $\Gamma _{k_{2}i_{2}}^{r-1/2}\Gamma
_{k_{1}i_{1}}^{r-1/2}$ and summing over $i_{1}$ and $i_{2}$ gives%
\begin{eqnarray}
\frac{\partial \Gamma _{k_{1}k_{2}}^{r-1}}{\partial t}+a_{k_{2}l}\Gamma
_{k_{1}l}^{r-1}+a_{k_{1}l}\Gamma _{k_{2}l}^{r-1} &=&n_{r}^{-1}\int d%
\overrightarrow{V}^{r}\sum_{I_{2}}\Gamma _{k_{2}i_{2}}^{r-1/2}\Gamma
_{k_{1}i_{1}}^{r-1/2}H_{I_{2}}\left( \overrightarrow{C}^{r}\right) \sum_{s}J%
\left[ f_{r},f_{s}\right] \\
&=&2n_{r}{}^{-1}\int d\overrightarrow{V}^{r}V_{k_{1}}V_{k_{2}}\sum_{s}J\left[
f_{r},f_{s}\right] .  \notag
\end{eqnarray}%
For the second order SME, one has%
\begin{equation}
\frac{\partial }{\partial t^{\prime }}A_{I_{2}}^{r}+\frac{\partial \ln T_{r}%
}{\partial t}%
A_{i_{1}i_{2}}^{r}+a_{i_{2}k}A_{ki_{1}}^{r}+a_{i_{1}k}A_{ki_{2}}^{r}+\frac{%
\partial \ln T_{r}}{\partial t}\delta
_{i_{2}i_{1}}+a_{i_{2}i_{1}}+a_{i_{1}i_{2}}=n_{r}^{-1}\frac{m_{r}}{k_{B}T_{r}%
}\int d\overrightarrow{V}^{r}V_{k_{1}}V_{k_{2}}\sum_{s}J\left[ f_{r},f_{s}%
\right]  \notag
\end{equation}

\bigskip

\section{The generating function}

\label{GeneratingFunction}

To evaluate the various kinetic integrals, we need the generating function%
\begin{eqnarray}
Z_{[n]}^{r_{1}r_{2}} &=&-\pi ^{-D}\int d\overrightarrow{V}_{1}d%
\overrightarrow{V}_{2}d\overrightarrow{q}_{2}\;\left( \prod_{j=1}^{n}%
\overrightarrow{q}_{12i}\right) \exp \left( \overrightarrow{\Lambda }\cdot 
\overrightarrow{V}_{1}\right) \overline{T}_{-}(12)\exp \left( -%
\overrightarrow{V}_{1}\cdot \overleftrightarrow{\Gamma }^{r_{1}}\cdot 
\overrightarrow{V}_{1}-\overrightarrow{V}_{2}\cdot \overleftrightarrow{%
\Gamma }^{r_{2}}\cdot \overrightarrow{V}_{2}\right)  \label{A1} \\
&=&\pi ^{-D}\int d\overrightarrow{V}_{1}d\overrightarrow{V}_{2}d%
\overrightarrow{q}_{2}\;\left( \prod_{j=1}^{n}\overrightarrow{q}%
_{12i}\right) \exp \left( -\overrightarrow{V}_{1}\cdot \overleftrightarrow{%
\Gamma }^{r_{1}}\cdot \overrightarrow{V}_{1}-\overrightarrow{V}_{2}\cdot 
\overleftrightarrow{\Gamma }^{r_{2}}\cdot \overrightarrow{V}_{2}\right)
T_{+}(12)\exp \left( \overrightarrow{\Lambda }\cdot \overrightarrow{V}%
_{1}\right)  \notag
\end{eqnarray}%
where the negative adjoint of the collision operator is 
\begin{equation}
T_{+}\left( 12\right) =-\delta \left( q_{12}-\sigma _{r_{1}r_{2}}\right)
\Theta \left( -\overrightarrow{v}_{12}\cdot \widehat{q}_{12}\right) 
\overrightarrow{v}_{12}\cdot \widehat{q}_{12}\left[ \widehat{b}_{12}-1\right]
\label{A2}
\end{equation}%
and in this appendix, I continue the generalization of the first appendix
and allow for an arbitrary flow state so that $\overrightarrow{V}_{i}=%
\overrightarrow{v}_{i}-\overrightarrow{u}^{r_{i}}\left( \overrightarrow{q}%
_{i}\right) $. Using 
\begin{equation}
\widehat{b}_{12}\overrightarrow{V}_{1}=\overrightarrow{V}_{1}-\left(
1+\alpha _{r_{1}r_{2}}\right) \frac{\mu _{r_{1}r_{2}}}{m_{r_{1}}}\left( 
\overrightarrow{v}_{12}\cdot \widehat{q}_{12}\right) \widehat{q}_{12}
\label{A3}
\end{equation}%
the generating function is%
\begin{align}
Z_{[n]}^{r_{1}r_{2}}& =-\pi ^{-D}\sigma _{r_{1}r_{2}}^{D-1}\int d%
\overrightarrow{V}_{1}d\overrightarrow{V}_{2}d\widehat{q}\;\left(
\prod_{j=1}^{n}\sigma _{r_{1}r_{2}}\widehat{q}_{i}\right) \exp \left( -%
\overrightarrow{V}_{1}\cdot \overleftrightarrow{\Gamma }^{r_{1}}\cdot 
\overrightarrow{V}_{1}-\overrightarrow{V}_{2}\cdot \overleftrightarrow{%
\Gamma }^{r_{2}}\cdot \overrightarrow{V}_{1}\right)  \label{A4} \\
& \times \Theta \left( -\overrightarrow{v}_{12}\cdot \widehat{q}\right)
\left( \overrightarrow{v}_{12}\cdot \widehat{q}\right) \left( \exp \left( 
\overrightarrow{\Lambda }\cdot \overrightarrow{V}_{1}-\left( 1+\alpha
_{r_{1}r_{2}}\right) \frac{\mu _{r_{1}r_{2}}}{m_{r_{1}}}\left( 
\overrightarrow{v}_{12}\cdot \widehat{q}\right) \overrightarrow{\Lambda }%
\cdot \widehat{q}_{12}\right) -\exp \left( \overrightarrow{\Lambda }\cdot 
\overrightarrow{V}_{1}\right) \right) .  \notag
\end{align}%
It is enough to restrict attention to the function%
\begin{eqnarray}
\widetilde{Z}^{r_{1}r_{2}}\left( x\right) &=&-2\det \left( 
\overleftrightarrow{\Gamma }^{r_{1}}\overleftrightarrow{\Gamma }%
^{r_{2}}\right) ^{1/2}\pi ^{-D}\int d\overrightarrow{V}_{1}d\overrightarrow{V%
}_{2}\;\Theta \left( -\overrightarrow{v}_{12}\cdot \widehat{q}\right) \left( 
\overrightarrow{v}_{12}\cdot \widehat{q}\right) \exp \left( -g_{12}\right)
\label{A5} \\
g_{12} &=&-\overrightarrow{V}_{1}\cdot \overleftrightarrow{\Gamma }%
^{r_{1}}\cdot \overrightarrow{V}_{1}-\overrightarrow{V}_{2}\cdot 
\overleftrightarrow{\Gamma }^{r_{2}}\cdot \overrightarrow{V}_{2}+%
\overrightarrow{\Lambda }\cdot \overrightarrow{V}_{1}-x\left( 
\overrightarrow{v}_{12}\cdot \widehat{q}\right) \overrightarrow{\Lambda }%
\cdot \widehat{q}_{12}.  \notag
\end{eqnarray}%
in terms of which the full generating function is%
\begin{equation}
Z_{[n]}^{r_{1}r_{2}}=\frac{1}{2}\sigma _{r_{1}r_{2}}^{D-1}\det \left( 
\overleftrightarrow{\Gamma }^{r_{1}}\overleftrightarrow{\Gamma }%
^{r_{2}}\right) ^{-1/2}\int d\widehat{q}\;\left( \prod_{j=1}^{n}\sigma
_{r_{1}r_{2}}\widehat{q}_{i}\right) \left( \widetilde{Z}^{rs}\left( \left(
1+\alpha _{r_{1}r_{2}}\right) \frac{\mu _{r_{1}r_{2}}}{m_{r_{1}}}\right) -%
\widetilde{Z}^{rs}\left( 0\right) \right)
\end{equation}

The velocity integrals are performed by switching to relative and center of
mass (CM) coordinates%
\begin{eqnarray}
\overrightarrow{v} &=&\overrightarrow{V}_{1}-\overrightarrow{V}_{2}
\label{A6} \\
\overrightarrow{V} &=&\frac{m_{r_{1}}}{m_{r_{1}}+m_{r_{2}}}\overrightarrow{V}%
_{1}+\frac{m_{r_{2}}}{m_{r_{1}}+m_{r_{2}}}\overrightarrow{V}_{2}  \notag
\end{eqnarray}%
so that%
\begin{eqnarray}
\overrightarrow{V}_{1} &=&\overrightarrow{V}+\frac{\mu _{r_{1}r_{2}}}{%
m_{r_{1}}}\overrightarrow{v}  \label{A7} \\
\overrightarrow{V}_{2} &=&\overrightarrow{V}-\frac{\mu _{r_{1}r_{2}}}{%
m_{r_{2}}}\overrightarrow{v}.  \notag
\end{eqnarray}%
In terms of the CM\ variables, the argument of the exponential is expanded
by first using%
\begin{eqnarray}
-\overrightarrow{V}_{1}\cdot \overleftrightarrow{\Gamma }^{r_{1}}\cdot 
\overrightarrow{V}_{1}-\overrightarrow{V}_{2}\cdot \overleftrightarrow{%
\Gamma }^{r_{2}}\cdot \overrightarrow{V}_{2} &=&-\overrightarrow{V}\cdot
\left( \overleftrightarrow{\Gamma }^{r_{1}}+\overleftrightarrow{\Gamma }%
^{r_{2}}\right) \cdot \overrightarrow{V}-\overrightarrow{v}\cdot \left(
\left( \frac{\mu _{r_{1}r_{2}}}{m_{r_{1}}}\right) ^{2}\overleftrightarrow{%
\Gamma }^{r_{1}}-\left( \frac{\mu _{r_{1}r_{2}}}{m_{r_{2}}}\right) ^{2}%
\overleftrightarrow{\Gamma }^{r_{2}}\right) \cdot \overrightarrow{v}
\label{A8} \\
&&-2\overrightarrow{V}\cdot \left( \frac{\mu _{r_{1}r_{2}}}{m_{r_{1}}}%
\overleftrightarrow{\Gamma }^{r_{1}}-\frac{\mu _{r_{1}r_{2}}}{m_{r_{2}}}%
\overleftrightarrow{\Gamma }^{r_{2}}\right) \cdot \overrightarrow{v}  \notag
\end{eqnarray}%
and the remaining terms become%
\begin{eqnarray}
&&\overrightarrow{\Lambda }\cdot \overrightarrow{V}_{1}-x\left( 
\overrightarrow{v}_{12}\cdot \widehat{q}_{12}\right) \overrightarrow{\Lambda 
}\cdot \widehat{q}  \label{A9} \\
&=&\overrightarrow{\Lambda }\cdot \overrightarrow{V}+\frac{\mu _{r_{1}r_{2}}%
}{m_{r_{1}}}\overrightarrow{\Lambda }\cdot \overrightarrow{v}-x\left( 
\overrightarrow{\Lambda }\cdot \widehat{q}\right) \overrightarrow{v}\cdot 
\widehat{q}-xw_{r_{1}r_{2}}\overrightarrow{\Lambda }\cdot \widehat{q}  \notag
\end{eqnarray}%
where $w_{r_{1}r_{2}}\equiv \left( \overrightarrow{u}^{r_{1}}\left( 
\overrightarrow{q}_{1}\right) -\overrightarrow{u}^{r_{2}}\left( 
\overrightarrow{q}_{1}-\sigma _{r_{1}r_{2}}\widehat{q}\right) \right) \cdot 
\widehat{q}$ (in USF, $w_{r_{1}r_{2}}=\sigma _{r_{1}r_{2}}\widehat{q}\cdot 
\overleftrightarrow{a}\cdot \widehat{q}$). The first step is to complete the
square in $\overrightarrow{V}$ 
\begin{eqnarray}
&&-\overrightarrow{V}\cdot \left( \overleftrightarrow{\Gamma }^{r_{1}}+%
\overleftrightarrow{\Gamma }^{r_{2}}\right) \cdot \overrightarrow{V}-2%
\overrightarrow{V}\cdot \left( \frac{\mu _{r_{1}r_{2}}}{m_{r_{1}}}%
\overleftrightarrow{\Gamma }^{r_{1}}-\frac{\mu _{r_{1}r_{2}}}{m_{r_{2}}}%
\overleftrightarrow{\Gamma }^{r_{2}}\right) \cdot \overrightarrow{v}+%
\overrightarrow{\Lambda }\cdot \overrightarrow{V}  \label{A10} \\
&=&-\left( \overrightarrow{V}+\overrightarrow{A}\right) \cdot \left( 
\overleftrightarrow{\Gamma }^{r_{1}}+\overleftrightarrow{\Gamma }%
^{r_{2}}\right) \cdot \left( \overrightarrow{V}+\overrightarrow{A}\right) +%
\overrightarrow{A}\cdot \left( \overleftrightarrow{\Gamma }^{r_{1}}+%
\overleftrightarrow{\Gamma }^{r_{2}}\right) \cdot \overrightarrow{A}  \notag
\end{eqnarray}%
with%
\begin{equation}
\overrightarrow{A}=\overrightarrow{v}\cdot \left( \frac{\mu _{r_{1}r_{2}}}{%
m_{r_{1}}}\overleftrightarrow{\Gamma }^{r_{1}}-\frac{\mu _{r_{1}r_{2}}}{%
m_{r_{2}}}\overleftrightarrow{\Gamma }^{r_{2}}\right) \cdot \left( 
\overleftrightarrow{\Gamma }^{r_{1}}+\overleftrightarrow{\Gamma }%
^{r_{2}}\right) ^{-1}-\frac{1}{2}\overrightarrow{\Lambda }\cdot \left( 
\overleftrightarrow{\Gamma }^{r_{1}}+\overleftrightarrow{\Gamma }%
^{r_{2}}\right) ^{-1}.  \label{A11}
\end{equation}%
giving%
\begin{eqnarray}
g_{12} &=&-\left( \overrightarrow{V}+\overrightarrow{A}\right) \cdot \left( 
\overleftrightarrow{\Gamma }^{r_{1}}+\overleftrightarrow{\Gamma }%
^{r_{2}}\right) \cdot \left( \overrightarrow{V}+\overrightarrow{A}\right) +%
\overrightarrow{A}\cdot \left( \overleftrightarrow{\Gamma }^{r_{1}}+%
\overleftrightarrow{\Gamma }^{r_{2}}\right) \cdot \overrightarrow{A}
\label{A12} \\
&&-\overrightarrow{v}\cdot \left( \left( \frac{\mu _{r_{1}r_{2}}}{m_{r_{1}}}%
\right) ^{2}\overleftrightarrow{\Gamma }^{r_{1}}-\left( \frac{\mu
_{r_{1}r_{2}}}{m_{r_{2}}}\right) ^{2}\overleftrightarrow{\Gamma }%
^{r_{2}}\right) \cdot \overrightarrow{v}  \notag \\
&&+\frac{\mu _{r_{1}r_{2}}}{m_{r_{1}}}\overrightarrow{\Lambda }\cdot 
\overrightarrow{v}-x\left( \overrightarrow{\Lambda }\cdot \widehat{q}\right) 
\overrightarrow{v}\cdot \widehat{q}-xw_{r_{1}r_{2}}\overrightarrow{\Lambda }%
\cdot \widehat{q}  \notag
\end{eqnarray}%
This can be simplified by expanding the second term and using%
\begin{eqnarray}
&&\left( \left( \frac{\mu _{r_{1}r_{2}}}{m_{r_{1}}}\overleftrightarrow{%
\Gamma }^{r_{1}}-\frac{\mu _{r_{1}r_{2}}}{m_{r_{2}}}\overleftrightarrow{%
\Gamma }^{r_{2}}\right) \cdot \overrightarrow{v}\right) \cdot \left( 
\overleftrightarrow{\Gamma }^{r_{1}}+\overleftrightarrow{\Gamma }%
^{r_{2}}\right) ^{-1}\cdot \left( \left( \frac{\mu _{r_{1}r_{2}}}{m_{r_{1}}}%
\overleftrightarrow{\Gamma }^{r_{1}}-\frac{\mu _{r_{1}r_{2}}}{m_{r_{2}}}%
\overleftrightarrow{\Gamma }^{r_{2}}\right) \cdot \overrightarrow{v}\right)
\label{A13} \\
&&-\overrightarrow{v}\cdot \left( \left( \frac{\mu _{r_{1}r_{2}}}{m_{r_{1}}}%
\right) ^{2}\overleftrightarrow{\Gamma }^{r_{1}}-\left( \frac{\mu
_{r_{1}r_{2}}}{m_{r_{2}}}\right) ^{2}\overleftrightarrow{\Gamma }%
^{r_{2}}\right) \cdot \overrightarrow{v}  \notag \\
&=&-\overrightarrow{v}\cdot \overleftrightarrow{\Gamma }^{r_{2}}\cdot \left( 
\overleftrightarrow{\Gamma }^{r_{1}}+\overleftrightarrow{\Gamma }%
^{r_{2}}\right) ^{-1}\cdot \overleftrightarrow{\Gamma }^{r_{1}}\cdot 
\overrightarrow{v}  \notag
\end{eqnarray}%
so that%
\begin{eqnarray}
g_{12} &=&-\left( \overrightarrow{V}+\overrightarrow{A}\right) \cdot \left( 
\overleftrightarrow{\Gamma }^{r_{1}}+\overleftrightarrow{\Gamma }%
^{r_{2}}\right) \cdot \left( \overrightarrow{V}+\overrightarrow{A}\right)
\label{A14} \\
&&-\overrightarrow{v}\cdot \overleftrightarrow{\Gamma }^{r_{2}}\cdot \left( 
\overleftrightarrow{\Gamma }^{r_{1}}+\overleftrightarrow{\Gamma }%
^{r_{2}}\right) ^{-1}\cdot \overleftrightarrow{\Gamma }^{r_{1}}\cdot 
\overrightarrow{v}  \notag \\
&&-\overrightarrow{\Lambda }\cdot \left( \overleftrightarrow{\Gamma }%
^{r_{1}}+\overleftrightarrow{\Gamma }^{r_{2}}\right) ^{-1}\cdot \left( \frac{%
\mu _{r_{1}r_{2}}}{m_{r_{1}}}\overleftrightarrow{\Gamma }^{r_{1}}-\frac{\mu
_{r_{1}r_{2}}}{m_{r_{2}}}\overleftrightarrow{\Gamma }^{r_{2}}\right) \cdot 
\overrightarrow{v}  \notag \\
&&+\frac{1}{4}\overrightarrow{\Lambda }\cdot \left( \overleftrightarrow{%
\Gamma }^{r_{1}}+\overleftrightarrow{\Gamma }^{r_{2}}\right) ^{-1}\cdot 
\overrightarrow{\Lambda }  \notag \\
&&+\frac{\mu _{r_{1}r_{2}}}{m_{r_{1}}}\overrightarrow{\Lambda }\cdot 
\overrightarrow{v}-x\left( \overrightarrow{\Lambda }\cdot \widehat{q}\right) 
\overrightarrow{v}\cdot \widehat{q}-xw_{r_{1}r_{2}}\overrightarrow{\Lambda }%
\cdot \widehat{q}.  \notag
\end{eqnarray}%
Furthermore,%
\begin{eqnarray}
&&\overrightarrow{\Lambda }\cdot \left( \overleftrightarrow{\Gamma }^{r_{1}}+%
\overleftrightarrow{\Gamma }^{r_{2}}\right) ^{-1}\cdot \left( \frac{\mu
_{r_{1}r_{2}}}{m_{r_{1}}}\overleftrightarrow{\Gamma }^{r_{1}}-\frac{\mu
_{r_{1}r_{2}}}{m_{r_{2}}}\overleftrightarrow{\Gamma }^{r_{2}}\right) \cdot 
\overrightarrow{v}  \label{A15} \\
&=&\left( \frac{\mu _{r_{1}r_{2}}}{m_{r_{1}}}\right) \overrightarrow{\Lambda 
}\cdot \overrightarrow{v}-\overrightarrow{\Lambda }\cdot \left( 
\overleftrightarrow{\Gamma }^{r_{1}}+\overleftrightarrow{\Gamma }%
^{r_{2}}\right) ^{-1}\cdot \overleftrightarrow{\Gamma }^{r_{2}}\cdot 
\overrightarrow{v}  \notag
\end{eqnarray}%
so%
\begin{eqnarray}
g_{12} &=&-\left( \overrightarrow{V}+\overrightarrow{A}\right) \cdot \left( 
\overleftrightarrow{\Gamma }^{r_{1}}+\overleftrightarrow{\Gamma }%
^{r_{2}}\right) \cdot \left( \overrightarrow{V}+\overrightarrow{A}\right)
\label{A16} \\
&&-\overrightarrow{v}\cdot \overleftrightarrow{\Gamma }^{r_{2}}\cdot \left( 
\overleftrightarrow{\Gamma }^{r_{1}}+\overleftrightarrow{\Gamma }%
^{r_{2}}\right) ^{-1}\cdot \overleftrightarrow{\Gamma }^{r_{1}}\cdot 
\overrightarrow{v}  \notag \\
&&+\overrightarrow{\Lambda }\cdot \left( \overleftrightarrow{\Gamma }%
^{r_{1}}+\overleftrightarrow{\Gamma }^{r_{2}}\right) ^{-1}\cdot 
\overleftrightarrow{\Gamma }^{r_{2}}\cdot \overrightarrow{v}  \notag \\
&&+\frac{1}{4}\overrightarrow{\Lambda }\cdot \left( \overleftrightarrow{%
\Gamma }^{r_{1}}+\overleftrightarrow{\Gamma }^{r_{2}}\right) ^{-1}\cdot 
\overrightarrow{\Lambda }  \notag \\
&&-x\left( \overrightarrow{\Lambda }\cdot \widehat{q}\right) \overrightarrow{%
v}\cdot \widehat{q}-xw_{r_{1}r_{2}}\overrightarrow{\Lambda }\cdot \widehat{q}%
.  \notag
\end{eqnarray}%
Next, we complete the square in $\overrightarrow{v}$ using%
\begin{eqnarray}
&&-\overrightarrow{v}\cdot \overleftrightarrow{\Gamma }^{r_{2}}\cdot \left( 
\overleftrightarrow{\Gamma }^{r_{1}}+\overleftrightarrow{\Gamma }%
^{r_{2}}\right) ^{-1}\cdot \overleftrightarrow{\Gamma }^{r_{1}}\cdot 
\overrightarrow{v}+\overrightarrow{\Lambda }\cdot \left( \overleftrightarrow{%
\Gamma }^{r_{1}}+\overleftrightarrow{\Gamma }^{r_{2}}\right) ^{-1}\cdot 
\overleftrightarrow{\Gamma }^{r_{2}}\cdot \overrightarrow{v}-x\left( 
\overrightarrow{\Lambda }\cdot \widehat{q}\right) \overrightarrow{v}\cdot 
\widehat{q}  \label{A17} \\
&=&-\overrightarrow{v}\cdot \left( \overleftrightarrow{\Gamma }^{r_{1}-1}+%
\overleftrightarrow{\Gamma }^{r_{2-1}}\right) ^{-1}\cdot \overrightarrow{v}%
+\left( \overrightarrow{\Lambda }\cdot \left( \overleftrightarrow{\Gamma }%
^{r_{1}}+\overleftrightarrow{\Gamma }^{r_{2}}\right) ^{-1}\cdot 
\overleftrightarrow{\Gamma }^{r_{2}}-x\left( \overrightarrow{\Lambda }\cdot 
\widehat{q}\right) \widehat{q}\right) \cdot \overrightarrow{v}  \notag \\
&=&-\left( \overrightarrow{v}+\overrightarrow{B}\right) \cdot \left( 
\overleftrightarrow{\Gamma }^{r_{1}-1}+\overleftrightarrow{\Gamma }%
^{r_{2-1}}\right) ^{-1}\cdot \left( \overrightarrow{v}+\overrightarrow{B}%
\right) +\overrightarrow{B}\cdot \left( \overleftrightarrow{\Gamma }%
^{r_{1}-1}+\overleftrightarrow{\Gamma }^{r_{2-1}}\right) ^{-1}\cdot 
\overrightarrow{B}  \notag
\end{eqnarray}%
where%
\begin{eqnarray}
\overrightarrow{B} &=&-\frac{1}{2}\overrightarrow{\Lambda }\cdot \left( 
\overleftrightarrow{\Gamma }^{r_{1}}+\overleftrightarrow{\Gamma }%
^{r_{2}}\right) ^{-1}\cdot \overleftrightarrow{\Gamma }^{r_{2}}\cdot \left( 
\overleftrightarrow{\Gamma }^{r_{1}-1}+\overleftrightarrow{\Gamma }%
^{r_{2-1}}\right) +\frac{1}{2}x\left( \overrightarrow{\Lambda }\cdot 
\widehat{q}\right) \widehat{q}\cdot \left( \overleftrightarrow{\Gamma }%
^{r_{1}-1}+\overleftrightarrow{\Gamma }^{r_{2-1}}\right)  \label{A18} \\
&=&-\frac{1}{2}\overrightarrow{\Lambda }\cdot \overleftrightarrow{\Gamma }%
^{r_{1}-1}+\frac{1}{2}x\left( \overrightarrow{\Lambda }\cdot \widehat{q}%
\right) \widehat{q}\cdot \left( \overleftrightarrow{\Gamma }^{r_{1}-1}+%
\overleftrightarrow{\Gamma }^{r_{2-1}}\right)  \notag
\end{eqnarray}%
giving%
\begin{eqnarray}
g_{12} &=&-\left( \overrightarrow{V}+\overrightarrow{A}\right) \cdot \left( 
\overleftrightarrow{\Gamma }^{r_{1}}+\overleftrightarrow{\Gamma }%
^{r_{2}}\right) \cdot \left( \overrightarrow{V}+\overrightarrow{A}\right)
\label{A19} \\
&&-\left( \overrightarrow{v}+\overrightarrow{B}\right) \cdot \left( 
\overleftrightarrow{\Gamma }^{r_{1}-1}+\overleftrightarrow{\Gamma }%
^{r_{2-1}}\right) ^{-1}\cdot \left( \overrightarrow{v}+\overrightarrow{B}%
\right)  \notag \\
&&+\overrightarrow{B}\cdot \left( \overleftrightarrow{\Gamma }^{r_{1}-1}+%
\overleftrightarrow{\Gamma }^{r_{2-1}}\right) ^{-1}\cdot \overrightarrow{B} 
\notag \\
&&+\frac{1}{4}\overrightarrow{\Lambda }\cdot \left( \overleftrightarrow{%
\Gamma }^{r_{1}}+\overleftrightarrow{\Gamma }^{r_{2}}\right) ^{-1}\cdot 
\overrightarrow{\Lambda }  \notag \\
&&-xw_{r_{1}r_{2}}\overrightarrow{\Lambda }\cdot \widehat{q}.  \notag
\end{eqnarray}%
Then, using%
\begin{eqnarray}
\int \exp \left( -\left( \overrightarrow{V}+\overrightarrow{A}\right) \cdot
\left( \overleftrightarrow{\Gamma }^{r_{1}}+\overleftrightarrow{\Gamma }%
^{r_{2}}\right) \cdot \left( \overrightarrow{V}+\overrightarrow{A}\right)
\right) d\overrightarrow{V} &=&\det \left( \overleftrightarrow{\Gamma }%
^{r_{1}}+\overleftrightarrow{\Gamma }^{r_{2}}\right) ^{-1/2}\int \exp \left(
-V^{2}\right) d\overrightarrow{V} \\
&=&\pi ^{D/2}\det \left( \overleftrightarrow{\Gamma }^{r_{1}}+%
\overleftrightarrow{\Gamma }^{r_{2}}\right) ^{-1/2}  \notag
\end{eqnarray}%
gives%
\begin{eqnarray}
\widetilde{Z}^{r_{1}r_{2}}\left( x\right) &=&-2\det \left( 
\overleftrightarrow{\Gamma }^{r_{1}}\overleftrightarrow{\Gamma }%
^{r_{2}}\right) ^{1/2}\pi ^{D/2}\det \left( \overleftrightarrow{\Gamma }%
^{r_{1}}+\overleftrightarrow{\Gamma }^{r_{2}}\right) ^{-1/2}  \label{A20} \\
&&\times \int d\overrightarrow{u}\;\Theta \left( -\overrightarrow{u}\cdot 
\widehat{q}+\overrightarrow{B}\cdot \widehat{q}-w_{r_{1}r_{2}}\right) \left( 
\overrightarrow{u}\cdot \widehat{q}-\overrightarrow{B}\cdot \widehat{q}%
+w_{r_{1}r_{2}}\right)  \notag \\
&&\times \exp \left( -\overrightarrow{u}\cdot \left( \overleftrightarrow{%
\Gamma }^{r_{1}-1}+\overleftrightarrow{\Gamma }^{r_{2-1}}\right) ^{-1}\cdot 
\overrightarrow{u}\right)  \notag \\
&&\times \exp \left( \overrightarrow{B}\cdot \left( \overleftrightarrow{%
\Gamma }^{r_{1}-1}+\overleftrightarrow{\Gamma }^{r_{2-1}}\right) ^{-1}\cdot 
\overrightarrow{B}+\frac{1}{4}\overrightarrow{\Lambda }\cdot \left( 
\overleftrightarrow{\Gamma }^{r_{1}}+\overleftrightarrow{\Gamma }%
^{r_{2}}\right) ^{-1}\cdot \overrightarrow{\Lambda }-xw_{r_{1}r_{2}}%
\overrightarrow{\Lambda }\cdot \widehat{q}\right) .  \notag
\end{eqnarray}%
Next, expanding%
\begin{equation}
\overrightarrow{B}\cdot \left( \overleftrightarrow{\Gamma }^{r_{1}-1}+%
\overleftrightarrow{\Gamma }^{r_{2-1}}\right) ^{-1}\cdot \overrightarrow{B}+%
\frac{1}{4}\overrightarrow{\Lambda }\cdot \left( \overleftrightarrow{\Gamma }%
^{r_{1}}+\overleftrightarrow{\Gamma }^{r_{2}}\right) ^{-1}\cdot 
\overrightarrow{\Lambda }=\frac{1}{4}\overrightarrow{\Lambda }\cdot 
\overleftrightarrow{G}^{r_{1}r_{2}}\left( x\right) \cdot \overrightarrow{%
\Lambda }
\end{equation}%
with%
\begin{eqnarray}
\overleftrightarrow{G}^{r_{1}r_{2}}\left( x\right) &=&\overleftrightarrow{%
\Gamma }^{r_{1}-1}+2\left( \overleftrightarrow{\Gamma }^{r_{1}}+%
\overleftrightarrow{\Gamma }^{r_{2}}\right) ^{-1}-x\left( 
\overleftrightarrow{\Gamma }^{r_{1}-1}\cdot \widehat{q}\widehat{q}+\widehat{q%
}\overleftrightarrow{\Gamma }^{r_{1}-1}\cdot \widehat{q}\right)
+x^{2}X_{r_{1}r_{2}}^{2}\widehat{q}\widehat{q}  \label{A22} \\
X_{r_{1}r_{2}}^{2} &=&\widehat{q}\cdot \left( \overleftrightarrow{\Gamma }%
^{r_{1}-1}+\overleftrightarrow{\Gamma }^{r_{2-1}}\right) \cdot \widehat{q} 
\notag
\end{eqnarray}%
gives%
\begin{eqnarray}
\widetilde{Z}^{r_{1}r_{2}}\left( x\right) &=&-2\det \left( 
\overleftrightarrow{\Gamma }^{r_{1}}\overleftrightarrow{\Gamma }%
^{r_{2}}\right) ^{1/2}\pi ^{-D/2}\det \left( \overleftrightarrow{\Gamma }%
^{r_{1}}+\overleftrightarrow{\Gamma }^{r_{2}}\right) ^{-1/2}  \label{A23} \\
&&\times \int d\overrightarrow{u}\;\Theta \left( -\overrightarrow{u}\cdot 
\widehat{q}+\overrightarrow{B}\cdot \widehat{q}-w_{r_{1}r_{2}}\right) \left( 
\overrightarrow{u}\cdot \widehat{q}-\overrightarrow{B}\cdot \widehat{q}%
+w_{r_{1}r_{2}}\right)  \notag \\
&&\times \exp \left( -\overrightarrow{u}\cdot \left( \overleftrightarrow{%
\Gamma }^{r_{1}-1}+\overleftrightarrow{\Gamma }^{r_{2-1}}\right) ^{-1}\cdot 
\overrightarrow{u}\right) \exp \left( \frac{1}{4}\overrightarrow{\Lambda }%
\cdot \overleftrightarrow{G}^{r_{1}r_{2}}\left( x\right) \cdot 
\overrightarrow{\Lambda }-xw_{r_{1}r_{2}}\overrightarrow{\Lambda }\cdot 
\widehat{q}\right) .  \notag
\end{eqnarray}%
The velocity integral is performed using%
\begin{eqnarray}
&&\int d\overrightarrow{u}\;\Theta \left( -\overrightarrow{u}\cdot \widehat{q%
}+\overrightarrow{B}\cdot \widehat{q}-w_{r_{1}r_{2}}\right) \left( 
\overrightarrow{u}\cdot \widehat{q}-\overrightarrow{B}\cdot \widehat{q}%
+w_{r_{1}r_{2}}\right) \exp \left( -\overrightarrow{u}\cdot 
\overleftrightarrow{M}\cdot \overrightarrow{u}\right)  \label{A24} \\
&=&\det \left( \overleftrightarrow{M}\right) ^{-1/2}\int d\overrightarrow{u}%
^{\prime }\;\Theta \left( -\overrightarrow{u}^{\prime }\cdot 
\overleftrightarrow{M}^{-1/2}\cdot \widehat{q}+\overrightarrow{B}\cdot 
\widehat{q}-w_{r_{1}r_{2}}\right) \left( \overrightarrow{u}^{\prime }\cdot 
\overleftrightarrow{M}^{-1/2}\cdot \widehat{q}-\overrightarrow{B}\cdot 
\widehat{q}+w_{r_{1}r_{2}}\right) \exp \left( -u^{\prime 2}\right)  \notag \\
&=&\det \left( \overleftrightarrow{M}\right) ^{-1/2}\pi ^{\frac{D-1}{2}%
}\left| \overleftrightarrow{M}^{-1/2}\cdot \widehat{q}\right| \int_{-\infty
}^{\frac{\overrightarrow{B}\cdot \widehat{q}-w_{r_{1}r_{2}}}{\left| 
\overleftrightarrow{M}^{-1/2}\cdot \widehat{q}\right| }}\;\left( u^{\prime }-%
\frac{\overrightarrow{B}\cdot \widehat{q}-w_{r_{1}r_{2}}}{\left| 
\overleftrightarrow{M}^{-1/2}\cdot \widehat{q}\right| }\right) \exp \left(
-u^{\prime 2}\right) du^{\prime }  \notag \\
&=&-\frac{1}{2}\det \left( \overleftrightarrow{M}\right) ^{-1/2}\pi ^{\frac{D%
}{2}}\left| \overleftrightarrow{M}^{-1/2}\cdot \widehat{q}\right|
F_{1}\left( \frac{w_{r_{1}r_{2}}-\overrightarrow{B}\cdot \widehat{q}}{\left| 
\overleftrightarrow{M}^{-1/2}\cdot \widehat{q}\right| }\right)  \notag
\end{eqnarray}%
where%
\begin{equation}
F_{n}(x)\equiv -\frac{2}{\sqrt{\pi }}\int_{-\infty }^{-x}\;\left( u^{\prime
}+x\right) \exp \left( -u^{\prime 2}\right) du^{\prime }
\end{equation}%
so that%
\begin{eqnarray}
F_{0}(x) &=&\text{erf}\left( x\right) -1  \notag \\
F_{1}(x) &=&\frac{1}{\sqrt{\pi }}e^{-x^{2}}+x\left( \text{erf}\left(
x\right) -1\right)
\end{eqnarray}%
Noting that $\left| \left( \overleftrightarrow{\Gamma }^{r_{1}-1}+%
\overleftrightarrow{\Gamma }^{r_{2-1}}\right) ^{1/2}\cdot \widehat{q}\right|
=X_{r_{1}r_{2}}$ and $\det \left( \left( \overleftrightarrow{\Gamma }%
^{r_{1}-1}+\overleftrightarrow{\Gamma }^{r_{2-1}}\right) ^{-1}\right)
^{-1/2}=\det \left( \overleftrightarrow{\Gamma }^{r_{1}}\right) ^{-1/2}\det
\left( \overleftrightarrow{\Gamma }^{r_{2}}\right) ^{-1/2}\det \left( 
\overleftrightarrow{\Gamma }^{r_{1}}+\overleftrightarrow{\Gamma }%
^{r_{2}}\right) ^{1/2}$ one has that%
\begin{equation}
\widetilde{Z}^{r_{1}r_{2}}\left( x\right) =F_{1}\left( \frac{w_{r_{1}r_{2}}-%
\overrightarrow{B}\cdot \widehat{q}}{\left| \overleftrightarrow{M}%
^{-1/2}\cdot \widehat{q}\right| }\right) \exp \left( \frac{1}{4}%
\overrightarrow{\Lambda }\cdot \overleftrightarrow{G}^{r_{1}r_{2}}\left(
x\right) \cdot \overrightarrow{\Lambda }-xw_{r_{1}r_{2}}\overrightarrow{%
\Lambda }\cdot \widehat{q}\right) .
\end{equation}

The final result is then summarized as

\begin{equation}
Z_{[n]}^{r_{1}r_{2}}=\frac{1}{2}\det \left( \overleftrightarrow{\Gamma }%
^{r_{1}}\overleftrightarrow{\Gamma }^{r_{2}}\right) ^{-1/2}\sigma
_{r_{1}r_{2}}^{D-1}\int d\widehat{q}\;\left( \prod_{j=1}^{n}\sigma
_{r_{1}r_{2}}\widehat{q}_{i_{j}}\right) \left[ \widetilde{Z}%
^{r_{1}r_{2}}\left( \left( 1+\alpha _{r_{1}r_{2}}\right) \frac{\mu
_{r_{1}r_{2}}}{m_{r_{1}}}\right) -\widetilde{Z}^{r_{1}r_{2}}\left( 0\right) %
\right]
\end{equation}%
with%
\begin{eqnarray}
\widetilde{Z}^{r_{1}r_{2}}\left( x\right) &=&X_{r_{1}r_{2}}F_{1}\left( \frac{%
2w_{r_{1}r_{2}}+\overrightarrow{\Lambda }\cdot \overleftrightarrow{\Gamma }%
^{r_{1}-1}\cdot \widehat{q}_{12}}{2X_{r_{1}r_{2}}}-\frac{1}{2}x\left( 
\overrightarrow{\Lambda }\cdot \widehat{q}\right) X_{r_{1}r_{2}}\right)
\label{A25} \\
&&\times \exp \left( \frac{1}{4}\overrightarrow{\Lambda }\cdot 
\overleftrightarrow{G}^{r_{1}r_{2}}\left( x\right) \cdot \overrightarrow{%
\Lambda }-xw_{r_{1}r_{2}}\overrightarrow{\Lambda }\cdot \widehat{q}\right) 
\notag \\
X_{r_{1}r_{2}} &=&\sqrt{\widehat{q}\cdot \left( \overleftrightarrow{\Gamma }%
^{r_{1-1}}+\overleftrightarrow{\Gamma }^{r_{2-1}}\right) \cdot \widehat{q}} 
\notag \\
\overleftrightarrow{G}^{r_{1}r_{2}}\left( x\right) &=&\overleftrightarrow{%
\Gamma }^{r_{1}-1}+2\left( \overleftrightarrow{\Gamma }^{r_{1}}+%
\overleftrightarrow{\Gamma }^{r_{2}}\right) ^{-1}-2x\overleftrightarrow{%
\Gamma }^{r_{1}-1}\cdot \widehat{q}\widehat{q}+x^{2}X_{r_{1}r_{2}}^{2}%
\widehat{q}\widehat{q}  \notag \\
w_{r_{1}r_{2}} &=&\left( \overrightarrow{u}^{r_{1}}\left( \overrightarrow{q}%
_{1}\right) -\overrightarrow{u}^{r_{2}}\left( \overrightarrow{q}_{1}-\sigma
_{r_{1}r_{2}}\widehat{q}\right) \right) \cdot \widehat{q}.  \notag
\end{eqnarray}%
In this calculation, it has been implicitly assumed that $%
\overleftrightarrow{\Gamma }^{r_{1}},\overleftrightarrow{\Gamma }^{r_{2}}$
and $n_{r_{1}},n_{r_{2}}$ are independent of position. However, this
assumption is unnecessary and the same result applies for spatially
dependent quantities provided the substitutions 
\begin{eqnarray}
\overleftrightarrow{\Gamma }^{r_{1}} &\rightarrow &\overleftrightarrow{%
\Gamma }^{r_{1}}\left( \overrightarrow{q}_{1}\right) \\
\overleftrightarrow{\Gamma }^{r_{2}} &\rightarrow &\overleftrightarrow{%
\Gamma }^{r_{2}}\left( \overrightarrow{q}_{1}-\sigma _{r_{1}r_{2}}\widehat{q}%
\right)  \notag
\end{eqnarray}%
etc., are made and quantities involving $\widehat{q}$ are brought under the
integrals.

\section{Evaluation of the collision integrals}

\label{Evaluations}

In this Appendix, the generating function is used to evaluate the
coefficients of the moment expansions.

\subsection{\protect\bigskip Evaluation of $E_{i_{1}i_{2}}^{r_{1}r_{2}}$}

We need 
\begin{equation}
E_{i_{1}i_{2}}^{r_{1}r_{2}}=2\det \left( \overleftrightarrow{\Gamma }^{r_{1}}%
\overleftrightarrow{\Gamma }^{r_{2}}\right) ^{1/2}\lim_{\overrightarrow{%
\Lambda }\rightarrow 0}\frac{\partial ^{2}}{\partial \Lambda
_{i_{1}}\partial \Lambda _{i_{2}}}\left[ \widetilde{Z}^{r_{1}r_{2}}\left(
x\right) -\widetilde{Z}^{r_{1}r_{2}}\left( 0\right) \right]
\end{equation}%
which is evaluated using%
\begin{eqnarray}
\lim_{\overrightarrow{\Lambda }\rightarrow 0}\frac{\partial ^{2}}{\partial
\Lambda _{i_{1}}\partial \Lambda _{i_{2}}}\widetilde{Z}^{r_{1}r_{2}}\left(
x\right) &=&X_{r_{1}r_{2}}F_{1}^{\prime \prime }\left( \frac{w_{r_{1}r_{2}}}{%
X_{r_{1}r_{2}}}\right) \left( \frac{\overleftrightarrow{\Gamma }%
^{r_{1}-1}\cdot \widehat{q}}{2X_{r_{1}r_{2}}}\right) _{i_{1}}\left( \frac{%
\overleftrightarrow{\Gamma }^{r_{1}-1}\cdot \widehat{q}}{2X_{r_{1}r_{2}}}%
\right) _{i_{2}}+X_{r_{1}r_{2}}F_{1}\left( \frac{w_{r_{1}r_{2}}}{%
X_{r_{1}r_{2}}}\right) \frac{1}{2}G_{i_{1}i_{2}}^{r_{1}r_{2}}\left( x\right)
\\
&&-\frac{1}{2}xX_{r_{1}r_{2}}\left[ \frac{1}{2}F_{1}^{\prime \prime }\left( 
\frac{w_{r_{1}r_{2}}}{X_{r_{1}r_{2}}}\right) +\left( \frac{w_{r_{1}r_{2}}}{%
X_{r_{1}r_{2}}}\right) F_{1}^{\prime }\left( \frac{w_{r_{1}r_{2}}}{%
X_{r_{1}r_{2}}}\right) \right] \left[ \left( \overleftrightarrow{\Gamma }%
^{r_{1}-1}\cdot \widehat{q}\right) _{i_{1}}\widehat{q}_{i_{2}}+\left( 
\overleftrightarrow{\Gamma }^{r_{1}-1}\cdot \widehat{q}\right) _{i_{2}}%
\widehat{q}_{i_{1}}\right]  \notag \\
&&+x^{2}X_{r_{1}r_{2}}^{3}\left[ \left( \frac{1}{2}\right) ^{2}F_{1}^{\prime
\prime }\left( \frac{w_{r_{1}r_{2}}}{X_{r_{1}r_{2}}}\right) +\left( \frac{%
w_{r_{1}r_{2}}}{X_{r_{1}r_{2}}}\right) F_{1}^{\prime }\left( \frac{%
w_{r_{1}r_{2}}}{X_{r_{1}r_{2}}}\right) +\left( \frac{w_{r_{1}r_{2}}}{%
X_{r_{1}r_{2}}}\right) ^{2}F_{1}\left( \frac{w_{r_{1}r_{2}}}{X_{r_{1}r_{2}}}%
\right) \right] \widehat{q}_{i_{1}}\widehat{q}_{i_{j}}  \notag
\end{eqnarray}%
Substituting the explicit expression for $\overleftrightarrow{G}\left(
x\right) $ gives%
\begin{eqnarray}
\lim_{\overrightarrow{\Lambda }\rightarrow 0}\frac{\partial ^{2}}{\partial
\Lambda _{i_{1}}\partial \Lambda _{i_{2}}}\left[ \widetilde{Z}%
^{r_{1}r_{2}}\left( x\right) -\widetilde{Z}^{r_{1}r_{2}}\left( 0\right) %
\right] &=&\widetilde{Z}^{r_{1}r_{2}}\left( x,\frac{w_{r_{1}r_{2}}}{%
X_{r_{1}r2}}\right) \\
\widetilde{Z}^{r_{1}r_{2}}\left( x,y\right) &=&-\frac{1}{2}xX_{r_{1}r_{2}}%
\left[ \frac{1}{2}F_{1}^{\prime \prime }\left( y\right) +yF_{1}^{\prime
}\left( y\right) +F_{1}\left( y\right) \right] \left[ \left( 
\overleftrightarrow{\Gamma }^{r_{1}-1}\cdot \widehat{q}\right) _{i_{1}}%
\widehat{q}_{i_{2}}+\left( \overleftrightarrow{\Gamma }^{r_{1}-1}\cdot 
\widehat{q}\right) _{i_{2}}\widehat{q}_{i_{1}}\right]  \notag \\
&&+x^{2}X_{r_{1}r_{2}}^{3}\left[ \left( \frac{1}{2}\right) ^{2}F_{1}^{\prime
\prime }\left( y\right) +yF_{1}^{\prime }\left( y\right) +y^{2}F_{1}\left(
y\right) +\frac{1}{2}F_{1}\left( y\right) \right] \widehat{q}_{i_{1}}%
\widehat{q}_{i_{j}}  \notag
\end{eqnarray}%
so that, using $F_{n}^{\prime }(x)=nF_{n-1}(x)+\delta _{n0}2\left(
F_{1}(x)-xF_{0}(x)\right) $, one has%
\begin{eqnarray}
\widetilde{Z}^{r_{1}r_{2}}\left( x,y\right) &=&-xX_{r_{1}r_{2}}F_{1}\left(
y\right) \left[ \left( \overleftrightarrow{\Gamma }^{r_{1}-1}\cdot \widehat{q%
}\right) _{i_{1}}\widehat{q}_{i_{2}}+\left( \overleftrightarrow{\Gamma }%
^{r_{1}-1}\cdot \widehat{q}\right) _{i_{2}}\widehat{q}_{i_{1}}\right] \\
&&+\frac{1}{2}x^{2}X_{r_{1}r_{2}}^{3}\left[ yF_{0}\left( y\right) +\left(
2y^{2}+2\right) F_{1}\left( y\right) \right] \widehat{q}_{i_{1}}\widehat{q}%
_{i_{j}}  \notag
\end{eqnarray}%
and%
\begin{eqnarray}
E_{i_{1}i_{2}}^{r_{1}r_{2}} &=&-\sigma _{r_{1}r_{2}}^{D-1}\left( 1+\alpha
_{r_{1}r_{2}}\right) \frac{\mu _{r_{1}r_{2}}}{m_{r_{1}}}\int d\widehat{q}%
\;X_{r_{1}r_{2}}F_{1}\left( \frac{w_{r_{1}r_{2}}}{X_{r_{1}r_{2}}}\right) %
\left[ \left( \overleftrightarrow{\Gamma }^{r_{1}-1}\cdot \widehat{q}\right)
_{i_{1}}\widehat{q}_{i_{2}}+\left( \overleftrightarrow{\Gamma }%
^{r_{1}-1}\cdot \widehat{q}\right) _{i_{2}}\widehat{q}_{i_{1}}\right] \\
&&+\frac{1}{2}\sigma _{r_{1}r_{2}}^{D-1}\left( 1+\alpha _{r_{1}r_{2}}\right)
^{2}\left( \frac{\mu _{r_{1}r_{2}}}{m_{r_{1}}}\right) ^{2}\int d\widehat{q}\;%
\widehat{q}_{i_{1}}\widehat{q}_{i_{2}}X_{r_{1}r_{2}}^{3}\left[ \left( \frac{%
w_{r_{1}r_{2}}}{X_{r_{1}r_{2}}}\right) F_{0}\left( \frac{w_{r_{1}r_{2}}}{%
X_{r_{1}r_{2}}}\right) +F_{1}\left( \frac{w_{r_{1}r_{2}}}{X_{r_{1}r_{2}}}%
\right) \left( 2\left( \frac{w_{r_{1}r_{2}}}{X_{r_{1}r_{2}}}\right)
^{2}+2\right) \right] .  \notag
\end{eqnarray}

\bigskip 

\subsection{Evaluation of $B_{i_{1}i_{2}}^{r_{1}r_{2}}$}

This follows by taking the appropriate limit of $E_{i_{1}i_{2}}^{r_{1}r_{2}}$%
: 
\begin{eqnarray}
B_{i_{1}i_{2}}^{r_{1}r_{2}} &=&\frac{m_{r_{1}}}{2k_{B}T_{r_{1}}}\lim_{%
\overleftrightarrow{\Gamma }^{x}\rightarrow \frac{m_{x}}{2k_{B}T_{x}}%
\overleftrightarrow{1}}E_{i_{1}i_{2}}^{r_{1}r_{2}} \\
&=&-2\sigma _{r_{1}r_{2}}^{D-1}\left( 1+\alpha _{r_{1}r_{2}}\right) \frac{%
\mu _{r_{1}r_{2}}}{m_{r_{1}}}Y_{r_{1}r_{2}}\int d\widehat{q}\;\widehat{q}%
_{i_{1}}\widehat{q}_{i_{2}}F_{1}\left( \frac{w_{r_{1}r_{2}}}{X_{r_{1}r_{2}}}%
\right)  \notag \\
&&+\frac{1}{4}\sigma _{r_{1}r_{2}}^{D-1}\left( 1+\alpha _{r_{1}r_{2}}\right)
^{2}\left( \frac{\mu _{r_{1}r_{2}}}{m_{r_{1}}}\right) ^{2}Y_{r_{1}r_{2}}^{3}%
\frac{m_{r_{1}}}{k_{B}T_{r_{1}}}\int d\widehat{q}\;\widehat{q}_{i_{1}}%
\widehat{q}_{i_{2}}\left[ \left( \frac{w_{r_{1}r_{2}}}{Y_{r_{1}r_{2}}}%
\right) F_{0}\left( \frac{w_{r_{1}r_{2}}}{Y_{r_{1}r_{2}}}\right)
+F_{1}\left( \frac{w_{r_{1}r_{2}}}{Y_{r_{1}r_{2}}}\right) \left( 2\left( 
\frac{w_{r_{1}r_{2}}}{Y_{r_{1}r_{2}}}\right) ^{2}+2\right) \right]  \notag
\end{eqnarray}%
with%
\begin{equation}
Y_{r_{1}r_{2}}=\sqrt{2\frac{k_{B}T_{r_{1}}}{m_{r_{1}}}+2\frac{k_{B}T_{r_{2}}%
}{m_{r_{2}}}}.
\end{equation}

\subsection{Evaluation of $D_{i_{1}i_{2},j_{1}j_{2}}^{r_{1}r_{2}}$\protect%
\bigskip}

We need to evaluate$D_{i_{1}i_{2},j_{1}j_{2}}^{r_{1}r_{2}}=-\frac{m_{r_{1}}}{%
2k_{B}T_{r_{1}}}\frac{m_{r_{2}}}{k_{B}T_{r_{2}}}\lim_{\overleftrightarrow{%
\Gamma }^{x}\rightarrow \frac{m_{x}}{2k_{B}T_{x}}\overleftrightarrow{1}}%
\frac{\partial }{\partial \Gamma _{j_{1}j_{2}}^{r_{2}}}%
E_{i_{1}i_{2}}^{r_{1}r_{2}}$

\begin{eqnarray}
D_{i_{1}i_{2},j_{1}j_{2}}^{r_{1}r_{2}} &=&-\frac{m_{r_{1}}}{2k_{B}T_{r_{1}}}%
\frac{m_{r_{2}}}{k_{B}T_{r_{2}}}\lim_{\overleftrightarrow{\Gamma }%
^{x}\rightarrow \frac{m_{x}}{2k_{B}T_{x}}\overleftrightarrow{1}}\frac{%
\partial }{\partial \Gamma _{j_{1}j_{2}}^{r_{2}}}\det \left( 
\overleftrightarrow{\Gamma }^{r_{1}}\overleftrightarrow{\Gamma }%
^{r_{2}}\right) ^{1/2}E_{i_{1}i_{2}}^{r_{1}r_{2}} \\
&=&\sigma _{r_{1}r_{2}}^{D-1}\left( 1+\alpha _{r_{1}r_{2}}\right) \frac{\mu
_{r_{1}r_{2}}}{m_{r_{1}}}\frac{m_{r_{1}}}{2k_{B}T_{r_{1}}}\frac{m_{r_{2}}}{%
k_{B}T_{r_{2}}}  \notag \\
&&\times \lim_{\overleftrightarrow{\Gamma }^{x}\rightarrow \frac{m_{x}}{%
2k_{B}T_{x}}\overleftrightarrow{1}}\frac{\partial }{\partial \Gamma
_{j_{1}j_{2}}^{r_{2}}}\int d\widehat{q}\;X_{r_{1}r_{2}}F_{1}\left( \frac{%
w_{r_{1}r_{2}}}{X_{r_{1}r_{2}}}\right) \left[ \left( \overleftrightarrow{%
\Gamma }^{r_{1}-1}\cdot \widehat{q}\right) _{i_{1}}\widehat{q}%
_{i_{2}}+\left( \overleftrightarrow{\Gamma }^{r_{1}-1}\cdot \widehat{q}%
\right) _{i_{2}}\widehat{q}_{i_{1}}\right]  \notag \\
&&-\frac{1}{4}\sigma _{r_{1}r_{2}}^{D-1}\left( 1+\alpha _{r_{1}r_{2}}\right)
^{2}\left( \frac{\mu _{r_{1}r_{2}}}{m_{r_{1}}}\right) ^{2}\frac{m_{r_{1}}}{%
k_{B}T_{r_{1}}}\frac{m_{r_{2}}}{k_{B}T_{r_{2}}}  \notag \\
&&\times \lim_{\overleftrightarrow{\Gamma }^{x}\rightarrow \frac{m_{x}}{%
2k_{B}T_{x}}\overleftrightarrow{1}}\frac{\partial }{\partial \Gamma
_{j_{1}j_{2}}^{r_{2}}}\int d\widehat{q}\;\widehat{q}_{i_{1}}\widehat{q}%
_{i_{2}}X_{r_{1}r_{2}}^{3}\left[ \left( \frac{w_{r_{1}r_{2}}}{X_{r_{1}r_{2}}}%
\right) F_{0}\left( \frac{w_{r_{1}r_{2}}}{X_{r_{1}r_{2}}}\right)
+F_{1}\left( \frac{w_{r_{1}r_{2}}}{X_{r_{1}r_{2}}}\right) \left( 2\left( 
\frac{w_{r_{1}r_{2}}}{X_{r_{1}r_{2}}}\right) ^{2}+2\right) \right]  \notag
\end{eqnarray}%
Then, using%
\begin{equation}
\lim_{\overleftrightarrow{\Gamma }^{x}\rightarrow \frac{m_{x}}{2k_{B}T_{x}}%
\overleftrightarrow{1}}\frac{\partial }{\partial \Gamma _{j_{1}j_{2}}^{r_{2}}%
}X_{r_{1}r_{2}}=-\frac{1}{2}Y_{r_{1}r_{2}}^{-1}\left( \frac{2k_{B}T_{r_{2}}}{%
m_{r_{2}}}\right) ^{2}\widehat{q}_{j_{1}}\widehat{q}_{j_{2}}
\end{equation}%
gives%
\begin{eqnarray}
D_{i_{1}i_{2},j_{1}j_{2}}^{r_{1}r_{2}} &=&\sigma _{r_{1}r_{2}}^{D-1}\left(
1+\alpha _{r_{1}r_{2}}\right) \frac{\mu _{r_{1}r_{2}}}{m_{r_{1}}}\frac{%
m_{r_{1}}}{2k_{B}T_{r_{1}}}\frac{m_{r_{2}}}{k_{B}T_{r_{2}}}\frac{%
4k_{B}T_{r_{1}}}{m_{r_{1}}} \\
&&\times \int d\widehat{q}\;\widehat{q}_{i_{1}}\widehat{q}_{i_{2}}\widehat{q}%
_{i_{1}^{\prime }}\widehat{q}_{i_{2}^{\prime }}\left( -\frac{1}{2}%
Y_{r_{1}r_{2}}^{-1}\left( \frac{2k_{B}T_{r_{2}}}{m_{r_{2}}}\right)
^{2}\right) \left( \lim_{\overleftrightarrow{\Gamma }^{x}\rightarrow \frac{%
m_{x}}{2k_{B}T_{x}}\overleftrightarrow{1}}\frac{\partial }{\partial
X_{r_{1}r_{2}}}X_{r_{1}r_{2}}F_{1}\left( \frac{w_{r_{1}r_{2}}}{X_{r_{1}r_{2}}%
}\right) \right)  \notag \\
&&-\frac{1}{4}\sigma _{r_{1}r_{2}}^{D-1}\left( 1+\alpha _{r_{1}r_{2}}\right)
^{2}\left( \frac{\mu _{r_{1}r_{2}}}{m_{r_{1}}}\right) ^{2}\frac{m_{r_{1}}}{%
k_{B}T_{r_{1}}}\frac{m_{r_{2}}}{k_{B}T_{r_{2}}}\int d\widehat{q}\;\widehat{q}%
_{i_{1}}\widehat{q}_{i_{2}}\widehat{q}_{j_{1}}\widehat{q}_{j_{2}}\left( -%
\frac{1}{2}Y_{r_{1}r_{2}}^{-1}\left( \frac{2k_{B}T_{r_{2}}}{m_{r_{2}}}%
\right) ^{2}\right)  \notag \\
&&\times \left( \lim_{\overleftrightarrow{\Gamma }^{x}\rightarrow \frac{m_{x}%
}{2k_{B}T_{x}}\overleftrightarrow{1}}\frac{\partial }{\partial X_{r_{1}r_{2}}%
}X_{r_{1}r_{2}}^{3}\left[ \left( \frac{w_{r_{1}r_{2}}}{X_{r_{1}r_{2}}}%
\right) F_{0}\left( \frac{w_{r_{1}r_{2}}}{X_{r_{1}r_{2}}}\right)
+F_{1}\left( \frac{w_{r_{1}r_{2}}}{X_{r_{1}r_{2}}}\right) \left( 2\left( 
\frac{w_{r_{1}r_{2}}}{X_{r_{1}r_{2}}}\right) ^{2}+2\right) \right] \right) 
\notag
\end{eqnarray}

Using%
\begin{eqnarray}
\frac{\partial }{\partial X}XF_{1}\left( \frac{w}{X}\right) &=&F_{1}\left( 
\frac{w}{X}\right) -\left( \frac{w}{X}\right) F_{0}\left( \frac{w}{X}\right)
\\
\frac{\partial }{\partial X}X^{3}\left[ \left( \frac{w}{X}\right)
F_{0}\left( \frac{w}{X}\right) +F_{1}\left( \frac{w}{X}\right) \left(
2\left( \frac{w}{X}\right) ^{2}+2\right) \right] &=&6X^{2}F_{1}\left( \frac{w%
}{X}\right)  \notag
\end{eqnarray}%
gives%
\begin{eqnarray}
D_{i_{1}i_{2},j_{1}j_{2}}^{r_{1}r_{2}} &=&-4\sigma _{r_{1}r_{2}}^{D-1}\left(
1+\alpha _{r_{1}r_{2}}\right) \frac{\mu _{r_{1}r_{2}}}{m_{r_{1}}}%
Y_{r_{1}r2}^{-1}\left( \frac{k_{B}T_{r_{2}}}{m_{r_{2}}}\right) \\
&&\times \int d\widehat{q}\;\widehat{q}_{i_{1}}\widehat{q}_{i_{2}}\widehat{q}%
_{j_{1}}\widehat{q}_{j_{2}}\left( F_{1}\left( \frac{w_{r_{1}r_{2}}}{%
Y_{r_{1}r_{2}}}\right) -\left( \frac{w_{r_{1}r_{2}}}{Y_{r_{1}r_{2}}}\right)
F_{0}\left( \frac{w_{r_{1}r_{2}}}{Y_{r_{1}r_{2}}}\right) \right)  \notag \\
&&+3\sigma _{r_{1}r_{2}}^{D-1}\left( 1+\alpha _{r_{1}r_{2}}\right)
^{2}\left( \frac{\mu _{r_{1}r_{2}}}{m_{r_{1}}}\right) ^{2}\frac{m_{r_{1}}}{%
k_{B}T_{r_{1}}}\left( \frac{k_{B}T_{r_{2}}}{m_{r_{2}}}\right)
Y_{r_{1}r_{2}}\int d\widehat{q}\;\widehat{q}_{i_{1}}\widehat{q}_{i_{2}}%
\widehat{q}_{j_{1}}\widehat{q}_{j_{2}}F_{1}\left( \frac{w_{r_{1}r_{2}}}{%
Y_{r_{1}r_{2}}}\right) .  \notag
\end{eqnarray}

\subsection{Evaluation of $C_{i_{1}i_{2},j_{1}j_{2}}^{r_{1}r_{2}}$}

This calculation is very similar to the preceding one. Noting that in the
previous calculation we had%
\begin{equation}
\left( \frac{m_{r_{1}}}{k_{B}T_{r_{1}}}\right) \left( \frac{m_{r_{2}}}{%
k_{B}T_{r_{2}}}\right) \lim_{\overleftrightarrow{\Gamma }^{x}\rightarrow 
\frac{m_{x}}{2k_{B}T_{x}}\overleftrightarrow{1}}\frac{\partial }{\partial
\Gamma _{j_{1}j_{2}}^{r_{2}}}X_{r_{1}r_{2}}=-2Y_{r_{1}r_{2}}^{-1}\widehat{q}%
_{j_{1}}\widehat{q}_{j_{2}}\left( \frac{m_{r_{1}}}{k_{B}T_{r_{1}}}\right)
\left( \frac{k_{B}T_{r_{2}}}{m_{r_{2}}}\right) 
\end{equation}%
whereas from the definition%
\begin{equation}
C_{i_{1}i_{2},j_{1}j_{2}}^{r_{1}r_{2}}=-\frac{1}{2}\left( \frac{m_{r_{1}}}{%
k_{B}T_{r_{1}}}\right) ^{2}\lim_{\overleftrightarrow{\Gamma }^{x}\rightarrow 
\frac{m_{x}}{2k_{B}T_{x}}\overleftrightarrow{1}}\frac{\partial }{\partial
\Gamma _{j_{1}j_{2}}^{r_{1}}}E_{i_{1}i_{2}}^{r_{1}r_{2}}
\end{equation}%
the present calculation will require 
\begin{equation}
\left( \frac{m_{r_{1}}}{k_{B}T_{r_{1}}}\right) ^{2}\lim_{\overleftrightarrow{%
\Gamma }^{x}\rightarrow \frac{m_{x}}{2k_{B}T_{x}}\overleftrightarrow{1}}%
\frac{\partial }{\partial \Gamma _{j_{1}j_{2}}^{r_{1}}}%
X_{r_{1}r_{2}}=-2Y_{r_{1}r_{2}}^{-1}\widehat{q}_{j_{1}}\widehat{q}_{j_{2}}
\end{equation}%
we can immediately write%
\begin{eqnarray}
C_{i_{1}i_{2},j_{1}j_{2}}^{r_{1}r_{2}} &=&\left( \frac{m_{r_{2}}}{%
k_{B}T_{r_{2}}}\right) \left( \frac{k_{B}T_{r_{1}}}{m_{r_{1}}}\right)
D_{i_{1}i_{2},j_{1}j_{2}}^{r_{1}r_{2}} \\
&&+\frac{1}{2}\left( \frac{m_{r_{1}}}{k_{B}T_{r_{1}}}\right) ^{2}\sigma
_{r_{1}r_{2}}^{D-1}\left( 1+\alpha _{r_{1}r_{2}}\right) \frac{\mu
_{r_{1}r_{2}}}{m_{r_{1}}}Y_{r_{1}r_{2}}  \notag \\
&&\times \int d\widehat{q}\;F_{1}\left( \frac{w_{r_{1}r_{2}}}{Y_{r_{1}r_{2}}}%
\right) \lim_{\overleftrightarrow{\Gamma }^{x}\rightarrow \frac{m_{x}}{%
2k_{B}T_{x}}\overleftrightarrow{1}}\frac{\partial }{\partial \Gamma
_{j_{1}j_{2}}^{r_{1}}}\left[ \left( \overleftrightarrow{\Gamma }%
^{r_{1}-1}\cdot \widehat{q}\right) _{i_{1}}\widehat{q}_{i_{2}}+\left( 
\overleftrightarrow{\Gamma }^{r_{1}-1}\cdot \widehat{q}\right) _{i_{2}}%
\widehat{q}_{i_{1}}\right] .  \notag
\end{eqnarray}%
Using%
\begin{equation}
\lim_{\overleftrightarrow{\Gamma }^{x}\rightarrow \frac{m_{x}}{2k_{B}T_{x}}%
\overleftrightarrow{1}}\frac{\partial }{\partial \Gamma _{j_{1}j_{2}}^{r_{1}}%
}\overleftrightarrow{\Gamma }_{i_{1}i_{1}^{\prime }}^{r_{1}-1}=-\lim_{%
\overleftrightarrow{\Gamma }^{x}\rightarrow \frac{m_{x}}{2k_{B}T_{x}}%
\overleftrightarrow{1}}\overleftrightarrow{\Gamma }_{i_{1}j_{1}}^{r_{1}-1}%
\overleftrightarrow{\Gamma }_{j_{2}i_{1}^{\prime }}^{r_{1}-1}=-\left( \frac{%
m_{1}}{2k_{B}T_{1}}\right) ^{-2}\delta _{i_{1}j_{1}}\delta
_{j_{2}i_{1}^{\prime }}
\end{equation}%
gives%
\begin{eqnarray}
C_{i_{1}i_{2},j_{1}j_{2}}^{r_{1}r_{2}} &=&\left( \frac{m_{r_{2}}}{%
k_{B}T_{r_{2}}}\right) \left( \frac{k_{B}T_{r_{1}}}{m_{r_{1}}}\right)
D_{i_{1}i_{2},j_{1}j_{2}}^{r_{1}r_{2}} \\
&&-2\sigma _{r_{1}r_{2}}^{D-1}\left( 1+\alpha _{r_{1}r_{2}}\right) \frac{\mu
_{r_{1}r_{2}}}{m_{r_{1}}}Y_{r_{1}r_{2}}\int d\widehat{q}\;\left( \delta
_{i_{1}j_{1}}\widehat{q}_{i_{2}}\widehat{q}_{j_{2}}+\delta _{i_{2}j_{1}}%
\widehat{q}_{i_{1}}\widehat{q}_{j_{2}}\right) F_{1}\left( \frac{%
w_{r_{1}r_{2}}}{Y_{r_{1}r_{2}}}\right)   \notag
\end{eqnarray}%
\begin{equation*}
\left[ \delta _{i_{1}j_{1}}\widehat{q}_{j_{2}}\widehat{q}_{i_{2}}+\delta
_{i_{2}j_{1}}\widehat{q}_{j_{2}}\widehat{q}_{i_{1}}\right] \Gamma
_{j_{1}j_{2}}^{r_{1}-1}
\end{equation*}

\subsection{Evaluation of the pressure}

Recall that the collisional part of the pressure is given by%
\begin{equation}
P_{i_{1}i_{2}}^{V}=\frac{1}{4}\sum_{r_{1}r_{2}}n_{r_{1}}n_{r_{2}}\chi
_{r_{1}r_{2}}\det \left( \overleftrightarrow{\Gamma }^{r_{1}}%
\overleftrightarrow{\Gamma }^{r_{2}}\right) ^{1/2}m_{r_{1}}\lim_{%
\overleftrightarrow{\Lambda }\rightarrow \overleftrightarrow{0}}\frac{%
\partial }{\partial \Lambda _{i_{2}}}Z_{i_{1}}^{r_{1}r_{2}}.
\end{equation}%
Starting with%
\begin{equation}
\lim_{\overleftrightarrow{\Lambda }\rightarrow \overleftrightarrow{0}}\frac{%
\partial }{\partial \Lambda _{i_{2}}}\widetilde{Z}^{r_{1}r_{2}}(x)=%
\;X_{r_{1}r_{2}}F_{1}^{\prime }\left( \frac{w_{r_{1}r_{2}}}{X_{r_{1}r_{2}}}%
\right) \left( \frac{\overleftrightarrow{\Gamma }^{r_{1}-1}\cdot \widehat{q}%
_{12}}{2X_{r_{1}r_{2}}}-\frac{1}{2}xX_{r_{1}r_{2}}\widehat{q}\right)
_{i_{2}}+X_{r_{1}r_{2}}F_{1}\left( \frac{w_{r_{1}r_{2}}}{X_{r_{1}r_{2}}}%
\right) \left( -xw_{r_{1}r_{2}}\widehat{q}_{i_{2}}\right)
\end{equation}%
gives%
\begin{equation}
\lim_{\overrightarrow{\Lambda }\rightarrow 0}\frac{\partial }{\partial
\Lambda _{i_{2}}}\left[ \widetilde{Z}^{r_{1}r_{2}}\left( x;%
\overleftrightarrow{\Gamma }^{r_{1}},\overleftrightarrow{\Gamma }^{r_{2}},%
\overrightarrow{\Lambda }\right) -\widetilde{Z}^{r_{1}r_{2}}\left( 0;%
\overleftrightarrow{\Gamma }^{r_{1}},\overleftrightarrow{\Gamma }^{r_{2}},%
\overrightarrow{\Lambda }\right) \right] =-\frac{1}{2}\widehat{q}%
_{i_{2}}xX_{r_{1}r_{2}}^{2}\left( F_{1}^{\prime }\left( \frac{w_{r_{1}r_{2}}%
}{X_{r_{1}r_{2}}}\right) +2\frac{w_{r_{1}r_{2}}}{X_{r_{1}r_{2}}}F_{1}\left( 
\frac{w_{r_{1}r_{2}}}{X_{r_{1}r_{2}}}\right) \right)
\end{equation}%
and%
\begin{equation}
P_{i_{1}i_{2}}^{V}=-\frac{1}{8}\sum_{r_{1}r_{2}}n_{r_{1}}n_{r_{2}}\sigma
_{r_{1}r_{2}}^{D}\chi _{r_{1}r_{2}}\left( 1+\alpha _{r_{1}r_{2}}\right) \mu
_{r_{1}r_{2}}\int d\widehat{q}\;\widehat{q}_{i_{1}}\widehat{q}%
_{i_{2}}X_{r_{1}r_{2}}^{2}\left( F_{0}\left( \frac{w_{r_{1}r_{2}}}{%
X_{r_{1}r_{2}}}\right) +2\frac{w_{r_{1}r_{2}}}{X_{r_{1}r_{2}}}F_{1}\left( 
\frac{w_{r_{1}r_{2}}}{X_{r_{1}r_{2}}}\right) \right) .
\end{equation}

\section{SME\ in the Boltzmann Limit}

\label{BoltzmannLimit}

The Boltzmann limit of the coefficients needed for the SME are%
\begin{eqnarray}
B_{i_{1}i_{2}}^{r_{1}r_{2}} &=&-\sigma _{r_{1}r_{2}}^{D-1}\left( 1+\alpha
_{r_{1}r_{2}}\right) \frac{\mu _{r_{1}r_{2}}}{m_{r_{1}}}\frac{1}{\sqrt{\pi }}%
Y_{r_{1}r_{2}}\left[ 2-\frac{1}{2}\left( 1+\alpha _{r_{1}r_{2}}\right)
\left( \frac{\mu _{r_{1}r_{2}}}{m_{r_{1}}}\right) Y_{r_{1}r_{2}}^{2}\frac{%
m_{r_{1}}}{k_{B}T_{r_{1}}}\right] \int d\widehat{q}\;\widehat{q}_{i_{1}}%
\widehat{q}_{i_{2}} \\
C_{i_{1}i_{2},i_{1}^{\prime }i_{2}^{\prime }}^{r_{1}r_{2}} &=&\frac{T_{r_{1}}%
}{T_{r_{2}}}\frac{m_{r_{2}}}{m_{r_{1}}}%
D_{i_{1}i_{2},j_{1}j_{2}}^{r_{1}r_{2}}-2\sigma _{r_{1}r_{2}}^{D-1}\left(
1+\alpha _{r_{1}r_{2}}\right) \frac{\mu _{r_{1}r_{2}}}{m_{r_{1}}}%
Y_{r_{1}r_{2}}\frac{1}{\sqrt{\pi }}\int d\widehat{q}\;\left( \delta
_{i_{1}i_{1}^{\prime }}\widehat{q}_{i_{2}}\widehat{q}_{i_{2}^{\prime
}}+\delta _{i_{2}i_{1}^{\prime }}\widehat{q}_{i_{1}}\widehat{q}%
_{i_{2}^{\prime }}\right)  \notag \\
D_{i_{1}i_{2},i_{1}^{\prime }i_{2}^{\prime }}^{r_{1}r_{2}} &=&-4\sigma
_{r_{1}r_{2}}^{D-1}\left( 1+\alpha _{r_{1}r_{2}}\right) \frac{\mu
_{r_{1}r_{2}}}{m_{r_{1}}}Y_{r_{1}r2}^{-1}\left( \frac{k_{B}T_{r_{2}}}{%
m_{r_{2}}}\right) \frac{1}{\sqrt{\pi }}\left[ 1-\frac{3}{4}\left( 1+\alpha
_{r_{1}r_{2}}\right) \left( \frac{\mu _{r_{1}r_{2}}}{m_{r_{1}}}\right) \frac{%
m_{r_{1}}}{k_{B}T_{r_{1}}}Y_{r_{1}r_{2}}^{2}\right] \int d\widehat{q}\;%
\widehat{q}_{i_{1}}\widehat{q}_{i_{2}}\widehat{q}_{i_{1}^{\prime }}\widehat{q%
}_{i_{2}^{\prime }}  \notag
\end{eqnarray}%
where%
\begin{equation}
Y_{r_{1}r_{2}}=\sqrt{2\frac{k_{B}T_{r_{1}}}{m_{r_{1}}}+2\frac{k_{B}T_{r_{2}}%
}{m_{r_{2}}}}.
\end{equation}%
Using the elementary integrals%
\begin{eqnarray}
\int d\widehat{q}\;\widehat{q}_{i_{1}}\widehat{q}_{i_{2}} &=&\frac{S_{D}}{D}%
\delta _{i_{1}i_{2}} \\
\int d\widehat{q}\;\widehat{q}_{i_{1}}\widehat{q}_{i_{2}}\widehat{q}%
_{i_{1}^{\prime }}\widehat{q}_{i_{2}^{\prime }} &=&\frac{S_{D}}{D^{2}+2D}%
\left( \delta _{i_{1}i_{2}}\delta _{i_{1}^{\prime }i_{2}^{\prime }}+\delta
_{i_{1}i_{1}^{\prime }}\delta _{i_{2}i_{2}^{\prime }}+\delta
_{i_{1}i_{2}^{\prime }}\delta _{i_{1}^{\prime }i_{2}}\right)  \notag
\end{eqnarray}%
where the area of a sphere in $D$ dimensions is 
\begin{equation}
S_{D}=\frac{2\pi ^{D/2}}{\Gamma \left( D/2\right) }
\end{equation}%
gives%
\begin{eqnarray}
B_{i_{1}i_{2}}^{r_{1}r_{2}} &=&B^{r_{1}r_{2}}\delta _{i_{1}i_{2}} \\
C_{i_{1}i_{2},i_{1}^{\prime }i_{2}^{\prime }}^{r_{1}r_{2}} &=&\frac{T_{r_{1}}%
}{T_{r_{2}}}\frac{m_{r_{2}}}{m_{r_{1}}}%
D_{i_{1}i_{2},j_{1}j_{2}}^{r_{1}r_{2}}+\;C^{r_{1}r_{2}}\left( \delta
_{i_{1}i_{1}^{\prime }}\delta _{i_{2}i_{2}^{\prime }}+\delta
_{i_{2}i_{1}^{\prime }}\delta _{i_{1}i_{2}^{\prime }}\right)  \notag \\
D_{i_{1}i_{2},i_{1}^{\prime }i_{2}^{\prime }}^{r_{1}r_{2}}
&=&D^{r_{1}r_{2}}\left( \delta _{i_{1}i_{2}}\delta _{i_{1}^{\prime
}i_{2}^{\prime }}+\delta _{i_{1}i_{1}^{\prime }}\delta _{i_{2}i_{2}^{\prime
}}+\delta _{i_{1}i_{2}^{\prime }}\delta _{i_{1}^{\prime }i_{2}}\right) 
\notag \\
B^{r_{1}r_{2}} &=&-\sigma _{r_{1}r_{2}}^{D-1}\left( 1+\alpha
_{r_{1}r_{2}}\right) \frac{\mu _{r_{1}r_{2}}}{m_{r_{1}}}\frac{1}{\sqrt{\pi }}%
Y_{r_{1}r_{2}}\left[ 2-\frac{1}{2}\left( 1+\alpha _{r_{1}r_{2}}\right)
\left( \frac{\mu _{r_{1}r_{2}}}{m_{r_{1}}}\right) Y_{r_{1}r_{2}}^{2}\frac{%
m_{r_{1}}}{k_{B}T_{r_{1}}}\right] \frac{S_{D}}{D}  \notag \\
C^{r_{1}r_{2}} &=&-2\sigma _{r_{1}r_{2}}^{D-1}\left( 1+\alpha
_{r_{1}r_{2}}\right) \frac{\mu _{r_{1}r_{2}}}{m_{r_{1}}}Y_{r_{1}r_{2}}\frac{1%
}{\sqrt{\pi }}\frac{S_{D}}{D}  \notag \\
D^{r_{1}r_{2}} &=&-4\sigma _{r_{1}r_{2}}^{D-1}\left( 1+\alpha
_{r_{1}r_{2}}\right) \frac{\mu _{r_{1}r_{2}}}{m_{r_{1}}}Y_{r_{1}r2}^{-1}%
\left( \frac{k_{B}T_{r_{2}}}{m_{r_{2}}}\right) \frac{1}{\sqrt{\pi }}\left[ 1-%
\frac{3}{4}\left( 1+\alpha _{r_{1}r_{2}}\right) \left( \frac{\mu
_{r_{1}r_{2}}}{m_{r_{1}}}\right) \frac{m_{r_{1}}}{k_{B}T_{r_{1}}}%
Y_{r_{1}r_{2}}^{2}\right] \frac{S_{D}}{D^{2}+2D}  \notag
\end{eqnarray}%
so that%
\begin{eqnarray}
D_{i_{1}i_{2},i_{1}^{\prime }i_{2}^{\prime }}^{r_{1}r_{2}}A_{i_{1}^{\prime
}i_{2}^{\prime }}^{r_{2}} &=&2D^{r_{1}r_{2}}A_{i_{1}i_{2}}^{r_{2}} \\
C_{i_{1}i_{2},i_{1}^{\prime }i_{2}^{\prime }}^{r_{1}r_{2}}A_{i_{1}^{\prime
}i_{2}^{\prime }}^{r_{1}} &=&2\frac{T_{r_{1}}}{T_{r_{2}}}\frac{m_{r_{2}}}{%
m_{r_{1}}}D^{r_{1}r_{2}}A_{i_{1}i_{2}}^{r_{1}}+2%
\;C^{r_{1}r_{2}}A_{i_{1}i_{2}}^{r_{1}}.  \notag
\end{eqnarray}%
Then, the moment equations become%
\begin{eqnarray}
2aA_{xy}^{r_{1}}\delta _{ix} &=&n\mathcal{B}^{r_{1}}+n%
\sum_{r_{2}}D^{r_{1}r_{2}}A_{ii}^{r_{2}}+n\mathcal{H}^{r_{1}}A_{ii}^{r_{1}}
\\
a+aA_{yy}^{r_{1}} &=&n\sum_{r_{2}}D^{r_{1}r_{2}}A_{xy}^{r_{2}}+n\mathcal{H}%
^{r_{1}}A_{xy}^{r_{1}}  \notag
\end{eqnarray}%
where%
\begin{eqnarray}
\mathcal{B}^{r_{1}} &=&\sum_{r_{2}}B^{r_{1}r_{2}}  \notag \\
\mathcal{H}^{r_{1}} &=&\sum_{r_{2}}\left[ \frac{T_{r_{1}}}{T_{r_{2}}}\frac{%
m_{r_{2}}}{m_{r_{1}}}D^{r_{1}r_{2}}+C^{r_{1}r_{2}}\right]  \notag
\end{eqnarray}%
Clearly all $A_{ii}^{r_{1}}$ are equal for $i\neq x$ and the tracelessness
means that $A_{xx}^{r_{1}}=-\left( D-1\right) A_{yy}^{r_{1}}.$ Then%
\begin{eqnarray}
\mathcal{H}^{r_{1}}A_{yy}^{r_{1}} &=&-\mathcal{B}^{r_{1}}-%
\sum_{r_{2}}D^{r_{1}r_{2}}A_{yy}^{r_{2}}  \label{lowdensity} \\
2aA_{xy}^{r_{1}} &=&nD\mathcal{B}^{r_{1}}  \notag \\
a^{2} &=&n^{2}D\frac{\sum_{r_{2}}D^{r_{1}r_{2}}\mathcal{B}^{r_{2}}+\mathcal{H%
}^{r_{1}}\mathcal{B}^{r_{1}}}{2+2A_{yy}^{r_{1}}}  \notag \\
T &=&\sum_{r}x_{r}T_{r}  \notag
\end{eqnarray}%
which constitute $n+n+n+1=3n+1$ equations for the unknowns $\left\{
A_{yy}^{r_{1}},A_{xy}^{r_{1}},T_{r_{1}}\right\} _{r_{1}=1}^{n},a$. For a
one-component fluid, one has that%
\begin{eqnarray}
Y_{r_{1}r_{2}} &\rightarrow &2\sqrt{\frac{k_{B}T}{m}} \\
\mathcal{B}^{r_{1}} &\rightarrow &-\left( 1-\alpha ^{2}\right) \frac{S_{D}}{D%
}\sigma ^{D-1}\sqrt{\frac{k_{B}T}{\pi m}}  \notag \\
D^{r_{1}r_{1}} &\rightarrow &\left( 1+\alpha \right) \frac{1+3\alpha }{2}%
\frac{S_{D}}{D^{2}+2D}\sigma ^{D-1}\sqrt{\frac{k_{B}T}{\pi m}}  \notag \\
\mathcal{H}^{r_{1}} &\rightarrow &\left( 1+\alpha \right) \left( \frac{%
1+3\alpha }{2}\frac{1}{D+2}-2\right) \frac{S_{D}}{D}\sigma ^{D-1}\sqrt{\frac{%
k_{B}T}{\pi m}}  \notag
\end{eqnarray}%
and eqs.(\ref{lowdensity}) can be solved explicitly with the result that%
\begin{eqnarray}
A_{yy} &=&-\frac{\left( 1-\alpha \right) \left( D+2\right) }{3-3\alpha +2D}
\\
a^{\ast }A_{xy}^{r_{1}} &=&-n^{\ast }\left( 1-\alpha ^{2}\right) \frac{S_{D}%
}{2\sqrt{\pi }}  \notag \\
a^{\ast } &=&n^{\ast }\frac{S_{D}}{D}\sqrt{\frac{D}{2\pi }}\left( 3-3\alpha
+2D\right) \sqrt{\frac{\left( 1-\alpha ^{2}\right) \left( 1+\alpha \right) }{%
\left( D+2\right) \left( D+1+\alpha \left( D-1\right) \right) }}  \notag
\end{eqnarray}%
where%
\begin{eqnarray}
a^{\ast } &=&a\sqrt{\frac{m\sigma ^{2}}{k_{B}T}} \\
n^{\ast } &=&n\sigma ^{D}.  \notag
\end{eqnarray}%
Recall that in this approximation%
\begin{equation}
P_{ij}^{K}=nk_{B}T\left( \overrightarrow{1}+\overleftrightarrow{A}^{r}\right)
\end{equation}%
so that%
\begin{eqnarray}
P_{yy} &=&nk_{B}T\left( \frac{1+D+\left( D-1\right) \alpha }{3+2D-3\alpha }%
\right) \\
P_{xy} &=&-nk_{B}T\left[ \frac{1}{\left( 3-3\alpha +2D\right) }\sqrt{D\left( 
\frac{D+2}{2}\right) \left( 1-\alpha \right) \left( D+1+\alpha \left(
D-1\right) \right) }\right]  \notag \\
\eta &=&-P_{xy}/a=\eta _{0}\left( \frac{1}{3-3\alpha +2D}\right) ^{2}\frac{%
4D\left( D+1+\alpha \left( D-1\right) \right) }{\left( 1+\alpha \right) } 
\notag
\end{eqnarray}%
with 
\begin{equation}
\eta _{0}=\sigma ^{1-D}\sqrt{mk_{B}T}\left( \frac{\left( D+2\right) \Gamma
\left( D/2\right) }{8\pi ^{\left( D-1\right) /2}}\right) .
\end{equation}

\bigskip

\bigskip

\bigskip

\bigskip

\bigskip 
\bibliographystyle{prsty}
\bibliography{physics}

\begin{thebibliography}{10}
\expandafter\ifx\csname bibnamefont\endcsname\relax
  \def\bibnamefont#1{#1}\fi
\expandafter\ifx\csname bibfnamefont\endcsname\relax
  \def\bibfnamefont#1{#1}\fi
\expandafter\ifx\csname url\endcsname\relax
  \def\url#1{\texttt{#1}}\fi
\expandafter\ifx\csname urlprefix\endcsname\relax\def\urlprefix{URL }\fi
\expandafter\ifx\csname bibinfo\endcsname\relax \def\bibinfo#1#2{#2}\fi
\expandafter\ifx\csname eprint\endcsname\relax \def\eprint#1{#1}\fi

\bibitem{GranPhysToday}
\bibinfo{author}{\bibfnamefont{H.~M.} \bibnamefont{Jaeger}},
  \bibinfo{author}{\bibfnamefont{S.~R.} \bibnamefont{Nagel}}, \bibnamefont{and}
  \bibinfo{author}{\bibfnamefont{R.~P.} \bibnamefont{Behringer}},
  \bibinfo{journal}{Phys. Today} \textbf{\bibinfo{volume}{49}},
  \bibinfo{pages}{32} (\bibinfo{year}{1996}).

\bibitem{GranRMP}
\bibinfo{author}{\bibfnamefont{H.~M.} \bibnamefont{Jaeger}},
  \bibinfo{author}{\bibfnamefont{S.~R.} \bibnamefont{Nagel}}, \bibnamefont{and}
  \bibinfo{author}{\bibfnamefont{R.~P.} \bibnamefont{Behringer}},
  \bibinfo{journal}{Rev. Mod. Phys.} \textbf{\bibinfo{volume}{68}},
  \bibinfo{pages}{1259} (\bibinfo{year}{1996}).

\bibitem{CampbellReview}
\bibinfo{author}{\bibfnamefont{C.~S.} \bibnamefont{Campbell}},
  \bibinfo{journal}{Ann. Rev. Fluid Mechanics} \textbf{\bibinfo{volume}{22}},
  \bibinfo{pages}{57} (\bibinfo{year}{1990}).

\bibitem{JenkinsRichman}
\bibinfo{author}{\bibfnamefont{J.~T.} \bibnamefont{Jenkins}} \bibnamefont{and}
  \bibinfo{author}{\bibfnamefont{M.~W.} \bibnamefont{Richman}},
  \bibinfo{journal}{J. Fluid Mech.} \textbf{\bibinfo{volume}{192}},
  \bibinfo{pages}{313} (\bibinfo{year}{1988}).

\bibitem{SelaGoldhirsch}
\bibinfo{author}{\bibfnamefont{N.}~\bibnamefont{Sel}},
  \bibinfo{author}{\bibfnamefont{I.}~\bibnamefont{Goldhirsch}},
  \bibnamefont{and} \bibinfo{author}{\bibfnamefont{S.~H.}
  \bibnamefont{Noskowicz}}, \bibinfo{journal}{Phys. Fluids}
  \textbf{\bibinfo{volume}{8}}, \bibinfo{pages}{2337} (\bibinfo{year}{1996}).

\bibitem{ChouRichman1}
\bibinfo{author}{\bibfnamefont{C.-S.} \bibnamefont{Chou}} \bibnamefont{and}
  \bibinfo{author}{\bibfnamefont{M.~W.} \bibnamefont{Richman}},
  \bibinfo{journal}{Physica A} \textbf{\bibinfo{volume}{259}},
  \bibinfo{pages}{430} (\bibinfo{year}{1998}).

\bibitem{ChouRichman2}
\bibinfo{author}{\bibfnamefont{C.-S.} \bibnamefont{Chou}},
  \bibinfo{journal}{Physica A} \textbf{\bibinfo{volume}{287}},
  \bibinfo{pages}{127} (\bibinfo{year}{2001}).

\bibitem{HCSLiouville}
\bibinfo{author}{\bibfnamefont{J.}~\bibnamefont{J.Brey}},
  \bibinfo{author}{\bibfnamefont{J.~W.} \bibnamefont{Dufty}}, \bibnamefont{and}
  \bibinfo{author}{\bibfnamefont{A.}~\bibnamefont{Santos}},
  \bibinfo{journal}{J. Stat. Phys.} \textbf{\bibinfo{volume}{87}},
  \bibinfo{pages}{1051} (\bibinfo{year}{1997}).

\bibitem{ErnstHCS}
\bibinfo{author}{\bibfnamefont{T.~P.~C.} \bibnamefont{van Noije}}
  \bibnamefont{and} \bibinfo{author}{\bibfnamefont{M.~H.} \bibnamefont{Ernst}},
  \bibinfo{journal}{Granular Matter} \textbf{\bibinfo{volume}{1}},
  \bibinfo{pages}{57} (\bibinfo{year}{1998}).

\bibitem{LutskoJCP}
\bibinfo{author}{\bibfnamefont{J.~F.} \bibnamefont{Lutsko}},
  \bibinfo{journal}{J. Chem. Phys.} \textbf{\bibinfo{volume}{120}},
  \bibinfo{pages}{6325} (\bibinfo{year}{2004}).

\bibitem{Lutsko96}
\bibinfo{author}{\bibfnamefont{J.~F.} \bibnamefont{Lutsko}},
  \bibinfo{journal}{Phys. Rev. Lett.} \textbf{\bibinfo{volume}{77}},
  \bibinfo{pages}{2225} (\bibinfo{year}{1996}).

\bibitem{LutskoHCS}
\bibinfo{author}{\bibfnamefont{J.~F.} \bibnamefont{Lutsko}},
  \bibinfo{journal}{Phys. Rev. E} \textbf{\bibinfo{volume}{63}},
  \bibinfo{pages}{061211} (\bibinfo{year}{2001}).

\bibitem{Lutsko_EnskogPRL}
\bibinfo{author}{\bibfnamefont{J.~F.} \bibnamefont{Lutsko}},
  \bibinfo{journal}{Phys. Rev. Lett.} \textbf{\bibinfo{volume}{78}},
  \bibinfo{pages}{243} (\bibinfo{year}{1997}).

\bibitem{LutskoEnskog}
\bibinfo{author}{\bibfnamefont{J.~F.} \bibnamefont{Lutsko}},
  \bibinfo{journal}{Phys. Rev. E} \textbf{\bibinfo{volume}{58}},
  \bibinfo{pages}{434} (\bibinfo{year}{1998}).

\bibitem{GarzoDiff}
\bibinfo{author}{\bibfnamefont{V.}~\bibnamefont{Garzo}},
  \bibinfo{journal}{Phys. Rev. E} \textbf{\bibinfo{volume}{66}},
  \bibinfo{pages}{021308} (\bibinfo{year}{2002}).

\bibitem{TanGold}
\bibinfo{author}{\bibfnamefont{M.-L.} \bibnamefont{Tan}} \bibnamefont{and}
  \bibinfo{author}{\bibfnamefont{I.}~\bibnamefont{Goldhirsch}},
  \bibinfo{journal}{Phys. Fluids} \textbf{\bibinfo{volume}{9}},
  \bibinfo{pages}{856} (\bibinfo{year}{1997}).

\bibitem{LeesEdwards}
\bibinfo{author}{\bibfnamefont{A.}~\bibnamefont{Lees}} \bibnamefont{and}
  \bibinfo{author}{\bibfnamefont{S.}~\bibnamefont{Edwards}},
  \bibinfo{journal}{J. Phys. C} \textbf{\bibinfo{volume}{5}},
  \bibinfo{pages}{1921} (\bibinfo{year}{1972}).

\bibitem{DSMC}
\bibinfo{author}{\bibfnamefont{J.~M.} \bibnamefont{Montanero}}
  \bibnamefont{and} \bibinfo{author}{\bibfnamefont{A.}~\bibnamefont{Santos}},
  \bibinfo{journal}{Phys. Rev. E} \textbf{\bibinfo{volume}{54}},
  \bibinfo{pages}{438} (\bibinfo{year}{1996}).

\bibitem{RET}
\bibinfo{author}{\bibfnamefont{H.}~\bibnamefont{van Beijeren}}
  \bibnamefont{and} \bibinfo{author}{\bibfnamefont{M.~H.} \bibnamefont{Ernst}},
  \bibinfo{journal}{Physica} \textbf{\bibinfo{volume}{68}},
  \bibinfo{pages}{437} (\bibinfo{year}{1973}).

\bibitem{ChapmanCowling}
\bibinfo{author}{\bibfnamefont{S.}~\bibnamefont{Chapman}} \bibnamefont{and}
  \bibinfo{author}{\bibfnamefont{T.~G.} \bibnamefont{Cowling}},
  \emph{\bibinfo{title}{Mathematical Theory of Nonuniform Gases}}
  (\bibinfo{publisher}{Cambridge University Press},
  \bibinfo{address}{Cambridge}, \bibinfo{year}{1970}).

\bibitem{Truesdell}
\bibinfo{author}{\bibfnamefont{C.}~\bibnamefont{Truesdell}} \bibnamefont{and}
  \bibinfo{author}{\bibfnamefont{R.}~\bibnamefont{Muncaster}},
  \emph{\bibinfo{title}{Fundamentals of Maxwell's Kinetic Theory of a Simple
  Monoatomic Gas}} (\bibinfo{publisher}{Academic}, \bibinfo{address}{New York},
  \bibinfo{year}{1980}).

\bibitem{GSL}
\emph{\bibinfo{title}{The GNU scientific library}},
  \eprint{http://sources.redhat.com/gsl}.

\bibitem{GoddardAlam}
\bibinfo{author}{\bibfnamefont{J.~D.} \bibnamefont{Goddard}} \bibnamefont{and}
  \bibinfo{author}{\bibfnamefont{M.}~\bibnamefont{Alam}},
  \bibinfo{journal}{Particulate Science and Technology}
  \textbf{\bibinfo{volume}{17}}, \bibinfo{pages}{69} (\bibinfo{year}{1999}).

\bibitem{LutskoShearInstab}
\bibinfo{author}{\bibfnamefont{J.~F.} \bibnamefont{Lutsko}},
  \bibinfo{journal}{cond-mat/0403551}  (\bibinfo{year}{2004}).

\bibitem{GarzoShearMixture}
\bibinfo{author}{\bibfnamefont{J.~M.} \bibnamefont{Montanero}}
  \bibnamefont{and} \bibinfo{author}{\bibfnamefont{V.}~\bibnamefont{Garzo}},
  \bibinfo{journal}{Physica A} \textbf{\bibinfo{volume}{310}},
  \bibinfo{pages}{17} (\bibinfo{year}{2002}).

\bibitem{SantosChi}
\bibinfo{author}{\bibfnamefont{A.}~\bibnamefont{Santos}},
  \bibinfo{author}{\bibfnamefont{S.~B.} \bibnamefont{Yuste}}, \bibnamefont{and}
  \bibinfo{author}{\bibfnamefont{M.~L.} \bibnamefont{de~Haro}},
  \bibinfo{journal}{J. Chem. Phys.} \textbf{\bibinfo{volume}{117}},
  \bibinfo{pages}{5785} (\bibinfo{year}{2002}).

\end{thebibliography}

\bigskip

\end{document}